\documentclass[11pt,a4paper]{article}
\pdfoutput=1 % if your are submitting a pdflatex (i.e. if you have
\usepackage{jheppub}
\usepackage{pdflscape}

\usepackage[utf8]{inputenc}
\usepackage[T1]{fontenc}
\usepackage{lmodern}
\usepackage{fontawesome}
\usepackage{hyperref}
\usepackage{tikz}
\usepackage{subcaption}
\usepackage{mathtools}
\usepackage{xcolor}
\usepackage{amsmath,amssymb}
\usepackage{units}
\usepackage{url}

\preprint{ \vbox{\hbox{IPARCOS-UCM-25-055}}}
 
\renewcommand{\arraystretch}{1.3}

\usepackage{pgf}
\usepackage{relsize}

\newcommand{\fabian}[1]{{\color{blue} \textbf{Fabian:}  #1}}

\newcommand{\VS}[1]{{\color{purple} \textbf{Vero:}  #1}}
\newcommand{\Bt}{\tilde{B}}
\newcommand{\Wt}{\tilde{W}}
\newcommand{\Gt}{\tilde{G}}

\usepackage{tikz-feynman} 

\usetikzlibrary{positioning}
\usetikzlibrary{automata,positioning,arrows,matrix,backgrounds,calc}
\usetikzlibrary{decorations.text}
\usetikzlibrary{decorations.pathmorphing}
\usetikzlibrary{shapes.geometric}
\usetikzlibrary{arrows.meta}

 \tikzset{
     ->-/.style={decoration={
   markings,
   mark=at position .65 with {\arrow{Latex}}},postaction={decorate}},
     -<-/.style={decoration={
   markings,
   mark=at position .6 with {\arrow{<}}},postaction={decorate}},
    ->/.style={decoration={
   markings,
   mark=at position .4 with {\arrow{>}}},postaction={decorate}},
 }

\title{A global analysis of ALP-mediated multiboson production at the LHC}

\author[a]{Fabian Esser,}
\author[b]{Alexandre Salas-Bern\'ardez,}
\author[c]{Veronica Sanz}
\author[d]{and Maria Ubiali}

\affiliation[a]{Institute of Particle and Nuclear Physics (IPNP), Faculty of Mathematics and Physics, Charles University Prague, V Hole\v{s}ovi\v{c}k\'{a}ch 2, 180 00 Praha 8, Czech Republic}
\affiliation[b]{Dept.  Análisis Matemático y Matemática Aplicada and IPARCOS, Univ. Complutense de Madrid, Plaza de las Ciencias 3, 28040 Madrid, Spain}
\affiliation[c]{Instituto de F\'isica Corpuscular (IFIC), Universidad de Valencia-CSIC, E-46980 Valencia, Spain}
\affiliation[d]{DAMTP, University of Cambridge, Wilberforce Road, Cambridge CB3 0WA, UK }

\emailAdd{fabian.esser@matfyz.cuni.cz}
\emailAdd{alexsala@ucm.es}
\emailAdd{veronica.sanz@uv.es}
\emailAdd{m.ubiali@damtp.cam.ac.uk}

\newcommand{\madgraph}{{\tt MadGraph5\_aMC@NLO}\,\,}

\abstract{Axion-like particles (ALPs) provide a well-motivated framework for physics beyond the Standard Model, coupling to gauge bosons through 
dimension-five operators protected by an approximate shift symmetry. At the LHC, such interactions lead to distinctive signatures in multiboson 
production, where the ALP appears as an off-shell mediator rather than a narrow resonance. In this work, we present the first global analysis of 
ALP-mediated multiboson processes, combining measurements of diphoton, $ZZ$, $W^+W^-$, dijet, and vector-boson–fusion final states. On the theory 
side, motivated from a UV perspective, we assume that the ALP couples only to the gauge sector of the SM, and classify the ALP–multiboson vertices 
that directly modify collider observables. Using Run-2 LHC measurements, we extract bounds on the Wilson coefficients $(c_{\Gt},c_{\Wt},c_{\Bt})$ 
that parametrise gluonic and electroweak ALP interactions. Our results show that the dijet channel dominates the sensitivity to ALP couplings 
and determines the limits on $c_{\Gt}$,  while diboson and VBF processes provide complementary constraints on the electroweak couplings. 
We further assess the validity of the EFT expansion at the multi-TeV scales probed by the data. This global study provides the most comprehensive 
picture to date of ALP–gauge interactions from multiboson production at the LHC, and highlights the opportunities for significant improvements with future high-luminosity 
measurements.}
\begin{document}

\maketitle

\section{Introduction}
\label{sec:intro}

Axion-like particles (ALPs) are among the most compelling candidates for physics beyond the Standard Model (SM). While the original Peccei–Quinn axion was proposed as a solution to the strong CP problem~\cite{Peccei:1977hh, Weinberg:1977ma, Wilczek:1977pj}, generic ALPs arise naturally in a broad class of ultraviolet completions, including string 
compactifications~\cite{Svrcek:2006yi, Arvanitaki:2009fg} and composite Higgs scenarios~\cite{, Contino:2011np, Ferretti:2013kya, Sanz:2015sua}. These pseudo-Nambu–Goldstone 
bosons inherit shift-symmetry protection, and couple to the SM gauge bosons through dimension-five operators suppressed by a high scale $f_a$. An up-to-date overview can be found in Ref.~\cite{Choi:2020rgn}, while a comprehensive public repository of constraints on ALP parameters — spanning high-energy, nuclear and atomic physics, as well as cosmology and 
astrophysics — is provided in Ref.~\cite{AxionLimits}.

At high-energy colliders, such couplings open up novel production mechanisms. In particular, ALP-mediated processes can lead to striking signatures involving multiple gauge bosons in the final state. Multiboson production is especially interesting for several reasons. First, these final states are rare within the SM, and therefore constitute sensitive probes of new physics. Second, the structure 
of ALP–gauge couplings leads to irreducible contact interactions that directly generate multi-gauge-boson vertices. Third, measurements of vector boson scattering and multi-boson production have become increasingly precise at the Large Hadron Collider (LHC), providing an ideal testing ground for this framework.
%~\cite{ATLAS:diboson, CMS:diboson, Bonilla:2022, ATLAS:2024ZZ}.

Previous work has analysed individual ALP off-shell signatures at the LHC, focusing on di-photon, massive di-boson, or vector-boson–scattering 
processes~\cite{Brivio:2017ije, Bauer:2017ris, Gavela:2019cmq,Feng:2025kof}, as well as ALP couplings to the top quark~\cite{Esser:2023fdo,Butterworth:2025szb, Hosseini:2024kuh, Bisal:2025jwv, Barbosa:2025zyn}, to gluons~\cite{Gavela:2019cmq,Butterworth:2025szb}, and to di-Higgs final states~\cite{Esser:2024pnc}.
 However, a comprehensive treatment of ALP-mediated multiboson production combining different channels has not yet been undertaken, or only partially as in Ref.~\cite{Bonilla:2022pxu} where vector boson fusion channels were explored. Moreover, in much of the existing literature, assumptions are often made about the gluon coupling, since gluon-induced production is typically dominant whenever it is present. Yet such assumptions can obscure the interplay between different effective operators. A global approach is therefore essential: only by combining complementary observables can one disentangle the underlying parameter space and truly probe the range of scenarios that are consistent with data.

In this paper we provide a systematic study of axion-mediated multiboson production at the LHC. Working in the linear ALP effective field theory, we classify the irreducible ALP–multiboson vertices, identify the relevant collider signatures, and confront them with existing LHC measurements. We present a global fit including di-photon, $ZZ$,  $W^+W^-$, dijet and vector-boson–fusion final states, and reinterpret the bounds in terms of the linear ALP EFT operators. Our analysis delivers the most complete picture to date of the constraints on ALP–gauge interactions from multiboson processes, and highlights the prospects for future improvements at the high-luminosity LHC.
Throughout this work we focus on a gauge-ALP benchmark scenario, which
allows us to isolate the irreducible multiboson signatures of ALP exchange
without committing to a specific ultraviolet completion.

The paper is organised as follows. In Section~\ref{sec:theory} we introduce the effective Lagrangian for ALPs, describe their couplings to the SM gauge bosons, and discuss possible UV origins of such interactions. Section~\ref{sec:global} presents our collider analysis: we first study the individual production channels — diphoton, $ZZ$, $W^+W^-$, dijet, and vector-boson–fusion final states — and then combine them into a global fit. Section~\ref{sec:results} reports the limits obtained under different statistical prescriptions (projection, profiling, marginalisation) and discusses EFT validity. We summarise our findings and outline future directions in Section~\ref{sec:conclusions}. Finally, in Appendix \ref{sec:interference} we discuss interference effects and we list the selection cuts for our signal simulation in Appendix \ref{app:cuts}.

\section{Theoretical set-up}
\label{sec:theory}

We work within the framework of the linear ALP effective field theory, in which the axion-like particle $a$ is introduced as a pseudo-Nambu–Goldstone 
boson coupling to the SM gauge bosons through dimension-five operators. We consider an ALP coupled exclusively at leading order to the gauge sector 
of the SM, a scenario that we will motivate from an explicit UV completion in Sec.~\ref{sec:modelbuilding}. 
In this scenario and at leading order, the relevant Lagrangian reads
\begin{equation}
\mathcal{L}_{\rm ALP} \;=\; 
\frac{a}{f_a} \Big[ 
c_{\tilde G}\, G^a_{\mu\nu} \tilde{G}^{a\,\mu\nu} 
+ c_{\tilde W}\, W^i_{\mu\nu} \tilde{W}^{i\,\mu\nu} 
+ c_{\tilde B}\, B_{\mu\nu} \tilde{B}^{\mu\nu} 
\Big],
\label{eq:LALP}
\end{equation}
where $f_a$ denotes the ALP decay constant, $c_{\tilde X}$ the dimensionless Wilson coefficients, 
and $\tilde{X}_{\mu\nu} = \tfrac{1}{2}\,\epsilon_{\mu\nu\rho\sigma} X^{\rho\sigma}$ the dual field-strength tensor for $X=G,W,B$. 
%\fabian{add references?}\VS{ we already did in the intro}

After electroweak symmetry breaking, these interactions translate into couplings of the ALP to gluons, photons, and massive electroweak bosons,
\begin{align}
\label{eq:Lgauge_broken}
\mathcal{L}_{\rm ALP} = - \frac{a}{f_a} \Big(
& g_{agg} \, G_{\mu\nu} \tilde{G}^{\mu\nu}
+ g_{aWW} \, W_{\mu\nu} \tilde{W}^{\mu\nu}
+ g_{aZZ} \, Z_{\mu\nu} \tilde{Z}^{\mu\nu} \nonumber \\
&+ g_{a\gamma Z} \, F_{\mu\nu} \tilde{Z}^{\mu\nu}
+ g_{a\gamma\gamma} \, F_{\mu\nu} \tilde{F}^{\mu\nu}
\Big) \ .
\end{align}
Explicitly, one obtains
\begin{align}\label{eq:EWSB}
g_{a\gamma\gamma} &= \frac{4}{f_a}\left(c_{\tilde B}\cos^2\theta_W + c_{\tilde W}\sin^2\theta_W\right),\qquad
g_{aZ\gamma} = \frac{4}{f_a}\sin\theta_W\cos\theta_W\left(c_{\tilde W}-c_{\tilde B}\right), \nonumber\\
g_{aZZ} &= \frac{4}{f_a}\left(c_{\tilde W}\cos^2\theta_W + c_{\tilde B}\sin^2\theta_W\right),\qquad
g_{aWW} = \frac{4}{f_a}\,c_{\tilde W},\qquad
g_{agg} = \frac{4}{f_a}\,c_{\tilde G},
\end{align}

In this work, we restrict ourselves to bosonic ALP couplings and neglect direct couplings to fermions. For most SM fermions these operators are phenomenologically 
uninteresting at the LHC, as their effects are either suppressed or redundant. The exception is the top quark, whose large Yukawa coupling and role in loop-induced 
processes make it especially relevant. We have studied ALP–top interactions in previous works~\cite{Esser:2023fdo,Butterworth:2025szb}, and here we neglect them 
at tree level but note that they can contribute at loop level. In the next section, we describe a UV scenario in which no direct couplings 
to fermions arise, consistently with the possibility that the ALP does not couple to the electroweak symmetry-breaking sector.

The interactions above generate contact interaction terms involving both two and three gauge bosons, as listed in Table~\ref{tab:ALP_multiboson_linear}. 
In our analysis we focus on the two-boson vertices, which currently provide the strongest experimental sensitivity. Channels with three gauge 
bosons — akin to those targeted in anomalous quartic gauge coupling (aQGC) searches — are not yet competitive, but they are expected to gain 
relevance with the increased luminosity of Run~3 and beyond, and would represent a natural extension of our present study.

%Here, ``irreducible’’ refers to genuine contact interactions that arise directly from the dimension-five operators in Eq.~\eqref{eq:LALP}, where all external gauge bosons couple at a single vertex. This is in contrast to reducible diagrams, in which multiboson final states are produced through intermediate propagators connecting two or more lower-point interactions (for example, two ALP–diboson vertices connected by an off-shell ALP) \fabian{I'm not sure if I understand the sentence in parentheses correctly. The process with an intermediate off-shell ALP is exactly the signal process that we want. What is excluded by irreducible are processes with intermediate SM propagators, e.g. $a \rightarrow g g g$ via $a \rightarrow g g \rightarrow g g g$ with an internal triple gluon vertex}. The irreducible vertices therefore represent the minimal building blocks of ALP-induced multiboson production in the linear EFT.

\begin{table}[th!]
\centering
\renewcommand{\arraystretch}{2.3}
\begin{tabular}{|c|c|}
\hline
Vertex & Feynman rule (linear ALP EFT) \\
\hline\hline
$a\gamma\gamma$ & $-\dfrac{4i}{f_a} \left( \cos^2\theta_W\,c_{\tilde B} + \sin^2\theta_W\,c_{\tilde W} \right) \epsilon^{\mu \nu \rho \sigma}  (p_{\gamma1})_{\rho} (p_{\gamma2})_{\sigma}$ \\
$a\gamma Z$ & $\dfrac{2i}{f_a} \sin(2\theta_W) ( c_{\tilde B} -  c_{\tilde W} ) \epsilon^{\mu \nu \rho \sigma}  (p_{Z})_{\rho} (p_{\gamma})_{\sigma}$ \\
$aZZ$ & $-\dfrac{4i}{f_a} \left( \sin^2\theta_W\,c_{\tilde B} + \cos^2\theta_W\,c_{\tilde W} \right) \epsilon^{\mu \nu \rho \sigma}  (p_{Z1})_{\rho} (p_{Z2})_{\sigma}$\\
$aW^+W^-$ & $-\dfrac{4i}{f_a} c_{\tilde W}\, \epsilon^{\mu \nu \rho \sigma}  (p_{W^+})_{\rho} (p_{W^-})_{\sigma}$ \\
$agg$ & $-\dfrac{4i}{f_a} c_{\tilde G}\, \delta_{ab} \epsilon^{\mu \nu \rho \sigma}  (p_{g1})_{\rho} (p_{g2})_{\sigma}$ \\
\hline 
$a\gamma W^+W^-$ & $\dfrac{4ie}{f_a} c_{\tilde W}\, \epsilon^{\mu\nu\rho\sigma} (p_{WW\gamma})_{\sigma}$\\
$aZW^+W^-$ & $\dfrac{4ie}{f_a} \dfrac{\cos\theta_W}{\sin\theta_W} c_{\tilde W}\, \epsilon^{\mu\nu\rho\sigma} (p_{WWZ})_{\sigma}$\\ 
$aggg$ & $\dfrac{4}{f_a} g_s c_{\tilde G} f_{abc}\, \epsilon^{\mu\nu\rho\sigma} ( p_{ggg})_{\sigma}$ \\
\hline
\end{tabular}
\caption{Irreducible tree-level ALP–multiboson vertices in the linear EFT. For the three-boson final states we define $p_{WW\gamma} = p_{W^+} + p_{W^-} + p_{\gamma}$, $p_{WWZ} = p_{W^+} + p_{W^-} + p_{Z}$ and $p_{ggg} = p_{g1} + p_{g2} + p_{g3}$.
%\fabian{Do we want to keep the ALP-VVV couplings? We do not use them (explicitely) in the analysis.}\VS{why not keep them, and then in the conclusions we talk about quartic anomalous as a possible next direction?}
}
\label{tab:ALP_multiboson_linear}
\end{table}

\subsection{Model-building a gauge--ALP coupling}
\label{sec:modelbuilding}

We consider an axion-like particle (ALP), $a$, as a light degree of freedom emerging from the spontaneous breaking of an approximate global $U(1)_X$ at a scale $f_a$. 
The associated shift symmetry $a \to a + c$ (with $c$ constant) protects the mass and interactions of $a$, and anomalous couplings to the SM gauge 
fields arise when the $U(1)_X$ current has mixed anomalies with $SU(3)_C\times SU(2)_L \times U(1)_Y$. After integrating out heavy states responsible for the anomaly, 
the low-energy effective interaction takes the form given in Eq.~(\ref{eq:LALP}) 
with coefficients determined by anomaly matching, which schematically can be written as $c_{\tilde X}\propto \text{Tr}[Q_X\,T_X^2]$ 
for heavy fermions carrying the global charge $Q_X$ and gauge generators $T_X$. This structure is the hallmark of UV completions where the ALP couples to the 
SM \emph{only} through gauge anomalies. 

Specifically, let $U(1)_X$ be an approximate global symmetry spontaneously broken below the scale $f_a$, with $a$ the corresponding pseudo-Nambu-Goldstone boson (pNGB). 
Take a set of heavy vector-like fermions $\Psi$ carrying $U(1)_X$ charge $q_X^\Psi$ and transforming under 
$SU(3)_C\times SU(2)_L\times U(1)_Y$ in representations $R_3^\Psi$, $R_2^\Psi$ and hypercharge $Y_\Psi$. 
In the UV the ALP couples derivatively to the $U(1)_X$ current,
\begin{equation}
\mathcal{L} \supset \frac{\partial_\mu a}{f_a}\, J_X^\mu, 
\qquad 
J_X^\mu = \sum_{\Psi} q_X^\Psi\, \bar\Psi \gamma^\mu \Psi \;.
\end{equation}
The $U(1)_X$
current is anomalous with the SM gauge groups,
\begin{eqnarray}
\partial_\mu J_X^\mu 
&=& \sum_{\Psi}
\Bigg[
\frac{g_s^2}{16\pi^2}\, 2\, q_X^\Psi\, T\!\left(R_3^\Psi\right)\, G^a_{\mu\nu}\tilde G^{a\,\mu\nu} \nonumber \\
& & +\frac{g^2}{16\pi^2}\, 2\, q_X^\Psi\, T\!\left(R_2^\Psi\right)\, W^i_{\mu\nu}\tilde W^{i\,\mu\nu}  
 +\frac{g'^2}{16\pi^2}\, 2\, q_X^\Psi\, \mathcal{Y}_\Psi^2\, B_{\mu\nu}\tilde B^{\mu\nu}
\Bigg],
\label{eq:divJ}
\end{eqnarray}
where $T(R)$ is the Dynkin index 
%\textcolor{red}{is this just the trace of two matrices of the group representation?}\VS{if they are in a diagonal rep, yes.} 
(\(T(\mathbf{3})=\tfrac12\), \(T(\mathbf{2})=\tfrac12\), \(T(\mathbf{1})=0\)) and 
\(\mathcal{Y}_\Psi\) denotes the hypercharge with the convention \(Q=T_3+Y\).
%if you use GUT normalisation, insert the usual \(3/5\) factor accordingly).

\begin{figure}[t]
\centering
\begin{tikzpicture}
  \begin{feynman}
    % external legs
    \vertex (a) at (-2,0) {\(a\)};         % ALP
    \vertex (x1) at (4, 1.2) {\(V_1\)};      % gauge boson 1
    \vertex (x2) at (4,-1.2) {\(V_2\)};      % gauge boson 2

    % triangle loop vertices
    \vertex (v1) at (0,0);
    \vertex (v2) at (2, 1.2);
    \vertex (v3) at (2,-1.2);

    % draw diagram
    \diagram*{
      (a)  -- [scalar] (v1),

      % closed fermion loop with consistent arrow flow
      (v1) -- [fermion, edge label=\(\Psi\)] (v2)
           -- [fermion, edge label=\(\Psi\)] (v3)
           -- [fermion, edge label=\(\Psi\)] (v1),

      % gauge boson attachments (pick one style)
      (v2) -- [boson, momentum=\(p_1\)] (x1),
      (v3) -- [boson, momentum=\(p_2\)] (x2),
    };
  \end{feynman}
\end{tikzpicture}
\caption{Representative anomaly triangle generating a \(a\,F\tilde F\) term in the effective Lagrangian
 with heavy fermions \(\Psi\).}
\label{fig:ALPtriangle}
\end{figure}
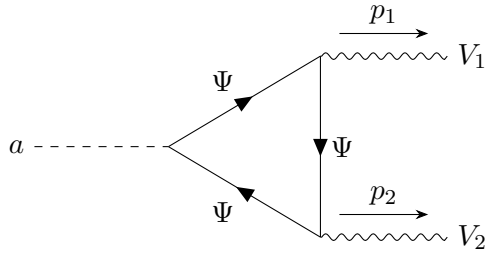

Integrating out the heavy fermions as in Fig.~\ref{fig:ALPtriangle} or, equivalently, performing a chiral rotation that removes the derivative coupling 
\(\big(\partial_\mu a/f_a\big)\bar\Psi\gamma^\mu\Psi\),\footnote{Start from
\(\mathcal{L}\supset \frac{\partial_\mu a}{f_a}\, \bar\Psi \gamma^\mu (q_X)\Psi\).
Perform the field redefinition 
\(\Psi\to \exp\!\big[i(q_X a/f_a)\gamma_5\big]\Psi\).
This removes the derivative term but generates, via the anomalous Jacobian,
\[
\Delta \mathcal{L} 
= \frac{a}{f_a}\sum_X \frac{g_X^2}{16\pi^2}\, 2 q_X\, \mathcal{I}_X\, X_{\mu\nu}\tilde X^{\mu\nu},
\]
which is precisely Eq.~\eqref{eq:WZ}.} the Fujikawa Jacobian generates the Wess-Zumino term
\begin{equation}
\mathcal{L}_{\rm WZ}
= \frac{a}{f_a}
\left[
\frac{g_s^2}{16\pi^2}\, \mathcal{A}_3\, G^a_{\mu\nu}\tilde G^{a\,\mu\nu}
+\frac{g^2}{16\pi^2}\, \mathcal{A}_2\, W^i_{\mu\nu}\tilde W^{i\,\mu\nu}
+\frac{g'^2}{16\pi^2}\, \mathcal{A}_1\, B_{\mu\nu}\tilde B^{\mu\nu}
\right],
\mathcal{A}_X \equiv 2\sum_{\Psi} q_X^\Psi\, \mathcal{I}_X^\Psi,
\label{eq:WZ}
\end{equation}
with \(\mathcal{I}_3^\Psi=T(R_3^\Psi)\), \(\mathcal{I}_2^\Psi=T(R_2^\Psi)\), and \(\mathcal{I}_1^\Psi=\mathcal{Y}_\Psi^2\).
Comparing Eq.~\eqref{eq:WZ} with our low-energy parameterisation in Eq.~\eqref{eq:LALP},
\[
\mathcal{L}_{\rm ALP}=\frac{a}{f_a}\left[c_{\tilde G}\, G\tilde G+c_{\tilde W}\, W\tilde W+c_{\tilde B}\, B\tilde B\right],
\]
we identify
\begin{equation}
c_{\tilde G}=\frac{g_s^2}{16\pi^2}\,\mathcal{A}_3,
\qquad
c_{\tilde W}=\frac{g^2}{16\pi^2}\,\mathcal{A}_2,
\qquad
c_{\tilde B}=\frac{g'^2}{16\pi^2}\,\mathcal{A}_1.
\label{eq:cMatching}
\end{equation}
Equivalently, one may write the schematic relation stated in the text,
\(c_{\tilde X}\propto \text{Tr}\big[Q_X\,T_X^2\big]\), with the constants of proportionality fixed by Eq.~\eqref{eq:cMatching}.

For example, for a single Dirac fermion \(\Psi\) in \((R_3,R_2)_Y\) with charge \(q_X\) we get
\begin{equation}
c_{\tilde G}=\frac{g_s^2}{16\pi^2}\, 2 q_X\, T(R_3),\quad
c_{\tilde W}=\frac{g^2}{16\pi^2}\, 2 q_X\, T(R_2),\quad
c_{\tilde B}=\frac{g'^2}{16\pi^2}\, 2 q_X\, Y^2.
\end{equation}
If several vector-like fermions contribute, the \(\mathcal{A}_i\) simply add. 
Colour-neutral  heavy spectra can yield \(c_{\tilde G}=0\) while keeping \(c_{\tilde W},c_{\tilde B}\neq 0\); 
conversely, coloured heavy states naturally generate \(c_{\tilde G}\neq 0\). 
Alignment among charges/representations can suppress particular linear combinations, 
leading to photophobic-like patterns where \(g_{a\gamma\gamma}\propto c_{\tilde B}\cos^2\theta_W+c_{\tilde W}\sin^2\theta_W\) is small at leading order. See, e.g., Refs.~\cite{AlonsoAlvarez:2019cgw,Biekotter:2025fll} for discussions connecting UV charge assignments to the pattern of electroweak ALP couplings, and string/composite realisations with Wess–Zumino–Witten origins~\cite{Agrawal:2022lsp,Agrawal:2024ejr}. Systematic classifications of anomaly-induced electroweak ALP couplings and their dependence on the heavy fermion spectrum can be found in Ref.~\cite{Alonso-Alvarez:2018irt}.

In particular, a simple and economical realisation employs vector-like fermions $\Psi$ that are charged under the SM gauge group and carry $U(1)_X$ charge, while all {SM} fermions and the Higgs boson are neutral under $U(1)_X$. When the heavy $\Psi$ are integrated out, Eq.~\eqref{eq:LALP} is generated with calculable $c_{\tilde X}$, but there are \emph{no} tree-level derivative couplings $ (\partial_\mu a)\,\bar f\gamma^\mu\gamma_5 f$ to SM fermions, and no portal to the electroweak symmetry-breaking (EWSB) sector. In particular, there is no $a$–Higgs boson mixing and no direct $a\bar f f$ or $a H^\dagger H$ operators at leading order. This “gauge–ALP” set-up is the UV rationale for our working assumption in the analysis: we focus on bosonic couplings and neglect direct fermionic ones.
%\footnote{Even if vanishing at the UV scale, renormalisation-group running and electroweak loops can radiatively induce small fermion couplings; these effects are subleading for our collider observables and are discussed in Section~\ref{subsec:TOPloops}. See, e.g., \cite{Bauer:2021rge,Galda:2023gminus2}. \textcolor{red}{cannot find these}} 

Different UV choices for the heavy spectrum and charge assignments select distinctive patterns among $(c_{\tilde G},c_{\tilde W},c_{\tilde B})$. For instance, if the heavy fermions are coloured, one naturally obtains $c_{\tilde G}\neq 0$ together with electroweak couplings; if they are colour-neutral, gluonic interactions can be absent while $c_{\tilde W}$ and/or $c_{\tilde B}$ remain sizeable, leading to electroweak-dominated phenomenology. Alignment symmetries can further suppress particular combinations, such as the “photophobic” limit in which the $a\gamma\gamma$ coupling cancels at leading order while $aZ\gamma$, $aZZ$, and $aW^+W^-$ remain nonzero, a construction explored  in collider contexts~\cite{Craig:2018kne}. These scenarios emphasise the importance of combining several multiboson channels rather than assuming a fixed hierarchy among the couplings. 

The effective couplings to mass eigenstates after EWSB follow from Eq.~\eqref{eq:LALP} and electroweak mixing, Eq.~(\ref{eq:EWSB}),
which directly control the irreducible contact vertices and the off-shell ALP exchange entering our multiboson analysis. 

We emphasise that the gauge-ALP scenario adopted here should be understood
as a well-defined benchmark rather than as the most general ALP effective
theory. In generic ultraviolet completions, additional operators (such as
ALP-fermion couplings or Higgs-portal interactions) may be generated at
loop level. In the class of constructions discussed above, however, these
effects are parametrically suppressed and do not play a leading role in the
multiboson observables considered in this work, which are dominantly
controlled by the anomaly-induced gauge couplings.
They arise only radiatively and are typically suppressed by loop factors and/or
additional heavy-mass insertions. As a result, their impact on the high-energy
multiboson observables analysed in this work is expected to be subleading.

From a phenomenological perspective, these operators primarily affect
processes involving fermionic final states, Higgs production, or low-energy
precision observables, rather than the purely bosonic channels that dominate
the sensitivity of our global analysis. While they may play an important role
in complementary probes of ALP physics, they do not qualitatively modify the
interpretation of the multiboson constraints derived here, which are driven by
off-shell ALP exchange through gauge interactions.

A concrete illustration of the phenomenological impact of such loop-induced
operators arises at low energies. Even if direct ALP-fermion couplings are
absent at tree level, gauge-ALP interactions generate them radiatively under
renormalisation-group evolution. At scales below the electroweak scale, these
induced couplings can mediate flavour-changing processes such as
$K \to \pi a$ or $B \to K a$, contribute to anomalous magnetic moments and
electric dipole moments, or lead to signals in beam-dump and fixed-target
experiments. Those effects were discussed in a previous work \cite{Esser:2023fdo} and are only relevant in a small mass parameter space. 

These observables provide some of the most stringent constraints on light
ALPs in complementary regions of parameter space. At the same time, they probe
a different set of effective interactions than the high-energy multiboson
processes considered in this work. As a result, while loop-induced ALP--fermion
and Higgs-portal operators can be crucial for low-energy phenomenology, they do
not qualitatively affect the interpretation of LHC multiboson data, which is
dominated by off-shell ALP exchange through gauge couplings.

\section{Global analysis of ALP-mediated multiboson production} \label{sec:global}

An important feature of the multiboson signatures considered here is that the ALP is exchanged \emph{off-shell}. 
In contrast to standard resonant searches, where one looks for a narrow peak in an invariant-mass distribution, 
our analysis does not rely on kinematic cuts around the ALP mass $m_a$. Instead, the ALP enters as a virtual mediator 
that modifies the tails of SM distributions, as depicted in Fig.~\ref{fig:ALPoffshell}. 
\begin{figure}[h!]

\centering
% needs: \usepackage{tikz-feynman} in preamble
\tikzfeynmanset{compat=1.1.0}

\begin{minipage}{0.47\textwidth}
\centering
\begin{tikzpicture}[scale=1.1]
  \begin{feynman}
    % Left panel: SM continuum VV -> V'V' (schematic box)
    \vertex (i1) at (-1.8,  1.0) {\(V\)};
    \vertex (i2) at (-1.8, -1.0) {\(V\)};
    \vertex (o1) at ( 1.8,  1.0) {\(V'\)};
    \vertex (o2) at ( 1.8, -1.0) {\(V'\)};

    \vertex (a1) at (-0.6,  0.6);
    \vertex (a2) at ( 0.6,  0.6);
    \vertex (a3) at ( 0.6, -0.6);
    \vertex (a4) at (-0.6, -0.6);

    \diagram*{
      (i1) -- [boson, momentum=\(p_1\)] (a1) -- [fermion, edge label'=\(\mathrm{SM}\)] (a2) -- [boson, momentum=\(k_1\)] (o1),
      (i2) -- [boson, momentum'=\(p_2\)] (a4) -- [fermion] (a3) -- [boson, momentum'=\(k_2\)] (o2),
      (a1) -- [fermion] (a4),
      (a2) -- [fermion] (a3),
    };
  \end{feynman}
\end{tikzpicture}

\vspace{1ex}

\small (a) SM continuum (no ALP resonance)
\end{minipage}
\hfill
\begin{minipage}{0.47\textwidth}
\centering
\begin{tikzpicture}[scale=1.1]
  \begin{feynman}
    % Right panel: off-shell ALP in s-channel V V -> a* -> V' V'
    \vertex (i1) at (-1.8,  1.0) {\(V\)};
    \vertex (i2) at (-1.8, -1.0) {\(V\)};
    \vertex (o1) at ( 1.8,  1.0) {\(V'\)};
    \vertex (o2) at ( 1.8, -1.0) {\(V'\)};

    \vertex (c1) at (-0.4, 0);
    \vertex (c2) at ( 0.4, 0);

    \diagram*{
      (i1) -- [boson, momentum=\(p_1\)] (c1),
      (i2) -- [boson, momentum'=\(p_2\)] (c1),
      (c1) -- [scalar, edge label=\(\,a^\ast\)] (c2),
      (c2) -- [boson, momentum=\(k_1\)] (o1),
      (c2) -- [boson, momentum'=\(k_2\)] (o2),
    };
  \end{feynman}
\end{tikzpicture}

\vspace{1ex}

\small (b) Off-shell ALP exchange in the \(s\)-channel
\end{minipage}

\caption{Schematic parton-level topologies for multiboson production. 
Left: SM continuum (box/topology example). Right: virtual ALP exchange with no resonant cut. Our analysis targets this non-resonant regime). 
Here \(V,V'\in\{g,\gamma,Z,W\}\).}
\label{fig:ALPoffshell}
\end{figure}

This distinction is crucial as it allows us to probe light ALPs that 
would otherwise evade conventional resonance searches due to selection cuts. In particular, for $m_a \ll \sqrt{\hat{s}}$ with $\hat{s}$ the squared centre 
of mass energy of the parton-level process, the ALP exchange leads 
to boosted gauge-boson final states, whose kinematic features can still be exploited experimentally even though no 
resonant structure is present.

Because of this off-shell character, the relevant observables are inclusive multiboson production rates and 
differential distributions, rather than resonance bumps. In practice, the absence 
of resonance cuts means that the sensitivity depends primarily on the effective couplings $c_{\tilde X}$ and only 
weakly on the ALP mass, provided $m_a \ll \sqrt{\hat{s}}$. 

For example, at the parton level, the $s$–channel amplitude for $V_1 V_2 \to V_3 V_4$ through an intermediate ALP takes the schematic form
\begin{equation}
\mathcal{M}_a \;\sim\; 
\frac{1}{\hat s - m_a^2 + i m_a \Gamma_a}\,
\Big(\tfrac{4}{f_a}c_{V_1V_2}\,\epsilon_{\mu\nu\rho\sigma}\,p_1^\rho p_2^\sigma \varepsilon_1^\mu \varepsilon_2^\nu\Big)\,
\Big(\tfrac{4}{f_a}c_{V_3V_4}\,\epsilon_{\alpha\beta\lambda\kappa}\,k_1^\lambda k_2^\kappa \varepsilon_3^\alpha \varepsilon_4^\beta\Big),
\end{equation}
 here $\Gamma_a$ is the ALP decay width.
Squaring and averaging over polarisations one finds, %\textcolor{red}{should we include the Appendix?}, 
up to angular factors,
\begin{equation}
|\mathcal{M}_a|^2 \;\propto\;
\frac{(c_{V_1V_2}\,c_{V_3V_4})^2}{f_a^4}\,
\frac{\hat s^4}{(\hat s - m_a^2)^2 + m_a^2 \Gamma_a^2}.
\label{eq:offshellamp}
\end{equation}
In the regime of interest, $m_a^2 \ll \hat s$, the propagator reduces to $1/\hat s^2$ and the explicit $m_a$ and $\Gamma_a$ dependence drop out.
The scaling with $\hat s$ in the numerator of the amplitude can be traced back to the momentum factors in the $\epsilon_{\mu\nu\rho\sigma}$ tensors of the ALP–gauge vertices. 
This behaviour is generic as long as $m_{V_i}^2, m_a^2 \ll \hat{s}$ and follows directly from the Feynman rules in Table~\ref{tab:ALP_multiboson_linear}.
The cross section therefore scales as
\begin{equation}
\hat\sigma(V_1 V_2 \to V_3 V_4) \;\propto\;
\frac{(c_{V_1V_2}\,c_{V_3V_4})^2}{f_a^4}\,\hat s,
\label{eq:cross_section_ci}
\end{equation}
with the growth with $\hat{s}$ already discussed in Ref. \cite{Gavela:2019cmq}. 
We stress that this growth with $\hat{s}$ is a generic feature of ALP-mediated
processes in the linear EFT and reflects the derivative nature of the
$aVV$ couplings. As a consequence, the sensitivity of the LHC measurements
considered in this work is driven by the high-energy tails of kinematic
distributions. The resulting bounds should therefore be interpreted as
constraints on the effective couplings $c_{\tilde X}/f_a$ within the domain
of validity of the EFT, rather than as exclusions of specific ultraviolet
completions.

%\color{red}{where?}.
% \fabian{The growth with $\hat{s}$ from the momentum factors refers to $M_a$, i.e. the numerator in Eq.~\ \ref{eq:offshellamp}. I would say the growth in $\sigma$ comes from the phase space integration? } \textcolor{red}{phase space for two-final state particles has no $s$, the squared amplitude goes as $s^2$ (each two boson final or initial state gives $s$ and the propagator gives $s^{-1}$) divided by the flux factor we are left with $s$ as in (3.3). Did we check this scaling?} \fabian{Fair point. Then the amplitude should already scale with $\hat{s}$}

In the following subsections we explore in detail the individual production channels, including 
$\gamma\gamma$, $ZZ$, $W^+W^-$, $\gamma Z$, and multiboson final states, and discuss the current LHC 
measurements that are most sensitive to each of them. Each channel probes a different linear combination 
of the underlying couplings $c_{\tilde G}, c_{\tilde W}, c_{\tilde B}$, and their complementarity is essential 
for a consistent interpretation. In the next section we then combine these ingredients and present the 
results of our global fit, which constrains the full parameter space of ALP–gauge interactions.
Note that our global analysis combines different channels without 
requiring assumptions on a specific resonance region, making it complementary to direct ALP resonance searches. \\
In Table~\ref{tab:expdata_sm} we list a compilation of experimental measurements in multi-boson final states, the provided 
distributions and their key features. We highlight in bold the analyses that we reinterpreted in terms of the signal processes detailed below.

In all the presented analyses we set $m_a = 1$ MeV and $f_a = 1$ TeV to generate events. As discussed before, for $m_a \ll \sqrt{\hat{s}}$ 
the signal cross section is roughly independent of the ALP mass. 
Note that we are only sensitive to the ratios $\tfrac{c_{\Gt}}{f_a}$, $\tfrac{c_{\Wt}}{f_a}$ and $\tfrac{c_{\Bt}}{f_a}$, cf.\ Eq.\ (\ref{eq:offshellamp}). 
Limits on $c_{\tilde{X}}$ in this article are reported for $f_a =1$ TeV, but any other value can be obtained by using the simple rescaling $\tfrac{c_{\tilde{X}}}{f_a}$.

Throughout this work, ALP-mediated signal events are generated at leading
order and consistently normalised across all channels. Since the ALP
contribution enters as a deformation of the high-energy tails of kinematic
distributions rather than as an absolute rate measurement, the sensitivity of
the analysis is dominated by shape information and by experimental
uncertainties, rather than by higher-order corrections to the overall signal
normalisation.

\begin{landscape}
\begin{table}[h]
\small
\begin{tabular}{c|c|c|c|c|l}
\hline
final state & experiment & $\sqrt{s}$ (TeV) & luminosity  & observables & reference and data\\
\hline
\hline
$\gamma \gamma$ & CMS & 7 & $\unit[5]{fb^{-1}}$ & $d\sigma_{\gamma\gamma}/dm_{\gamma\gamma}$ &\cite{CMS:2014mvm} \href{https://www.hepdata.net/record/ins1298393}{HepData}  \\
& \textbf{ATLAS} & 13 & $\unit[139]{fb^{-1}}$ & $d\sigma_{\gamma\gamma}/dm_{\gamma\gamma}$  
& \cite{ATLAS:2021mbt} \href{https://atlas.web.cern.ch/Atlas/GROUPS/PHYSICS/PAPERS/STDM-2017-30/}{Analysis} \href{https://www.hepdata.net/record/104925}{HepData} 
\href{https://rivet.hepforge.org/analyses/ATLAS_2021_I1887997.html}{Rivet}\\
\hline

$Z Z$ & CMS & 13 &  $\unit[137]{fb^{-1}}$ & $d \sigma/m_{ZZ}$, $d \sigma/p_T^l$ & \cite{CMS:2020gtj}\href{https://atlas.web.cern.ch/Atlas/GROUPS/PHYSICS/PAPERS/STDM-2022-17/}{Analysis} \href{https://www.hepdata.net/record/ins1814609}{HepData} \\
& \textbf{ATLAS} & 13.6 & $\unit[29]{fb^{-1}}$ & $d\sigma/dm_{4l}$  &\cite{ATLAS:2023dew} \href{https://www.hepdata.net/record/144768}{HepData}\\
\hline

$W^+ W^-$& CMS & 13 & $\unit[34.8]{fb^{-1}}$ & total cross section & \cite{CMS:2024hey} \href{https://www.hepdata.net/record/ins2796231}{HepData}\\
& \textbf{ATLAS} & 13 & $\unit[36.1]{fb^{-1}}$ & $d\sigma_{WW}/dm_{e\mu}$ &  \cite{ATLAS:2019rob} \href{https://www.hepdata.net/record/ins1734263?version=1}{HepData} \href{https://rivet.hepforge.org/analyses/ATLAS_2019_I1734263}{Rivet} \\
\hline

dijet (gg) & \textbf{ATLAS} & 13 & $\unit[3.2]{fb^{-1}}$ & $d^2\sigma_{jj}/dm_{jj}dy*$ & \cite{ATLAS:2017ble} \href{https://www.hepdata.net/record/ins1634970}{HepData} \href{https://rivet.hepforge.org/analyses/ATLAS_2018_I1634970}{Rivet}  \\    
& CMS & 13 & $\unit[2.3]{fb^{-1}}$ & $d^2\sigma_{jj}/dm_{jet}dp_T$ &  \cite{CMS:2018vzn} \href{https://www.hepdata.net/record/ins1682495}{HepData} \href{https://rivet.hepforge.org/analyses/CMS_2018_I1686000}{Rivet}\\
& CMS & 13 & $\unit[2.6]{fb^{-1}}$ & $1/\sigma d^2\sigma_{jj}/d\chi d m_{jj}$ &  \cite{CMS:2017caz} \href{https://www.hepdata.net/record/ins1519995}{HepData} \href{https://rivet.hepforge.org/analyses/CMS_2017_I1519995}{Rivet}\\
& CMS & 13 & $\unit[35.9]{fb^{-1}}$ & $1/\sigma d^2\sigma_{jj}/d\chi d m_{jj}$ &  \cite{CMS:2018ucw} \href{https://www.hepdata.net/record/ins1663452}{HepData} \\
\hline

$\gamma Zjj$ (VBF) & \textbf{CMS} & 13 & $\unit[139]{fb^{-1}}$ & $\mathbf{d\sigma/dp_T^{\gamma}}$, $d\sigma/dp_T^{\rm lead,l}$, $d\sigma/p_T^{\rm lead,j}$, $d^2\sigma/dy_{jj} dm_{jj}$ & \cite{CMS:2021gme} \href{https://www.hepdata.net/record/ins1869513}{HepData} \\
\hline

$ZZ jj, W^+ W^- jj$ (VBF) & \textbf{ATLAS} & 13 & $\unit[140]{fb^{-1}}$ & Events/ 200 GeV in $m_{VV}$ bins & \cite{ATLAS:2025omi} \href{https://atlas.web.cern.ch/Atlas/GROUPS/PHYSICS/PAPERS/STDM-2018-27/}{Additional material} \\
\hline
\hline
\end{tabular}
\caption{List of experimental analyses used in this work. The searches and distributions highlighted in bold are included in our global analysis of ALP-mediated multiboson final states.}
\label{tab:expdata_sm}
\end{table}
\end{landscape}

\subsection{Diphoton final state}
\label{sec:diphoton}

The diphoton channel provides one of the cleanest probes of ALP–gauge interactions at the LHC. 
We base our study on the ATLAS measurement of inclusive diphoton production at $\sqrt{s}=13$~TeV with 
139~fb$^{-1}$~\cite{ATLAS:2021mbt}, for which the WiKi for physics analysis, the \texttt{HepData} entry 
and the corresponding \texttt{Rivet} routine are listed in Table~\ref{tab:expdata_sm}. 
The most sensitive distribution is provided by the invariant mass of the photon pair $m_{\gamma \gamma}$ \cite{ATLAS:2021mbt}. 
The differential cross section, the experimental uncertainty and different SM background estimations are given in bins ranging from $13.5$ 
GeV up to $1020$ GeV and one overflow bin up to $13$ TeV in \texttt{HepData}. \\
In order to compare the data and background with an ALP-mediated signal, we generate signal events for the process 
\[
pp \;\to\; a^\ast \;\to\; \gamma\gamma
\] 
with \madgraph~\cite{Alwall:2014hca} at LO, using {\tt NNPDF4.0}~\cite{NNPDF:2021njg} in the four-flavour scheme and the {\tt ALP\_linear\_UFO}~\cite{ALPUFO}.
We choose a benchmark scenario with $c_{\Gt} =1, c_{\Wt} =1, c_{\Bt}=0$, and note that the distributions for any other choices of $(c_{\Gt}, c_{\Wt}, c_{\Bt})$ can be easily obtained from the Feynman Rules in Tab.\ \ref{tab:ALP_multiboson_linear} and Eq.\ (\ref{eq:cross_section_ci}). 
The event selection follows the fiducial cuts specified in Ref.~\cite{ATLAS:2021mbt}. 
 \begin{figure}[th!]
    \centering
    \includegraphics[width=0.8\linewidth]{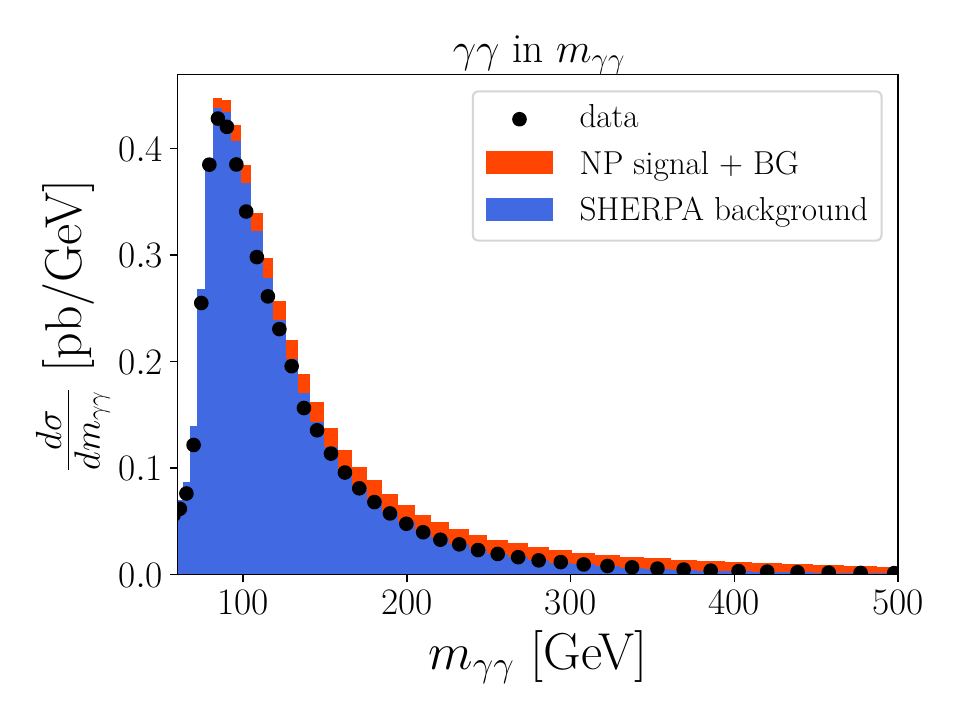}
    \caption{Diphoton invariant-mass distribution. The signal is shown for $c_{\tilde G}=1$, $c_{\tilde W}=1$ and $c_{\tilde B}=0$, compared to ATLAS data~\cite{ATLAS:2021mbt} 
    and the SM background (BG) prediction from \texttt{SHERPA}.}
    \label{fig:gammagamma}
\end{figure}
We then construct the $m_{\gamma\gamma}$ distribution using the same binning and fiducial selection cuts as in the ATLAS analysis, normalising the signal events to 
$139~\mathrm{fb}^{-1}$ with the cross section from the Les Houches Event (LHE) file  header. The SM background prediction from \texttt{SHERPA}, together with the measured data points 
and their statistical uncertainties, is taken from Ref.~\cite{ATLAS:2021mbt}, allowing for a direct, bin-by-bin comparison. 
Figure~\ref{fig:gammagamma} shows the signal contribution superimposed on the SM background and 
data, expressed in terms of the cross section per bin normalised by the bin width. The impact of the ALP is to modify 
the high-mass tail of the distribution, consistent with the off-shell character of the exchange discussed above.  

From Eq.~\eqref{eq:offshellamp}, the diphoton cross section depends on the product 
$g_{agg}\,g_{a\gamma\gamma}$. In terms of the underlying Wilson coefficients this corresponds to
\begin{equation}
g_{agg}\,g_{a\gamma\gamma} \;\propto\;
c_{\tilde G}\,\Big(c_{\tilde B}\cos^2\theta_W + c_{\tilde W}\sin^2\theta_W\Big).
\end{equation}
The diphoton channel therefore constrains only this specific combination of $c_{\tilde G},c_{\tilde W},c_{\tilde B}$, 
and cannot disentangle the individual coefficients. Independent information from other channels is required to 
lift this degeneracy, as it will be shown in the global analysis. This single channel 
cannot provide independent bounds on either coupling, but only on their product. 
This limitation highlights the importance of combining complementary channels.  

 \subsection{$ZZ$ final state}
\label{sec:ZZ}

\begin{figure}[ht!]
    \centering
    \includegraphics[width=0.8\linewidth]{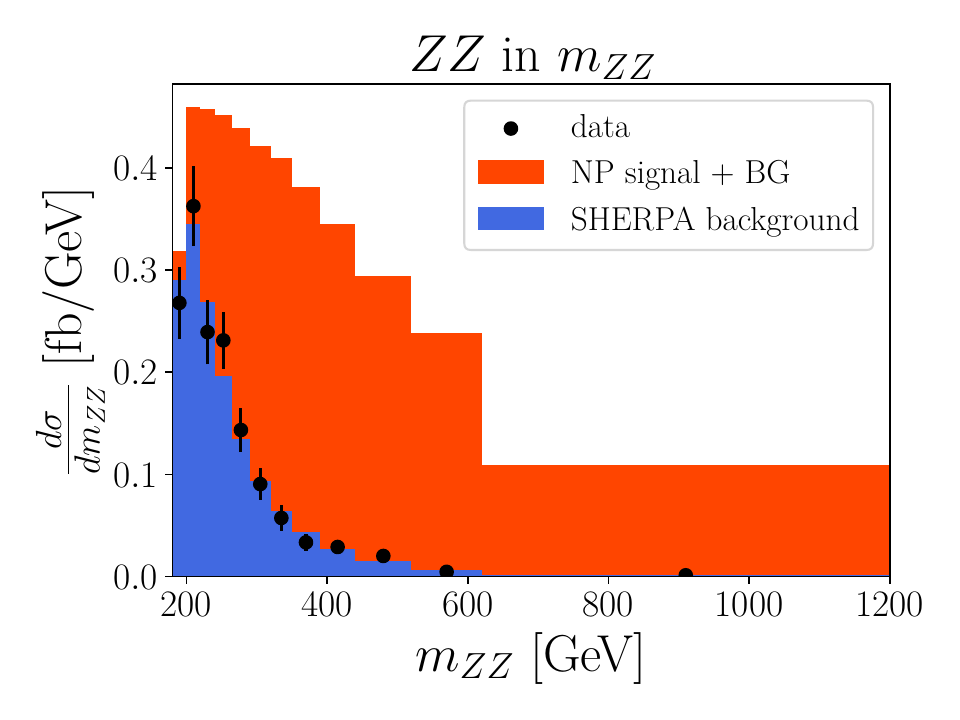}
    \caption{Differential cross section for the four-lepton invariant mass $m_{4\ell}$ in $ZZ$ production. 
    The signal is shown for $c_{\tilde G}=1$, $c_{\tilde W}=1$, $c_{\tilde B}=0$, compared to ATLAS data~\cite{ATLAS:2023dew}  
    and the SM prediction.}
    \label{fig:ZZ}
\end{figure}

The four-lepton final state provides a clean probe of $ZZ$ production, with excellent mass resolution 
and relatively low backgrounds. We base our study on the ATLAS measurement of $pp\to ZZ\to 4\ell$ at 
$\sqrt{s}=13$~TeV with 29~fb$^{-1}$~\cite{ATLAS:2023dew}. The measured differential cross section in $m_{4l}$, 
the uncertainties and the background simulation are taken from the \texttt{HepData} entry listed in Table \ref{tab:expdata_sm}, 
with bins extending from $180$ GeV to $1200$ GeV. The analysis requires both lepton pairs to fall in the $Z$-mass window, 
$50~\text{GeV}<m_{\ell\ell}<150~\text{GeV}$, and the four-lepton invariant mass in the range 
180~GeV $ <m_{4\ell}<1200$~GeV, with additional fiducial cuts on lepton rapidity and separation. \\
Signal events for the process
\[
pp \;\to\; a^\ast \;\to\; ZZ \;\to\; 4\ell
\]
and the benchmark setting $c_{\Gt} =1, c_{\Wt} =1, c_{\Bt}=0$
are generated with \madgraph, using the {\tt ALP\_linear\_UFO} model and normalised to the integrated 
luminosity of 29~fb$^{-1}$ with the cross section and acceptance derived from the LHE files. 
We reconstruct the distribution in the four-lepton invariant mass $m_{4\ell}$ using the same binning as 
the ATLAS measurement. Figure~\ref{fig:ZZ} shows the signal prediction compared to the ATLAS data and 
the SM background simulation. As expected from the off-shell ALP exchange, the main effect is a distortion 
of the high-mass tail of the $m_{4\ell}$ spectrum rather than a resonant structure.

\subsection{$W^+W^-$ final state}
\label{sec:WW}

\begin{figure}[h!]
    \centering
    \includegraphics[width=0.8\linewidth]{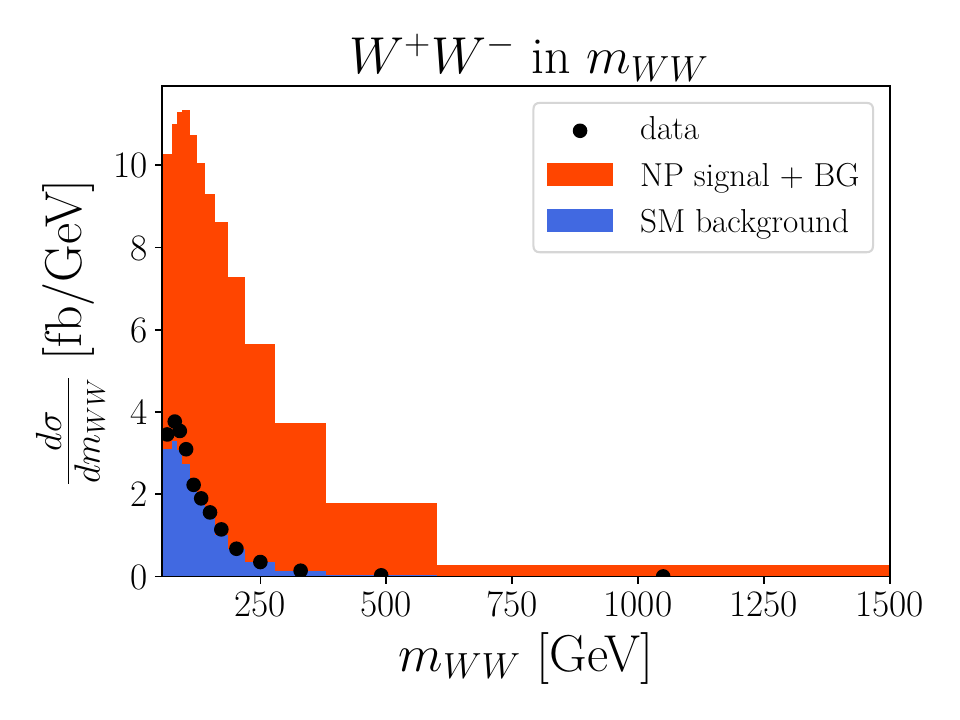}
    \caption{Differential cross section in the $WW$ final state as a function of the diboson invariant mass. The
    signal is shown for $c_{\tilde G}=1$, $c_{\tilde W}=1$ and $c_{\tilde B}=0$, compared to the ATLAS data
    and the SM background simulation in \cite{ATLAS:2019rob}.}
    \label{fig:WW}
\end{figure}

The $W^+W^-$ final state provides another clean probe of electroweak ALP couplings. 
We base our analysis on the ATLAS measurement of $pp\to W^+W^-\to \ell^+\nu \ell^-\bar\nu$ with different flavours for $\ell^-$ and $\ell^+$ at 
$\sqrt{s}=13$~TeV with an integrated luminosity of 36.1~fb$^{-1}$~\cite{ATLAS:2019rob}. 
The \texttt{HepData} entry linked in Table\ \ref{tab:expdata_sm} gives the differential cross section in the invariant mass of the final state 
leptons $m_{e\mu}$, the experimental uncertainty and the SM background simulation in bins from $55$ GeV to $1500$ GeV.
For comparison, we also note the more recent CMS analysis~\cite{CMS:2019jcb}, which we have not included in our study due to insufficient public information to reliably add the ALP signal on top of the SM results.\\ 
Signal events for $c_{\Gt} =1, c_{\Wt} =1, c_{\Bt}=0$ are generated with \madgraph, using the {\tt ALP\_linear\_UFO} model for the process
\[
pp \;\to\; a^\ast \;\to\; W^+W^- \;\to\; \ell^+\nu \ell^-\bar\nu
\]
for $\ell^+ \ell^- \in \{ e^+ \mu^-, \mu^+ e^- \} $.
The events are analysed with the same fiducial selection and $m_{WW}$ binning as the ATLAS measurement, 
with normalisation to $36.1~\text{fb}^{-1}$ using the cross section from the LHE file. 
The observable of interest, the invariant mass of the $W$-boson pair, is reconstructed from the dilepton system. \\ 
As in the $ZZ$ case, the effect of an off-shell ALP is to distort the high-mass tail of the $m_{WW}$ spectrum. 
Figure~\ref{fig:WW} shows the predicted signal superimposed on the ATLAS data and the SM background. 
This channel depends on the combination of  gluon fusion into an ALP off-shell state ($g_{agg}$) to produce a $WW$ final state, via 
\[
g_{aWW} = \frac{4}{f_a} \, c_{\tilde W},
\]
and therefore provides a direct and complementary constraint on the previous channels probing the electroweak sector of the ALP EFT.

\subsection{Dijet final state}
\label{sec:dijet}

The dijet channel probes the coupling of the ALP to gluons and provides a direct constraint on $c_{\tilde G}$. 
We base our analysis on the ATLAS measurement of the double-differential dijet cross section 
\[
\frac{d^2\sigma}{dm_{jj}\,dy^*}
\]
in the invariant mass of the jet pair, $m_{jj}$, and the half-rapidity 
separation $y^* = \tfrac{1}{2}|y_{j1}-y_{j2}|$ at $\sqrt{s}=13$~TeV with 3.2~fb$^{-1}$~\cite{ATLAS:2017ble}. 
The ATLAS analysis divides the phase space into six regions of $y^*$:
\begin{align}
&R1: 0<y^*<0.5, \quad  &R2: 0.5<y^*<1.0, \qquad &R3: 1.0<y^*<1.5, \nonumber \\
&R4: 1.5<y^*<2.0, \quad &R5: 2.0<y^*<2.5, \qquad &R6: 2.5<y^*<3.0. \nonumber
\end{align}
The $m_{jj}$ bins depend on the $y*$ region and extend up to several TeV in the most central region. 
The individual data points of the double differential distribution are extracted from the \texttt{HepData} entry 
linked in Table~\ref{tab:expdata_sm}. Note that the ATLAS analysis does not provide any background estimation. 
We take the simulation for the SM background in the same binning 
from NNLO QCD predictions in the leading-colour approximation, computed with the
{\sc NNLOjet} code~\cite{Gehrmann-DeRidder:2019ibf,NNLOJET:2025rno}. The central
factorisation and renormalisation scales where chosen as
$\mu_F=\mu_R=m_{jj}$. These predictions were
released~\cite{Britzger:2022lbf} as interpolation grids in the
{\sc APPLfast} format through the {\sc Ploughshare}
website~\cite{Ploughshareurl}, converted to the {\sc PineAPPL}
format~\cite{Carrazza:2020gss} and released along with the {\tt NNPDF} public code~\cite{NNPDF:2021uiq,NNPDF:2021njg}. 
We do not account for NLO electroweak corrections or photon-initiated contributions~\cite{Schwan:2021txc}.\\
Due to the double-differential character of the provided data, the analysis comprises 136 bins in total and taking 
into account only the diagonal uncertainties is not enough as will become clear below. Thus, we utilise the full experimental covariance 
matrix including experimental correlated systematic uncertainties, which is imported from the {\tt NNPDF} public code~\cite{NNPDF:2021uiq,NNPDF:2021njg}. 
Moreover, given that the NNLO QCD theory predictions in the large $m_{jj}$ region that most strongly constrain our bounds
are dominated by PDF uncertainties~\cite{Chen:2022tpk,Chiefa:2025loi}, we explicitly checked the impact of the inclusion of the main theory uncertainty 
of the SM background on the bounds we obtain. To do so, we added the PDF uncertainty (including full correlations among the various bins due to the PDFs) 
via the addition of a PDF theory covariance matrix to the likelihood~\cite{NNPDF:2019ubu,Ball:2021icz}. 

Signal events for $pp\to a^\ast\to gg$ and $c_{\Gt}=1$, $c_{\Wt}=0$, $c_{\Bt}=0$ are generated with \madgraph and the {\tt ALP\_linear\_UFO} model. Note that, 
since this channel depends only on $c_{\Gt}$, the choice of $c_{\Wt}$ and $c_{\Bt}$ does not matter here.
One million events are produced, which populate the $y^*$ regions unevenly: about half fall into the most 
central bin ($0<y^*<0.5$), while the most forward bin ($2.5<y^*<3$) contains only a few hundred events.  
This feature mirrors the expected kinematics of gluon fusion.  We reconstruct the double-differential cross section in the same binning as the experimental analysis (see App. \ref{app:cuts}).

\begin{figure}[h!]
    \centering
    \includegraphics[width=0.48\linewidth]{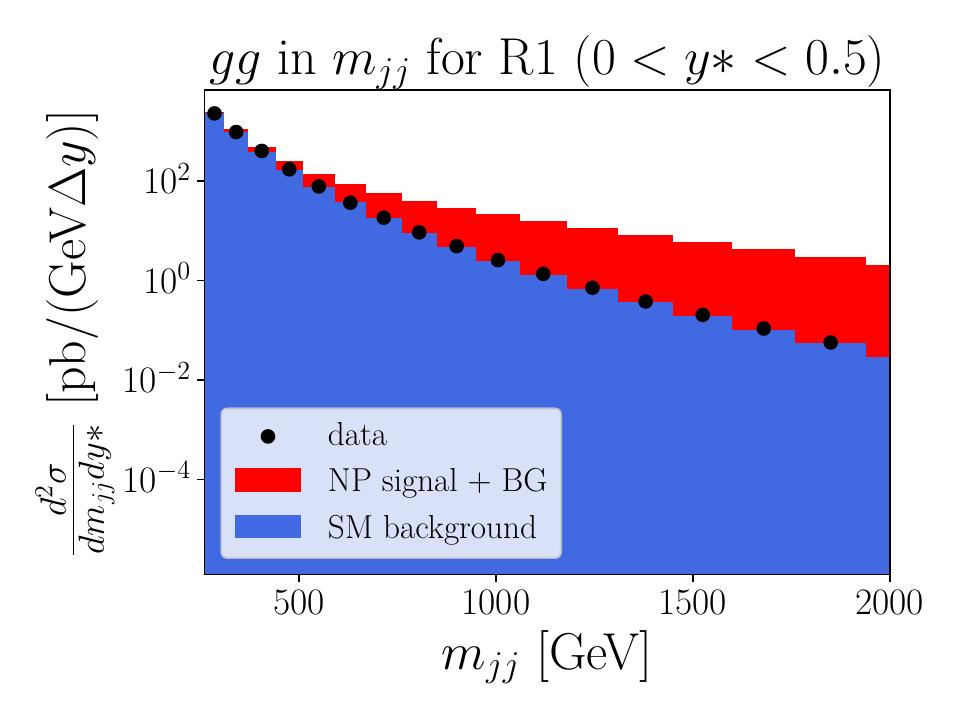}
    \includegraphics[width=0.48\linewidth]{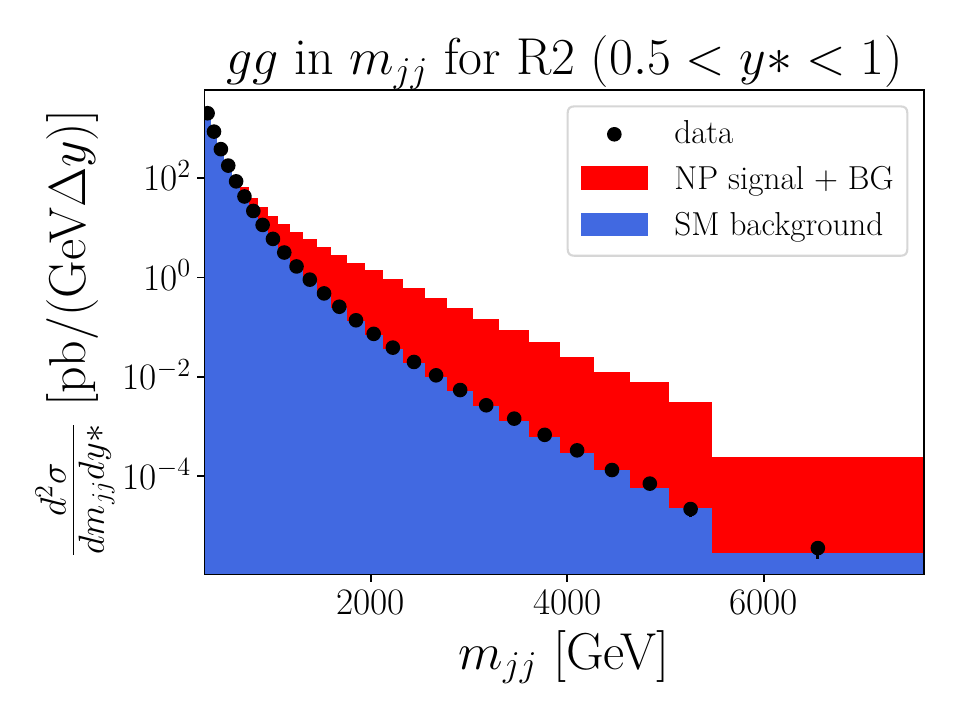}
    \includegraphics[width=0.48\linewidth]{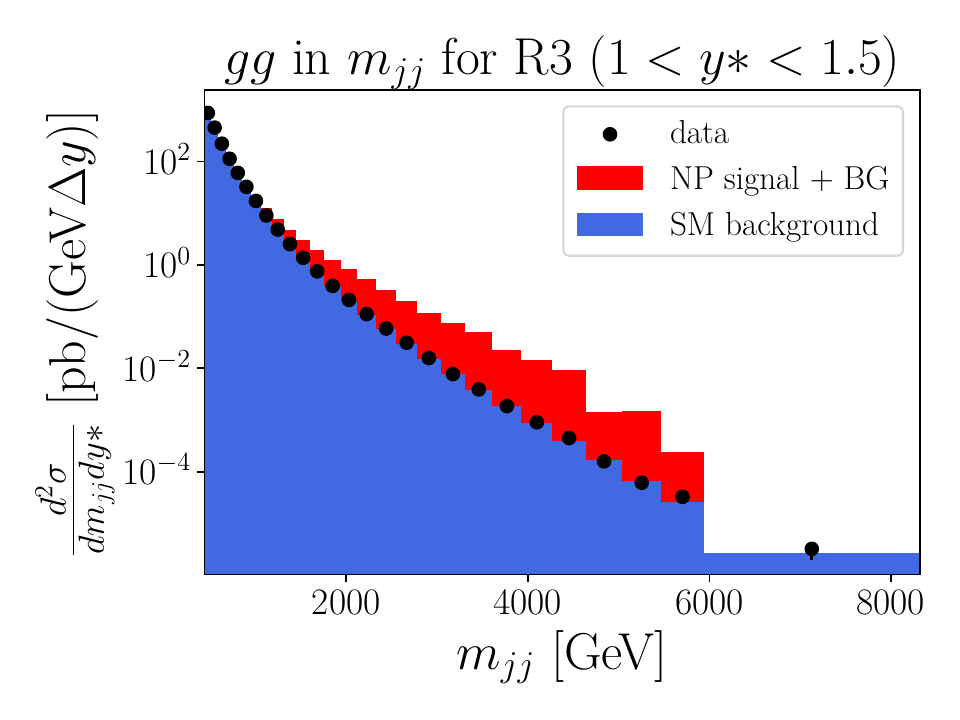}
    \includegraphics[width=0.48\linewidth]{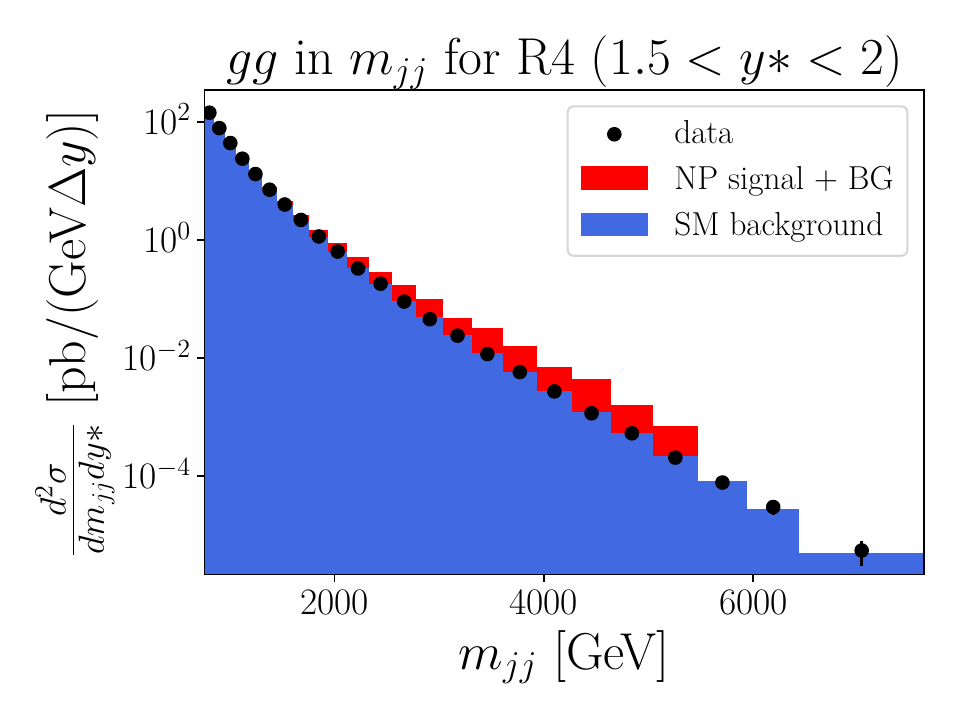}
    \includegraphics[width=0.48\linewidth]{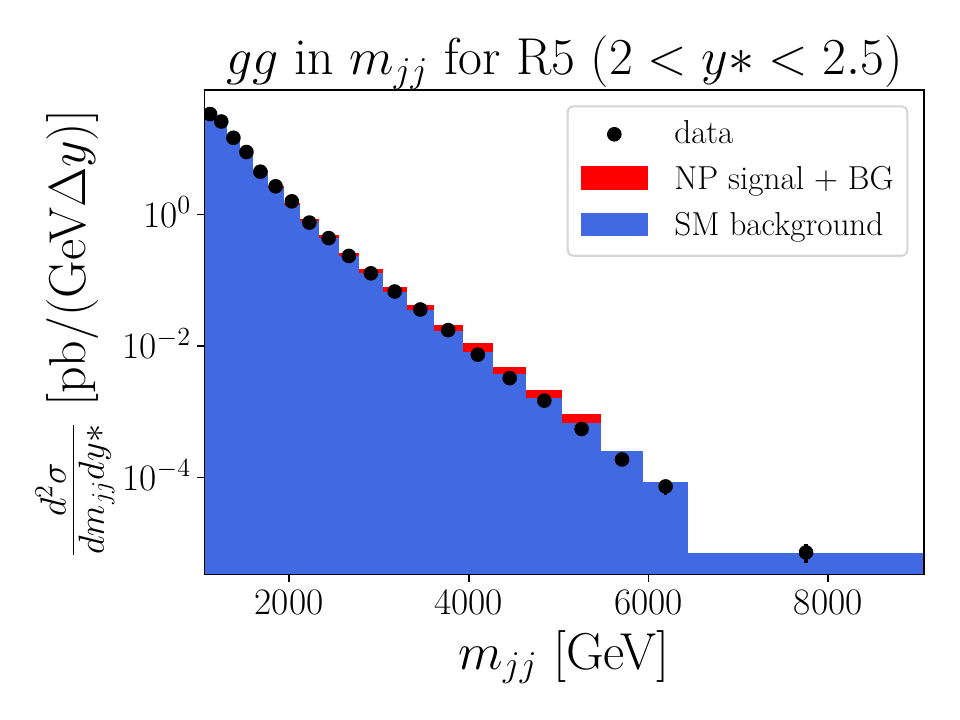}
    \includegraphics[width=0.48\linewidth]{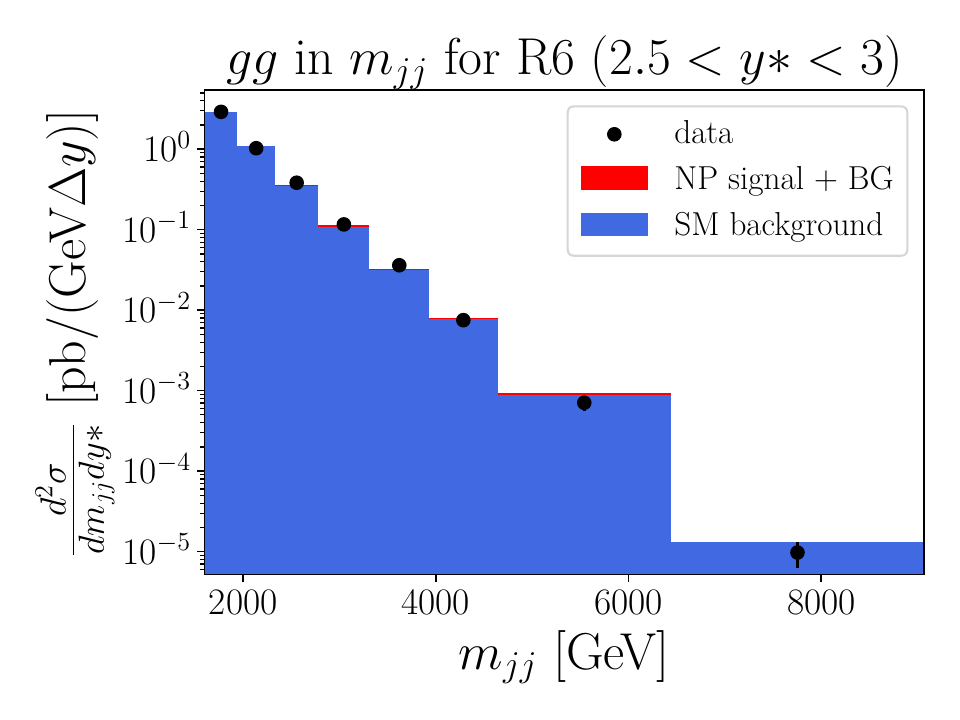}
    \caption{Double-differential distribution in $m_{jj}$ and $y^*$ for $pp\to a^\ast\to gg$, compared to 
    ATLAS data~\cite{ATLAS:2017ble} and the SM background. For better visibility, the signal is shown for $c_{\tilde G}=3$, $c_{\tilde B}=c_{\tilde W}=0$.}
    \label{fig:gg_dijet}
\end{figure}

To compare signal, data and background, we rescale the SM background simulation to the data by applying a 
$k$-factor in each $y^*$ region such that the total number of background events matches the observed yield. 
This procedure compensates for missing higher-order corrections in the generator and is applied consistently 
to both signal and background. \\  
Fig.~\ref{fig:gg_dijet} shows the distributions in the six different $y^*$ regions. As for the diphoton, $ZZ$ and $W^-W^+$ final states, 
the effect of the ALP-mediated signal becomes apparent in the tail of the invariant mass distribution. It is more visible in the more 
central regions with smaller $y^*$ simply because there are more signal events generated in these regions. 
For illustration purposes, we show the distributions with error bars corresponding to the diagonal entries of the 
covariance matrix. For the statistical analysis, however, we employ the full covariance.  

\begin{figure}[h!]
    \centering
    \includegraphics[width=\linewidth]{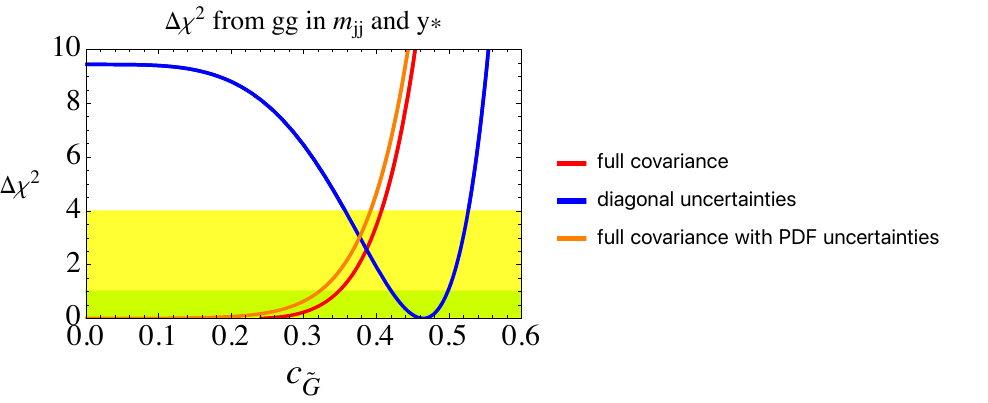}
    \caption{$\chi^2$ dependence on $c_{\tilde G}$ from the dijet double-differential spectrum. 
    The red curve includes the full experimental covariance matrix, while the blue curve assumes uncorrelated bin uncertainties. The orange curve adds the PDFs theory covariance matrix  
    to the experimental covariance matrix.}
    \label{fig:chi2_gg}
\end{figure}

The dependence of the $\chi^2$ on $c_{\tilde G}$ is shown in Fig.~\ref{fig:chi2_gg}. 
Here we define the test statistic as
\begin{equation}
\chi^2(c_{\tilde G},c_{\tilde B},c_{\tilde W}) \;=\;
\left(N^{\rm data} - N^{\rm th}(c_{\tilde G},c_{\tilde B},c_{\tilde W})\right)^T 
\, \Sigma^{-1} \,
\left(N^{\rm data} - N^{\rm th}(c_{\tilde G},c_{\tilde B},c_{\tilde W})\right),
\end{equation}
where $N^{\rm data}$ denotes the vector of observed event counts in each bin, 
$N^{\rm th}$ the corresponding theory prediction, and $\Sigma$ the covariance 
matrix of experimental uncertainties.
The theory prediction is given by
\begin{equation}
    N^{\rm th} = N^{BG} + c_{\Gt}^4 N^{\rm ALP} (c_{\Gt} = 1),
\end{equation}
where $N^{ALP} (c_{\Gt} =1)$ denotes the number of signal events for the benchmark scenario and $N^{BG}$ the number of background events. 
The scaling of the total cross section and thus the number of events with $c_{\Gt}$ follows directly from Eq.~(\ref{eq:offshellamp}).
In the plots, the error bars shown correspond to 
the diagonal entries of $\Sigma$, while the fits are performed using the full 
covariance, which captures correlations across bins and can substantially affect the 
resulting limits.

Using the full covariance matrix we find that values $c_{\tilde G}\gtrsim 0.4$ are excluded at the $2\sigma$ level. 
By contrast, an analysis based only on the diagonal entries would prefer a non-zero best-fit value around 
$c_{\tilde G}\simeq 0.4$--$0.5$, illustrating the importance of accounting for correlated uncertainties.
This highlights the importance of including bin-by-bin experimental correlations in the likelihood as soon as those are provided by 
experimentalists. On the other hand, by comparing the orange and the red curves in Fig. \ref{fig:chi2_gg}, we see that the inclusion of the fully correlated PDF uncertainty 
does not modify the bounds in any significant way. Rather counter-intuitively the inclusion of PDF uncertainties and PDF-induced correlations in the 
likelihood makes the bounds slightly tighter. This is due to the correlations among bins due to PDFs that have a negative sign. 
Given that the inclusion of the PDF uncertainties does not change our conclusions in any significant way, we do not include it systematically at this stage.  
Note that most of the other measurements that we include in this analysis do not provide yet the experimental correlations that are much more relevant 
than any correlated (or uncorrelated) theory uncertainties. Hence an inclusion of further correlated theory uncertainties without the inclusion of all experimental correlations would not be consistent. We leave the systematic inclusion of correlations 
and full theory uncertainties to future work, as soon as experimentalists release tables with data and full breakdown of systematics.

\subsection{Vector boson fusion $Z\gamma jj$ final state}
\label{sec:Zgammajj}

\begin{figure}[h!]
    \centering
    \includegraphics[width=0.4\linewidth]{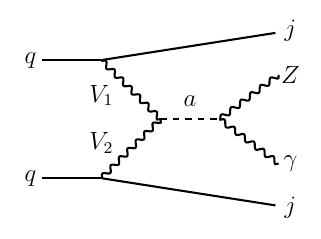}
    \includegraphics[width=0.4\linewidth]{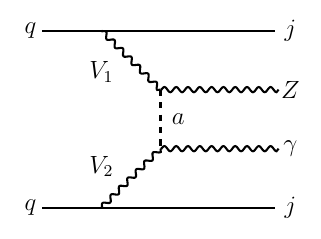}
    
    \caption{Diagrams contributing to $Z\gamma jj$: Vector boson fusion with an $s$-channel intermediate ALP for $V_1 V_2 \in \{W^+ W^-, ZZ, Z\gamma, \gamma \gamma \}$ (left) and vector boson scattering with a $t$-channel intermediate ALP for $V_1, V_2 \in \{ \gamma, Z\}$ (right).}
\label{fig:VBF_diagrams_gammaZ}
\end{figure}

Vector boson fusion provides sensitivity to ALP couplings in the electroweak sector and enables us to constrain the couplings in the $(c_{\Wt}, c_{\Bt})$ plane, 
independently of $c_{\Gt}$.\\ 
We consider the CMS measurement of $pp\to Z\gamma jj$ at $\sqrt{s}=13$~TeV with 
137~fb$^{-1}$~\cite{CMS:2021gme}, where the $Z$ boson is reconstructed in the 
dilepton channel. The corresponding \texttt{HepData} entry is listed in 
Table~\ref{tab:expdata_sm}.  
Signal events are generated with \madgraph for the process

\begin{equation}
pp \;\overset{\rm NP=2}{\longrightarrow}\; Z\gamma jj,
\label{eq:Zgammajj_generate}
\end{equation}
using the {\tt ALP\_linear\_UFO} model and setting $c_{\Gt} = 0$, $c_{\Wt} =1$ and $c_{\Bt}=0$. Note that it is important to set $c_{\Phi}$, 
which governs the ALP-fermion couplings, to zero in \madgraph. We use the syntax in Eq.\ (\ref{eq:Zgammajj_generate}), setting the New Physics (NP) 
order  $\text{NP}=2$, because not all diagrams that contribute to the signal process contain an $s$-channel ALP, which distinguishes this 
process significantly from the final states discussed before.
In addition to the typical vector boson fusion diagram with an $s$-channel ALP there are contributions from vector boson scattering (VBS) 
with a $t$-channel ALP, \textit{cf.}\ Fig.\ \ref{fig:VBF_diagrams_gammaZ}

\begin{figure}[h!]
    \centering
    \includegraphics[width=0.48\linewidth]{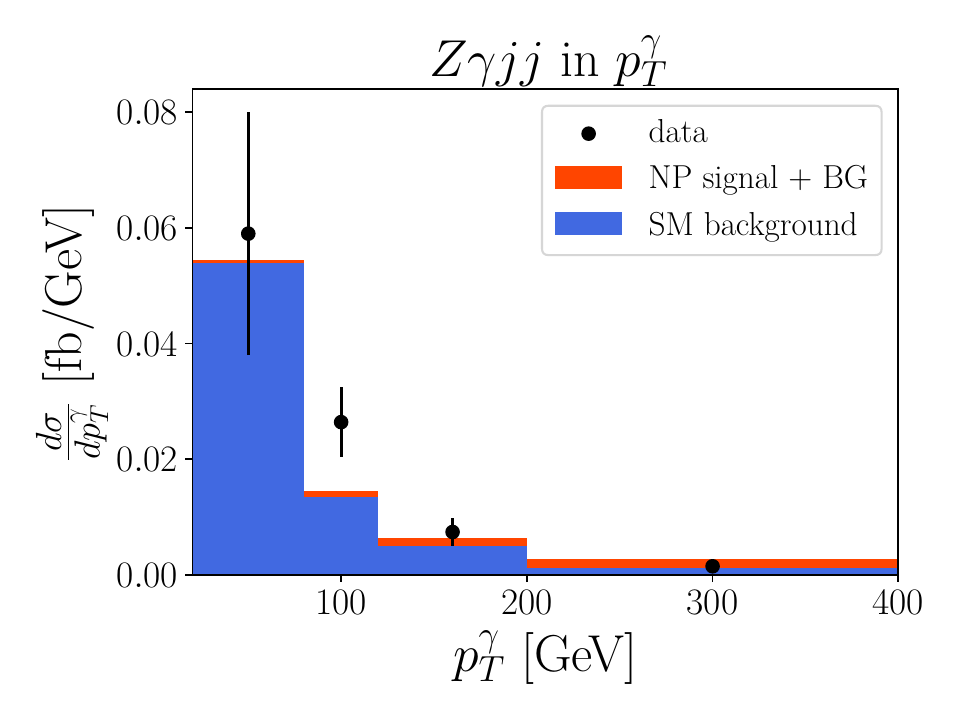}
    \includegraphics[width=0.48\linewidth]{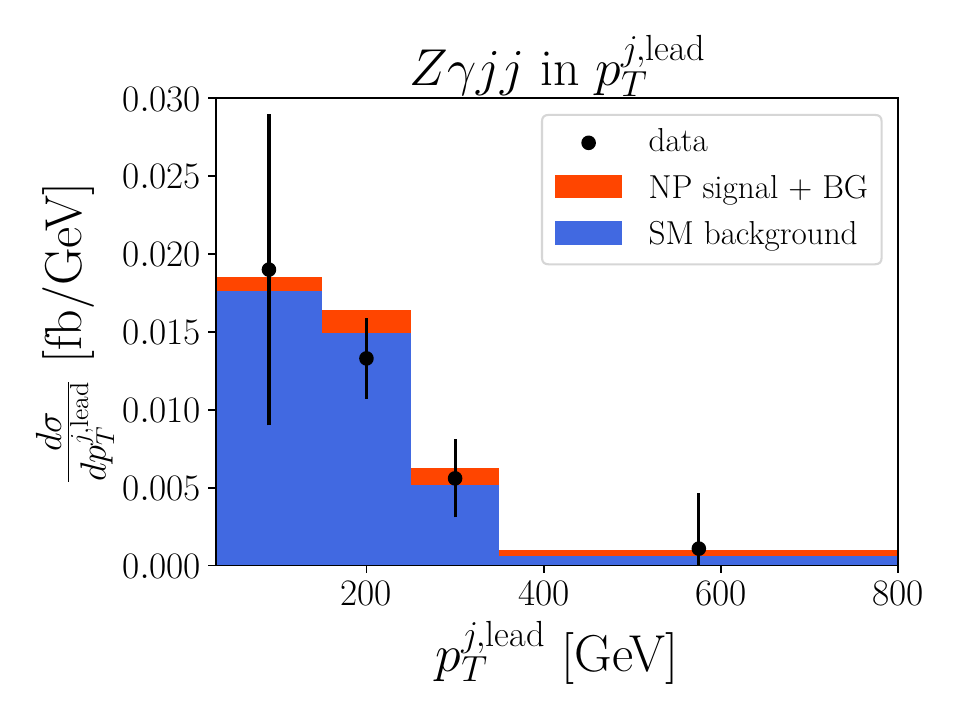}
    \caption{Differential distributions in $p_T^{\gamma}$ (left) and $p_T^{j,\text{lead}}$ (right) 
    for the $Z\gamma jj$ final state. Signal is shown for $c_{\tilde G}=1$$c_{\tilde W}=1$ and $c_{\tilde B}=0$, compared to CMS data~\cite{CMS:2021gme} and the SM background.  }
\label{fig:gammaZjj}
\end{figure}

Combining all possible diagrams, this process consists of 8 subprocesses. Since we do not know their relative frequency, 
we cannot rescale the cross section for different values of $c_{\Wt}$ and $c_{\Bt}$ as easily as for the previous analyses. 
Hence we calculate the cross section in \madgraph looping over various combinations for values of $c_{\Wt}$ and $c_{\Bt}$ and fit a quartic polynomial of the form
\begin{equation}
\sigma_{Z\gamma jj}(c_{\tilde W},c_{\tilde B}) \;=\; \sum_{m, n =0}^{4, n+m\leq 4}
\alpha_{n m}\,c_{\tilde W}^n \,c_{\tilde B}^m, 
\end{equation}
which covers all possible contributions from vector boson fusion and vector boson scattering diagrams
to the observed cross sections. With the fit at hand, the number of events for arbitrary $c_{\Wt}$ and $c_{\Bt}$ can be obtained from
\begin{equation}
    N^{\rm ALP}(c_{\Wt}, c_{\Bt}) = N^{\rm ALP}(c_{\Wt}=1, c_{\Bt}=0) \frac{\sigma_{Z \gamma jj}(c_{\Wt}, c_{\Bt})}{\sigma_{Z \gamma jj}(c_{\Wt}=1, c_{\Bt}=0)}
\end{equation}
with $N^{\rm ALP}(c_{\Wt}=1, c_{\Bt}=0)$  and $\sigma_{Z \gamma jj}(c_{\Wt}=1, c_{\Bt}=0)$ the distribution and cross section, respectively, for the benchmark scenario. \\
The fiducial event selection follows the CMS analysis, requiring two isolated leptons consistent 
with a $Z$ boson, one isolated photon, and at least two jets in the forward region. 
The signal events for $c_{\Wt} =1$ and $c_{\Bt}=0$ are normalised to $137~\text{fb}^{-1}$ using the cross section from \madgraph, the acceptance estimated from the 
LHE files for this benchmark scenario and multiplying with the branching ratio of Z into a lepton pair, $\text{BR}(Z \rightarrow l^+ l^-)$. We further assume 
that the acceptance to phase space cuts does not change when rescaling the total cross section. \\ 
We analyse two differential distributions: the photon transverse momentum $p_T^\gamma$ and 
the transverse momentum of the leading jet $p_T^{j,\text{lead}}$. The binning matches the 
CMS measurement, with $p_T^\gamma$ spanning 40--1000~GeV and $p_T^{j,\text{lead}}$ spanning 30--800~GeV. 
Figure~\ref{fig:gammaZjj} shows the signal predictions compared to the CMS data and the SM background. 
The characteristic effect of the ALP is an enhancement in the high-$p_T$ tails of both distributions, 
arising from the momentum dependence of the $aVV$ vertices. 

\begin{figure}[h!]
    \centering
    \includegraphics[width=0.48\linewidth]{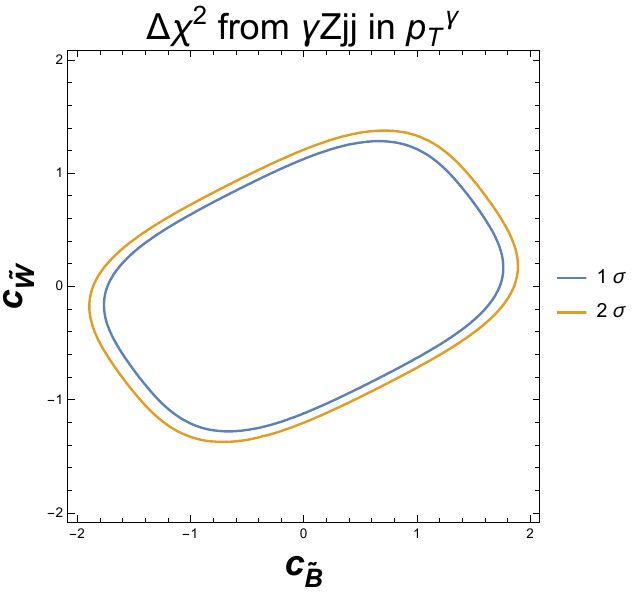}
    \includegraphics[width=0.48\linewidth]{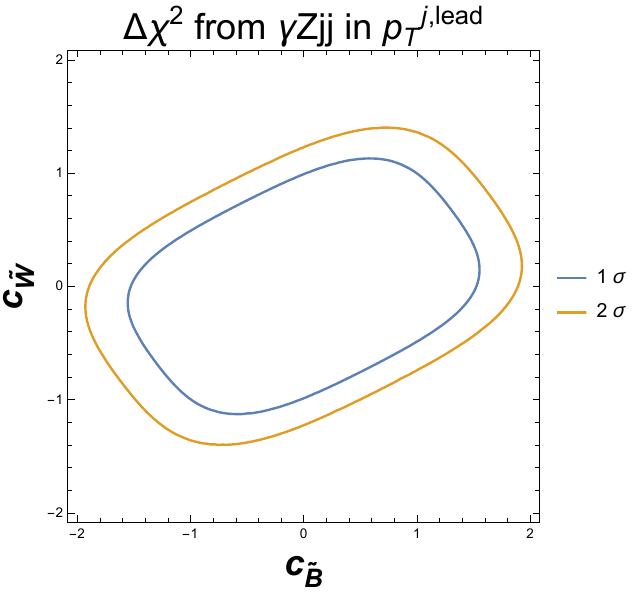}
    \caption{$\chi^2$ and exclusion limits from the $\gamma Z j j$ final state in $p_T^{\gamma}$ (left) and $p_T^{\rm j,lead}$ (right).  }
\label{fig:chi2_gammaZjj}
\end{figure}

Performing a $\chi^2$ analysis for both distributions leads to the $1 \sigma$ and $2 \sigma$ exclusion plots shown in Fig.~\ref{fig:chi2_gammaZjj}. The distribution in $p_T^{\gamma}$ gives the slightly stronger limits and thus we take this distribution into account for the combined $\chi^2$ fit.

\subsection{Vector boson fusion $ZZ jj$ and $W^+ W^- jj$ final states}
\label{sec:VBFVVjj}

Vector–boson fusion  provides direct sensitivity to the electroweak ALP couplings $c_{\Wt}$ and $c_{\Bt}$ via off–shell ALP exchange in $VV'jj$ topologies. Different $VV'$ states will probe a different combination of the two Wilson coefficient. Adding analyses for various combinations of $VV'$ therefore constrains complementary directions in the $(c_{\Bt}, c_{\Wt})$ plane.
The analysis in \cite{ATLAS:2025omi} measures the distributions of $VV'jj$ with $V,V' \in \{Z,W\}$. 
%Note that since the ALP is electrically neutral, processes with intermediate s-channel ALPs can only contribute to $ZZjj$ and $W^+ W^- jj$ production. \textcolor{red}{??do not see the meaning of this sentence} \VS{ just saying that the ALP } \\

\begin{figure}[h!]
    \centering
    \includegraphics[width=0.4\linewidth]{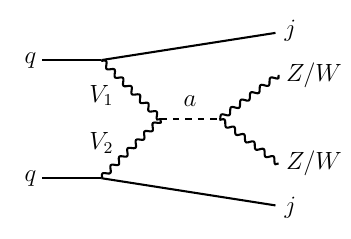}
    \includegraphics[width=0.4\linewidth]{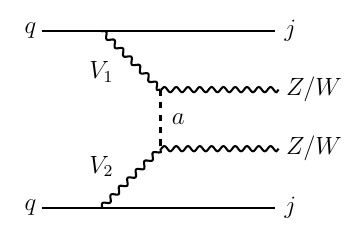}
    
    \caption{Diagrams contributing to $ZZ\gamma jj$ and $W^+ W^- jj$: Vector boson fusion with an $s$-channel intermediate ALP for $V_1 V_2 \in \{W^+ W^-, ZZ, Z\gamma, \gamma \gamma \}$ (left) and vector boson scattering with a $t$-channel intermediate ALP for $V_1, V_2 \in \{ \gamma, Z\}$ for $ZZjj$ and $V_1 V_2 \in \{ W^+W^-\}$ for $W^+W^-jj$(right).}
\label{fig:VBF_diagrams_ZZWW}
\end{figure}

The $VV'jj$ amplitudes receive contributions from both VBF and VBS with several ALP vertices, cf.\ Fig.\ \ref{fig:VBF_diagrams_ZZWW}, and an analytic closed form of the fiducial cross section in terms of $(c_{\tilde W}, c_{\tilde B})$ is therefore not available. Similarly to $Z \gamma jj$, we proceed by scanning a grid in $(c_{\tilde W},c_{\tilde B})$ with $c_{\tilde G}=0$ and fitting for each channel a quartic polynomial
\begin{equation}
\begin{split}
\sigma_{ZZjj}^{\rm fid}(c_{\tilde W},c_{\tilde B}) &= \sum_{m, n =0}^{4, n+m\leq 4}
\alpha_{n m}^{(ZZ)}\,c_{\tilde W}^n \,c_{\tilde B}^m \\
\sigma_{WWjj}^{\rm fid}(c_{\tilde W},c_{\tilde B}) &= \sum_{m, n =0}^{4, n+m\leq 4}
\alpha_{n m}^{(WW)}\,c_{\tilde W}^n \,c_{\tilde B}^m\;,
\end{split}
\end{equation}
which captures the SM–ALP interference (quadratic terms) and the pure-ALP contribution (quartic terms) from the processed depicted in Fig. \ref{fig:VBF_diagrams_ZZWW}.  \\
We consider three fiducial final states in which one boson ($V$) is reconstructed hadronically (two jets) and the other ($V'$) leptonically:
\begin{itemize}
    \item $ZZjj$ with $Z\!\to\!\nu\bar\nu$ (0$\ell$),
    \item $ZZjj$ with $Z\!\to\!\ell^+\ell^-$ (2$\ell$),
    \item $W^+W^-jj$ with $W^\pm\!\to\!\ell^\pm\nu$ (1$\ell$).
\end{itemize}
To avoid ambiguities in assigning the hadronic boson jets at generator level, we simulate the 
corresponding leptonic decay modes and then rescale the event yields with the appropriate 
branching ratios to obtain the correct fiducial cross sections. The events are subsequently 
normalised to $140~\mathrm{fb}^{-1}$ using the \madgraph cross section and the acceptance after 
fiducial selections.\\
We split events into {\it resolved} if $\Delta R_{jj}>0.4$, and  {\it merged } otherwise to account for boosted hadronic bosons reconstructed as 
either two resolved jets or a single fat jet. Note that the analyses in \cite{ATLAS:2025omi} do not provide any \texttt{HepData} and we have to 
digitise the measured distribution, the experimental uncertainty and the SM background simulation from the additional figures linked in Table~\ref{tab:expdata_sm}.
For each channel and each regions (resolved and merged) we analyse the differential distribution in the diboson invariant mass $m_{VV}$, using the binning of the 
experimental plots we digitise (ten bins up to $m_{VV}\sim 2$~TeV in the resolved category, slightly fewer in the merged category depending on the analysis's figure). 

Background templates and the data points (with upper/lower uncertainty envelopes) are read from the digitised CSV files from the public experimental results. Error bars in the figures correspond to the diagonal uncertainties, while the fit uses those same bin-by-bin uncertainties (no published covariance matrix is available for these plots).

\begin{figure}[ht!]
    \centering
    \includegraphics[width=0.48\linewidth]{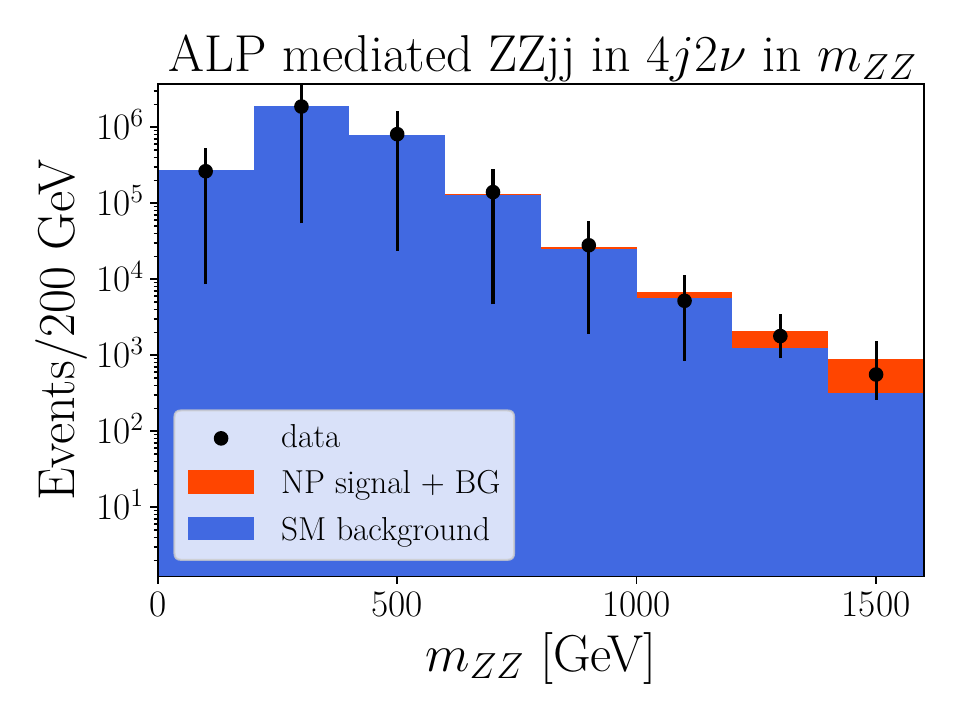}
    \includegraphics[width=0.48\linewidth]{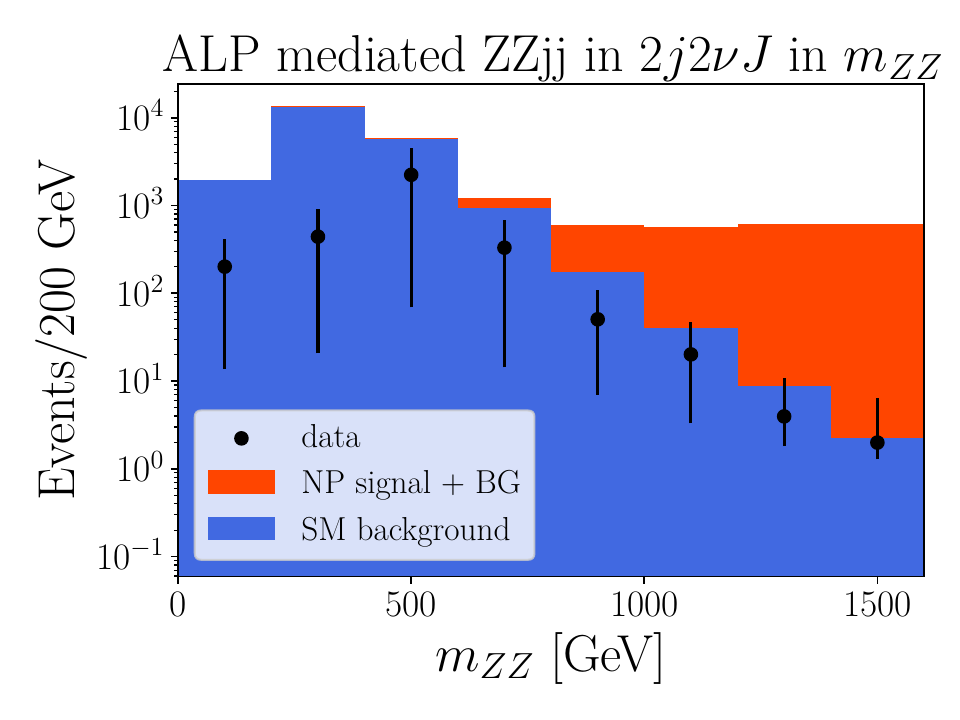}
    \includegraphics[width=0.48\linewidth]{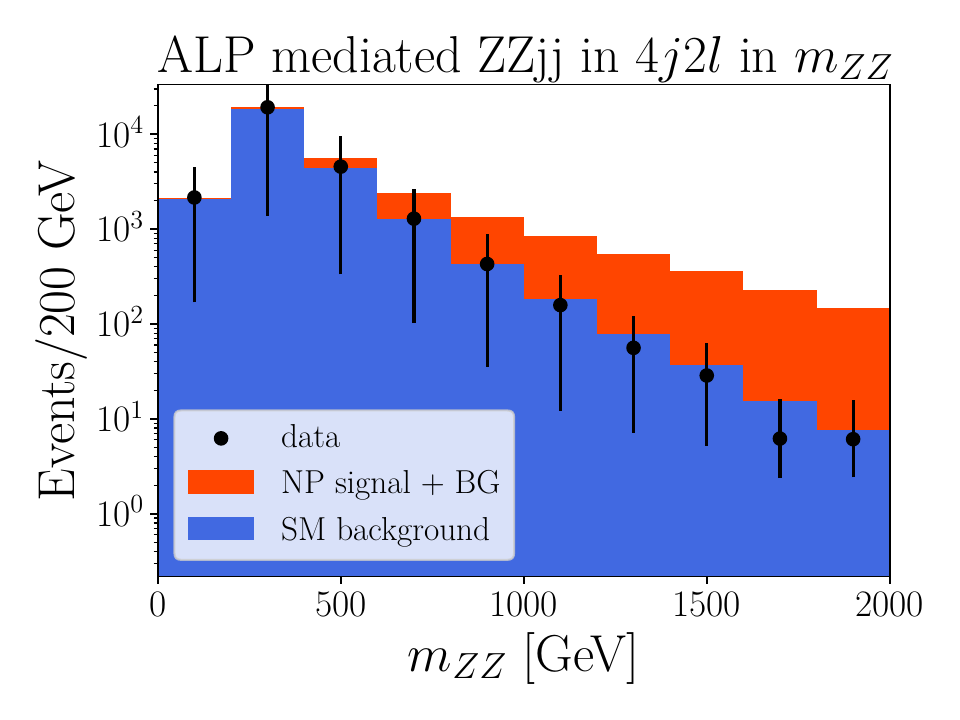}
    \includegraphics[width=0.48\linewidth]{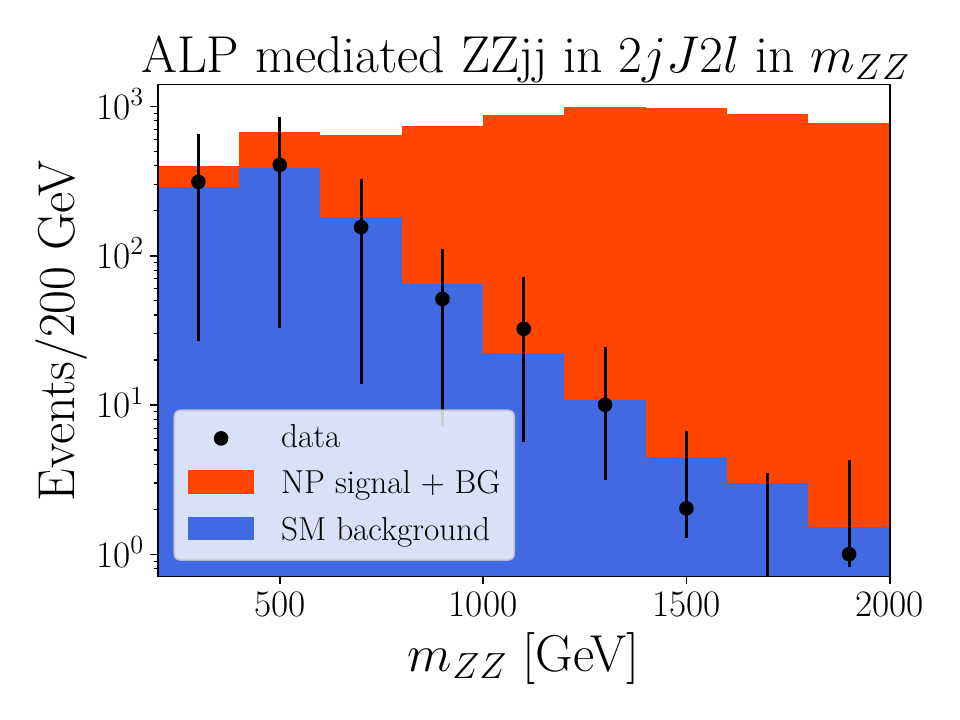}
    \includegraphics[width=0.48\linewidth]{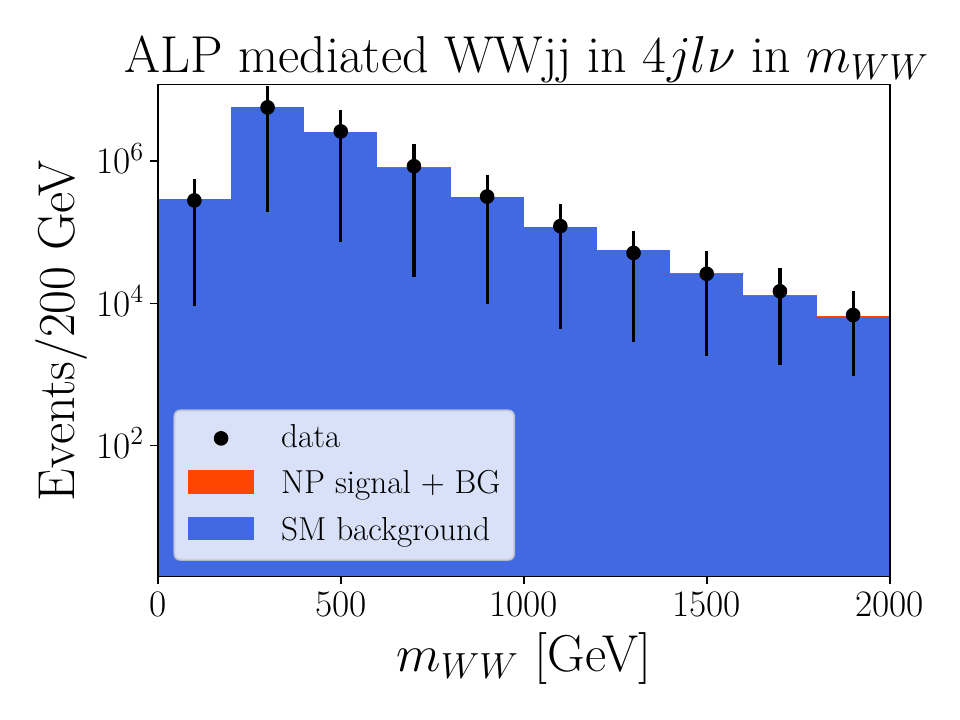}
    \includegraphics[width=0.48\linewidth]{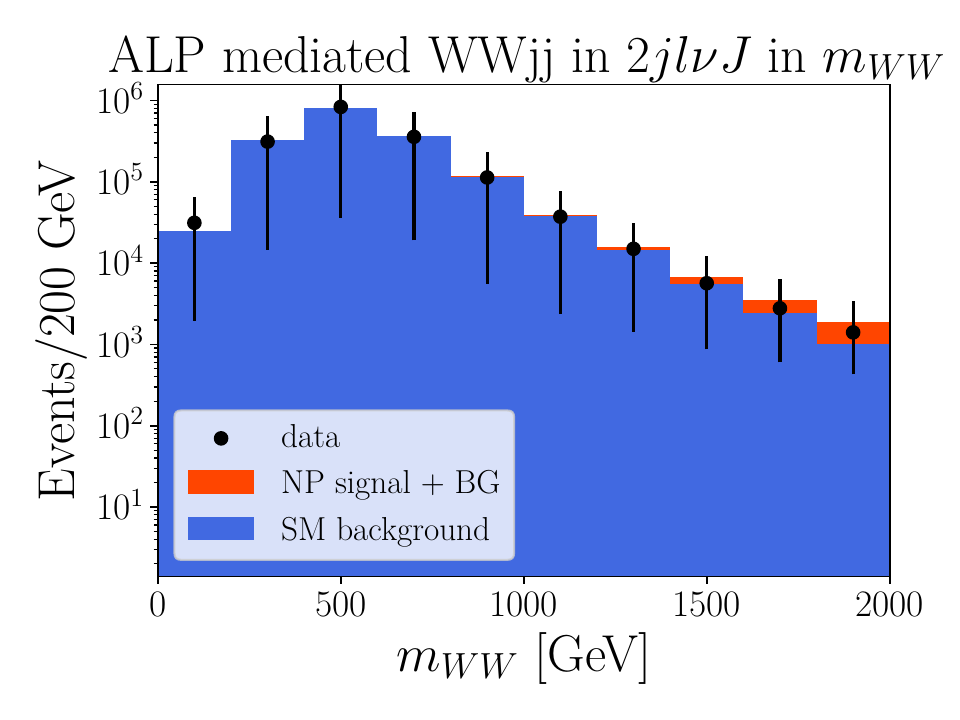}
    \caption{Differential distributions in $m_{VV}$ for $VVjj$ final states. One boson decays hadronically, reconstructed either as two resolved jets (left panels) or one merged fat jet (right panels), while the other boson decays leptonically: $Z\to\nu\bar\nu$ (0$\ell$), $Z\to\ell^+\ell^-$ (2$\ell$), $W^\pm\to\ell^\pm\nu$ (1$\ell$). For illustration we show the signal for $c_{\tilde G}=0$, $c_{\tilde W}=10$ and $c_{\tilde B}=0$, with vanishing fermionic ALP couplings. }
    \label{fig:VBP}
\end{figure}

\begin{figure}[t]
    \centering
    \includegraphics[width=0.48\linewidth]{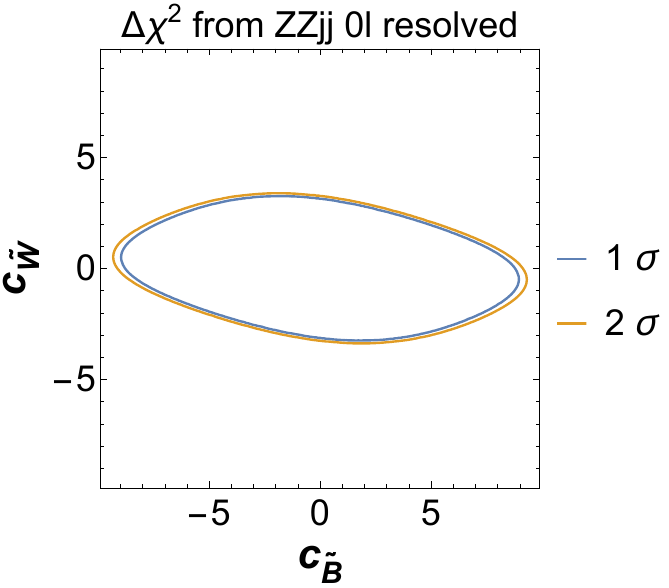}
    \includegraphics[width=0.48\linewidth]{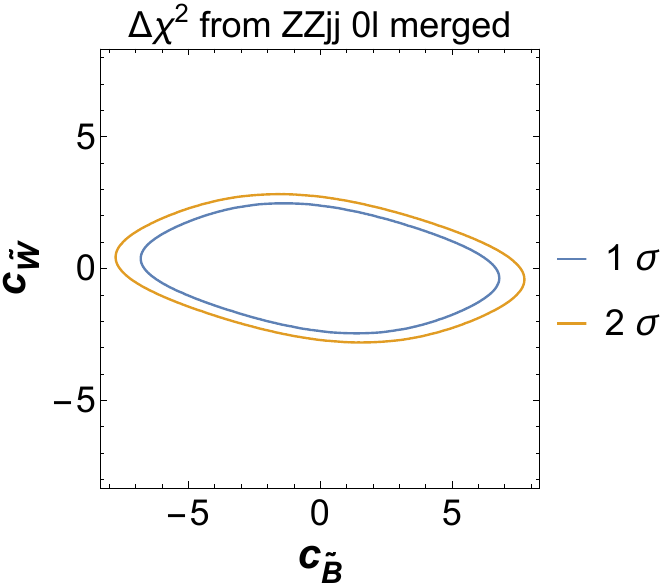}
    \includegraphics[width=0.48\linewidth]{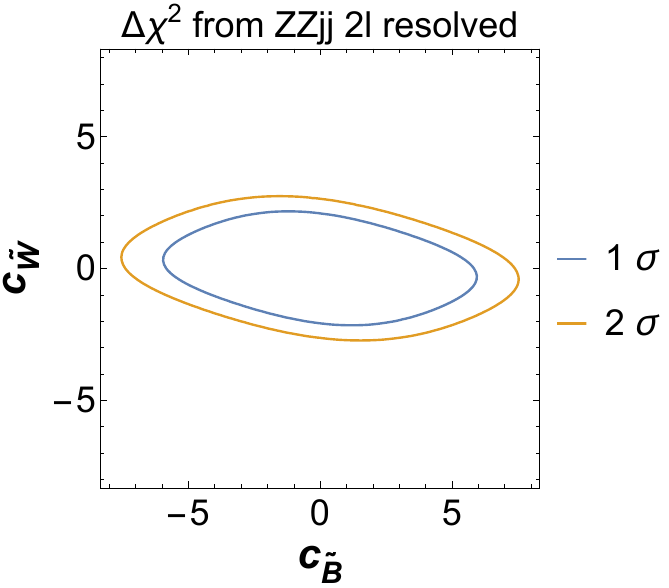}
    \includegraphics[width=0.48\linewidth]{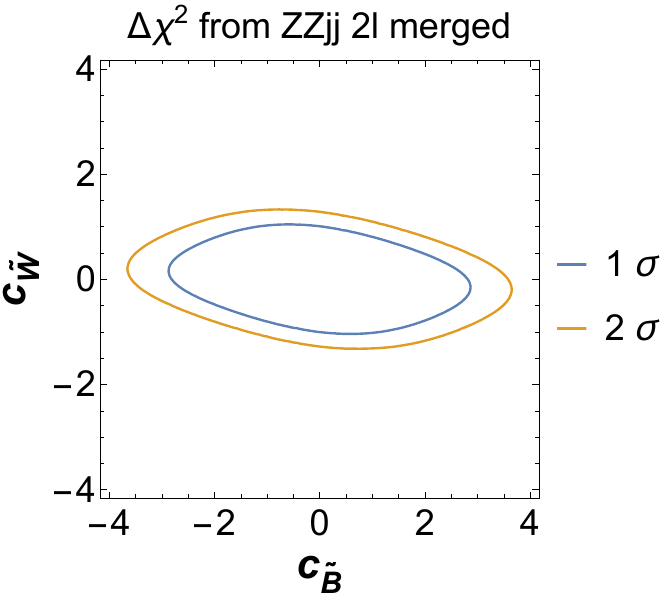}
    \includegraphics[width=0.48\linewidth]{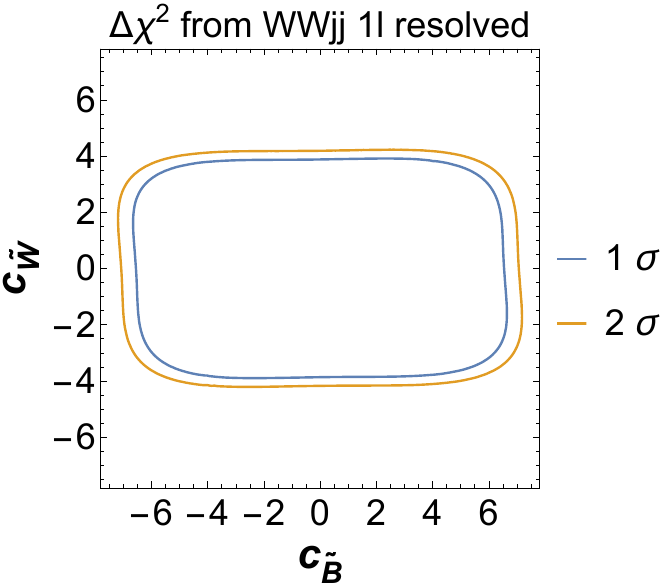}
    \includegraphics[width=0.48\linewidth]{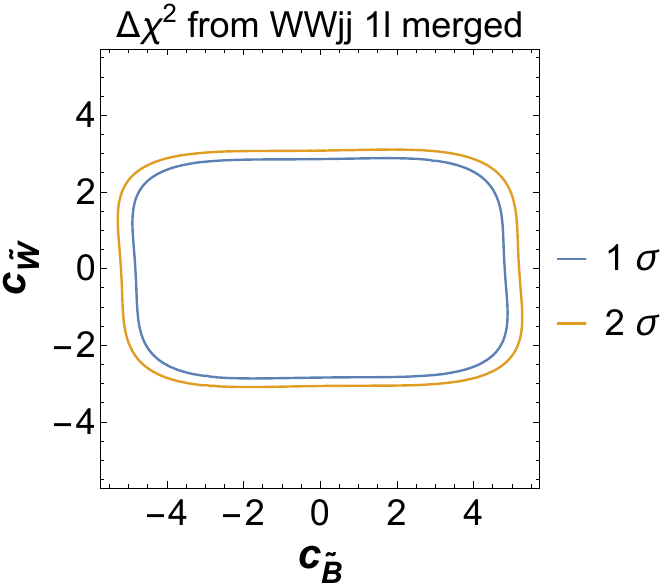}
    \caption{$\chi^2$ in the $(c_{\tilde B},c_{\tilde W})$ plane for the six $VVjj$ categories: 
    $Z\!\to\!0\ell$ (top), $Z\!\to\!2\ell$ (middle), $W\!\to\!1\ell$ (bottom); resolved (left) vs.\ merged (right).}
    \label{fig:chi2_VBP}
\end{figure}
\paragraph{Data treatment and uncertainties.}
Since only figure overlays are publicly available for these VBF selections, we digitise the published spectra for the $0\ell$, $2\ell$ ($ZZjj$) and $1\ell$ ($WWjj$) final states and propagate the quoted bin uncertainties. The plots in Fig.~\ref{fig:VBP} show the signal overlaid on the digitised data and background for both resolved and merged categories; error bars correspond to the diagonal uncertainties used later in the fit. As expected for off-shell ALP exchange, the sensitivity resides in the high-$m_{VV}$ bins, with the merged $2\ell$ category providing the single strongest constraint among the six subchannels.

\begin{figure}[h!]
    \centering
    \includegraphics[width=0.8\linewidth]{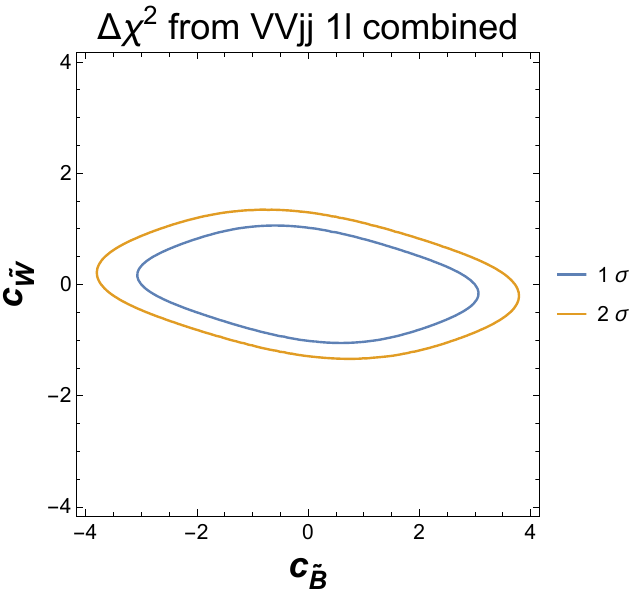}
    \caption{Combined $\chi^2$ in the $(c_{\tilde B},c_{\tilde W})$ plane from all $VVjj$ categories. 
    The contour is dominated by the $ZZjj$ 2$\ell$ merged selection, with complementary information from the other channels. }
    \label{fig:chi2_VBP_combined}
\end{figure}

\paragraph{Constraints in the $(c_{\tilde B},c_{\tilde W})$ plane.}
For each subchannel and category we build a binned $\chi^2(c_{\tilde B},c_{\tilde W})$ using the digitised data, SM background and the polynomial surrogate for the signal. Fig.~\ref{fig:chi2_VBP} displays the resulting contours; the $2\ell$ merged selection dominates the sensitivity, with the $0\ell$ resolved and $1\ell$ merged providing complementary directions through their different acceptance and background compositions. Combining all six subchannels yields the global constraint shown in Fig.~\ref{fig:chi2_VBP_combined}.

%%%%%%%%%%%%%%%%%%%%%%%%%%%%%%%%%%%%%%%%%%%%%%%
\section{Global fit results}
\label{sec:results}
%%%%%%%%%%%%%%%%%%%%%%%%%%%%%%%%%%%%%%%%%%%%%%%

The sensitivity of each experimental channel to ALP effects depends not only on the 
statistical power of the measurement, but also on the number of free parameters varied 
in the fit. A limit obtained by varying a single coupling while fixing all others to zero 
is necessarily stronger than one obtained in a simultaneous multi-parameter fit, where 
flat directions and correlations can dilute the constraints. To quantify these effects 
consistently, we perform combined fits to the three Wilson coefficients 
$(c_{\Gt}, c_{\Wt}, c_{\Bt})$, which control the ALP couplings to gluons and electroweak 
bosons. For definiteness we fix the ALP decay constant to $f_a = 1~\text{TeV}$, noting 
that all results trivially rescale as $1/f_a$.

The global likelihood is then explored 
under different statistical prescriptions (projection, profiling, and marginalisation)
which correspond to different ways of treating the parameter space when extracting 
one-dimensional limits. The resulting exclusion contours can therefore differ 
significantly, and the comparison between them provides insight into the robustness 
of the constraints and the complementarity among different final states.

Before presenting the results of the global fit, we emphasise that all
constraints derived in this work should be interpreted as bounds on the
effective couplings $c_{\tilde X}/f_a$. The choice $f_a = 1~\mathrm{TeV}$ is
made for convenience and does not carry physical significance. Owing to the
exact factorisation of the ALP contribution, cf.~Eq.~(3.2), all limits can be
trivially rescaled to any other value of $f_a$.\\

Note that a global on-shell ALP analysis was performed in Ref.~\cite{Bruggisser:2023npd}. Previous work based on the off-shell contributions from ALP-mediation  considered global analyses restricted to vector-boson scattering observables~\cite{Bonilla:2022pxu}. Here, by including also diphoton, diboson, dijet and VBF signatures, we provide the first global treatment of ALP-mediated multiboson production at the LHC.

\subsection{Global fits in two and three dimensions}

\begin{figure}
    \centering
    \includegraphics[width=\linewidth]{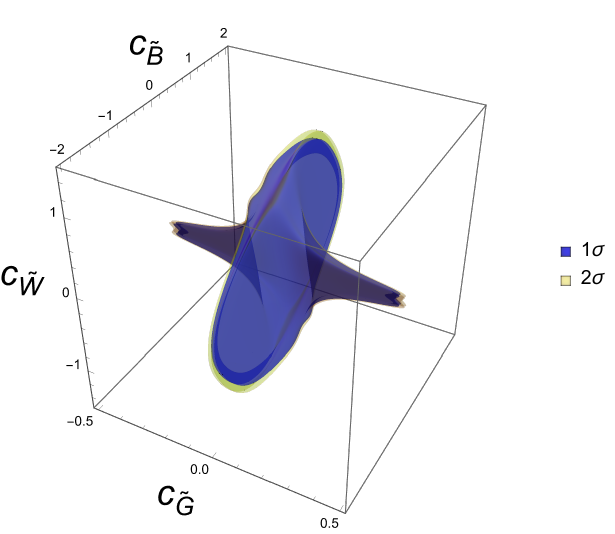}
    \caption{Three-dimensional global fit including all channels. The shaded regions indicate the 
    $1\sigma$ and $2\sigma$ confidence levels. The dijet analysis dominates the 
    constraint on $c_{\Gt}$, while the electroweak boson channels constrain $c_{\Bt}$ 
    and $c_{\Wt}$ once $c_{\Gt}$ is fixed.}
    \label{fig:global_fit_3D}
\end{figure}

Figure~\ref{fig:global_fit_3D} shows the full 3D fit including all channels. 
The dijet final state provides the dominant sensitivity to $c_{\Gt}$ and tightly 
constrains this direction. For $c_{\Gt} > 0$, the combination of diphoton, $ZZ$, 
and $WW$ channels further restricts the electroweak coefficients $c_{\Bt}$ and $c_{\Wt}$. 
In contrast, when $c_{\Gt}\simeq 0$ these channels lose sensitivity, and the inclusion of 
vector boson fusion topologies, such as $\gamma Z jj$ and $VVjj$, becomes essential to 
close the parameter space and obtain finite limits. 

\begin{figure}[h!]
    \centering
    \includegraphics[width=0.45\linewidth]{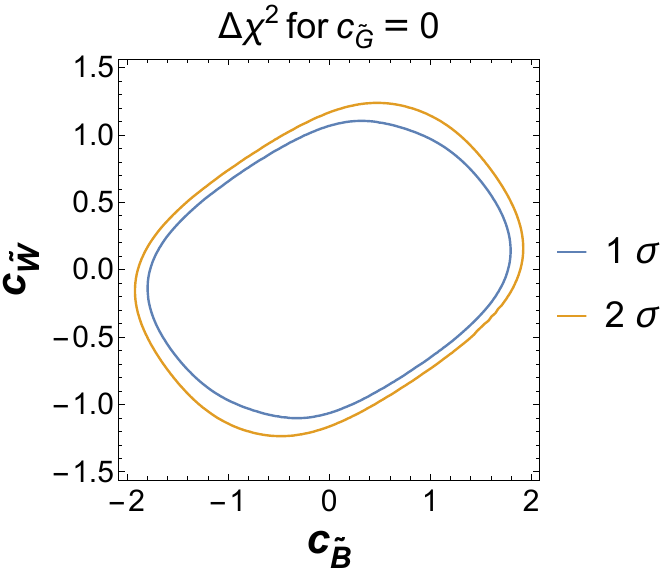}
    \includegraphics[width=0.45\linewidth]{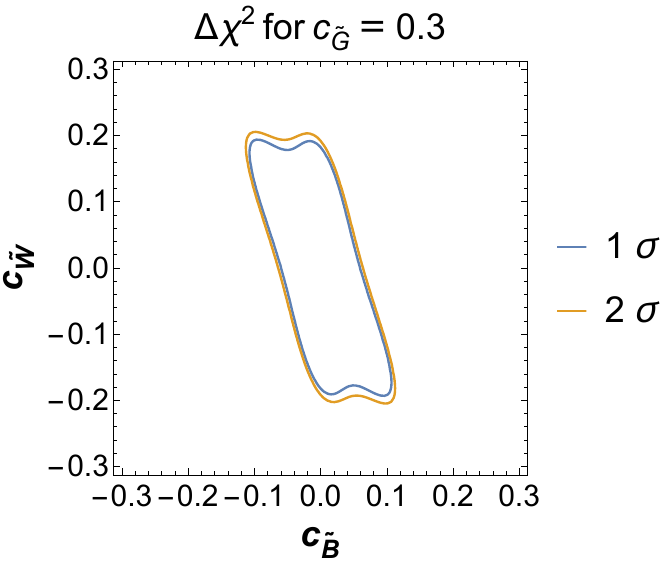}
    \includegraphics[width=0.45\linewidth]{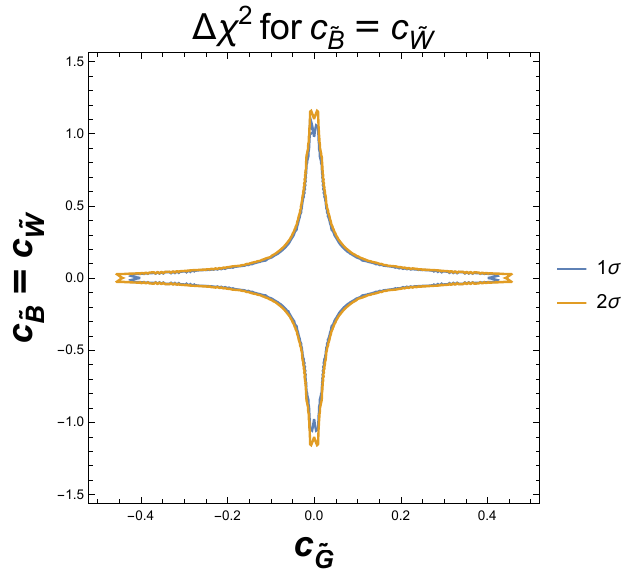}
    \includegraphics[width=0.45\linewidth]{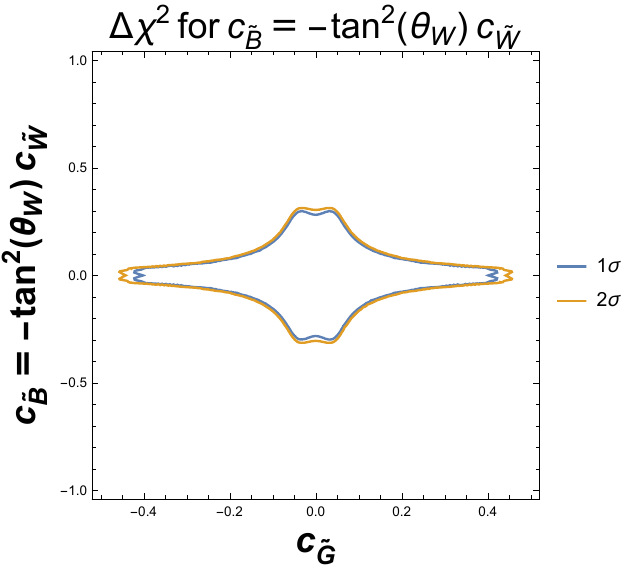}
    \caption{Two-dimensional projections of the global fit. 
    Top: $(c_{\Bt},c_{\Wt})$ plane for $c_{\Gt}=0$ (left) and $c_{\Gt}=0.3$ (right). 
    Bottom: projections onto the $(c_{\Gt},c_{\Bt}=c_{\Wt})$ and 
    $(c_{\Gt},{c_{\Bt}=- \tan^2(\theta_W) \, c_{\Wt}})$ planes. }
    \label{fig:global_fit_2D_projections}
\end{figure}

To visualise the interplay between parameters, we present in 
Fig.~\ref{fig:global_fit_2D_projections} several two-dimensional slices of the full 
3D likelihood. The top row shows projections onto the $(c_{\Bt},c_{\Wt})$ plane 
for $c_{\Gt}=0$ (left) and $c_{\Gt}=0.3$ (right). The bottom row shows
projections in the $(c_{\Gt},c_{\Bt}=c_{\Wt})$ and $(c_{\Gt},c_{\Bt}=-\tan^2(\theta_W) \, c_{\Wt})$ planes, respectively.
The first case with $c_{\Bt} = c_{\Wt}$ corresponds to the scenario in which the ALP coupling to one photon and one Z-boson, $g_{a\gamma Z}$, vanishes, meaning that the ALP can couple only to either two photons or two Z-bosons in the neutral electroweak sector. The second case with $c_{\Bt} = - \tan^2(\theta_W) \, c_{\Wt}$ mimics the photophobic scenario, in which the ALP does not couple to a pair of photons, $g_{a\gamma \gamma}=0$, cf.\ Table \ref{tab:ALP_multiboson_linear}. 
Note that $c_{\Gt}\gtrsim 0.41$ is excluded at the $2\sigma$ level by the dijet channel.

\begin{figure}[h!]
    \centering
    \includegraphics[width=0.8\linewidth]{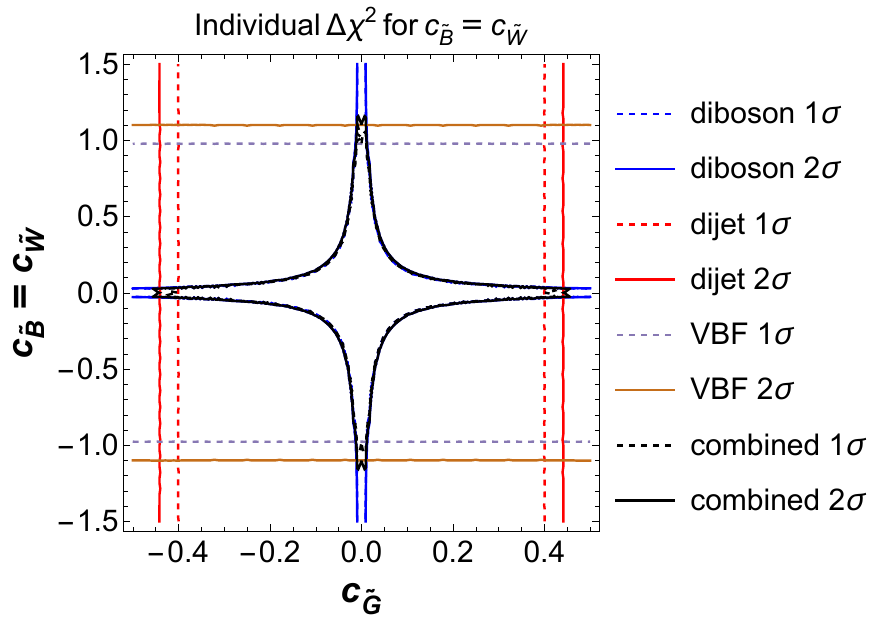}    \includegraphics[width=0.8\linewidth]{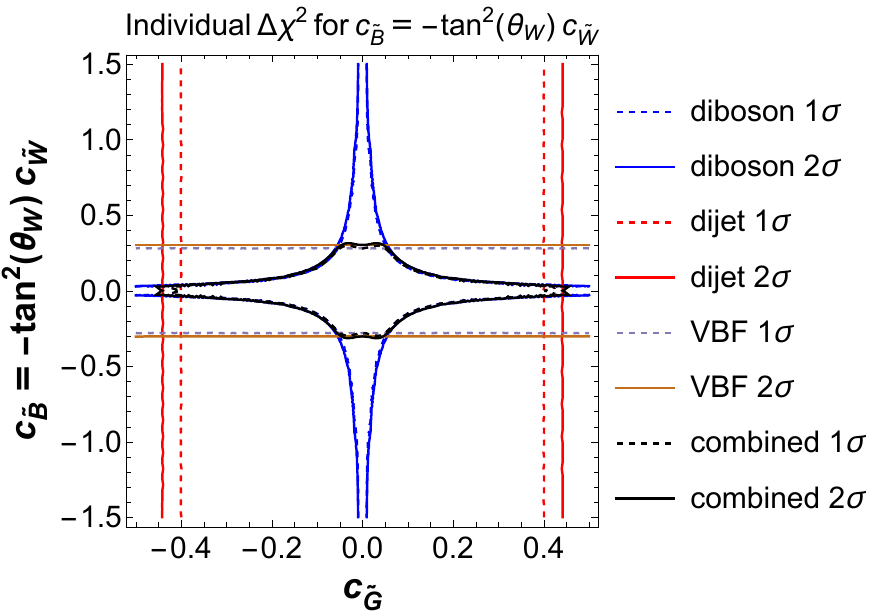}
    \caption{Individual contributions from diboson (blue), dijet (red) and VBF (green) channels, and combined limits (black), in the plane $c_{\Bt} = c_{\Wt}$ (top) and in the photophobic plane $c_{\Bt}= - \tan^2(\theta_W) \, c_{\Wt}$ (bottom). }
\label{fig:global_fit_2D_individual}
\end{figure}

Fig.\ \ref{fig:global_fit_2D_individual}  shows the contours that one obtains from subsets of the full set of final states shown for the previously discussed projections in the $c_{\Bt} = c_{\Wt}$ (top) and ${c_{\Bt} = -\tan^2(\theta_W) \, c_{\Wt}}$ (bottom) planes. As expected, diboson final states provide the tightest constraints for $c_{\Gt} \neq 0$ and $c_{\Bt} = c_{\Wt} \neq 0$ in both cases. Dijet final states depend only on $c_{\Gt}$ and close the contour for $c_{\Bt} = c_{\Wt} =0$.
VBF final states, on the other hand, depend on $c_{\Bt}$ and $c_{\Wt}$ but not $c_{\Gt}$ and close the limits for $c_{\Gt} \to 0$.

\subsection{Projected, profiled, and marginalised limits}

Table~\ref{tab:limits} summarises the numerical limits on $c_{\Gt}$, $c_{\Bt}$, and $c_{\Wt}$ 
under three different statistical prescriptions:  

\paragraph{Projected limits:} obtained by slicing the 3D likelihood along the axis of a single 
  coefficient while fixing the others to zero. These limits correspond to the 2$\sigma$ 
  threshold $\Delta \chi^2 = 8.02$ in three dimensions.  

\paragraph{Profiled limits:} obtained by minimizing $\chi^2$ with respect to the other two 
  coefficients. This produces an effective 1D likelihood $\chi^2_{\text{prof}}(c_i)$ 
  with 2$\sigma$ corresponding to $\Delta \chi^2 = 4$. As expected, the profiled limits 
  are stronger than the projected ones.  

\paragraph{Marginalised limits:} defined as the extremal values of each coefficient on the 
  $1\sigma$ and $2\sigma$ surfaces of the 3D fit, with the $2\sigma$ limit corresponding to $\Delta \chi^2 = 8.02$. They have a direct geometric 
  interpretation and facilitate comparison with the projected case, although they are 
  somewhat weaker due to the full volume of parameter space being taken into account.  \\

\begin{table}[h!]
    \centering
    \begin{tabular}{|l||c|c||c|c||c|c|}
    \hline
    & \multicolumn{2}{|c||}{Projected} & \multicolumn{2}{|c||}{Profiled} & \multicolumn{2}{|c|}{3D marginalised} \\
    \hline
    & $1\sigma$ & $2\sigma$ & $1\sigma$ & $2\sigma$ & $1\sigma$ & $2\sigma$ \\   
    \hline
    $c_{\Gt}$ & 0.40 & 0.44 & 0.35 & 0.41 & 0.46 & 0.48 \\
    $c_{\Bt}$ & 1.77 & 1.89 & 1.60 & 1.78 & 1.92 & 2.00 \\
    $c_{\Wt}$ & 1.07 & 1.16 & 0.85 & 1.09 & 1.26 & 1.33 \\ 
    \hline
    \end{tabular}
    \caption{Comparison of projected, profiled, and marginalised $1\sigma$ and $2\sigma$ 
    limits for $c_{\Gt}$, $c_{\Bt}$, and $c_{\Wt}$. All numerical values quoted for the
Wilson coefficients $c_{\tilde X}$ correspond to the reference choice
$f_a = 1~\mathrm{TeV}$ and should be interpreted as bounds on the ratios
$c_{\tilde X}/f_a$.
}
    \label{tab:limits}
\end{table}

As expected, the profiled bounds are tighter 
than the projected ones, since correlations with the other couplings are taken into 
account. The marginalised limits are somewhat weaker than the projected case, reflecting 
the volume of parameter space that remains allowed once all three coefficients vary 
simultaneously. 

Among the three operators, $c_{\Gt}$ is most strongly constrained, with 
$c_{\Gt}\lesssim 0.4$ at the $2\sigma$ level, consistent with the dijet analysis that 
dominates the $c_{\Gt}$ direction in Fig.~\ref{fig:global_fit_3D}. The $WW$ and $ZZ$ 
channels provide the leading sensitivity to $c_{\Wt}$, while the bounds on $c_{\Bt}$ 
remain weaker, consistent with the partial cancellations between $c_{\Bt}$ and $c_{\Wt}$ 
in the $a\gamma Z$ coupling, as seen in the 2D projections of 
Fig.~\ref{fig:global_fit_2D_projections}. 

Numerically, profiling improves the limits relative to projection by about 15\% for 
$c_{\Gt}$, 10\% for $c_{\Bt}$, and nearly 25\% for $c_{\Wt}$. This illustrates the 
importance of treating correlations correctly in a multi-parameter fit. All results 
assume $f_a=1~\text{TeV}$ and rescale as $1/f_a$. The numbers quoted in the table 
therefore provide a compact summary of the exclusion regions shown in 
Figs.~\ref{fig:global_fit_3D} and~\ref{fig:global_fit_2D_projections}.

\subsection{Leading new physics contributions and relative constraining power}
\label{sec:leading_contribution}
Focusing on off-shell intermediate ALPs, all processes contributing to an observable $\mathcal{O}^a$ necessarily contain two insertions of new physics vertices, meaning that we can parametrise the amplitude by
\begin{equation}
    \mathcal{A}^a = \mathcal{A}^{a,SM} + \sum_{ij} \mathcal{A}^{a,ALP}_{ij} c_i c_j
\end{equation}
for $c_i, c_j \in \{c_{\Gt}, c_{\Bt}, c_{\Wt} \}$.
Then, the observable, e.g.\ the differential cross section, is calculated from
\begin{equation}
    \mathcal{O}^a = \big|\mathcal{A}^{a,SM}\big|^2 + 2 \text{Re} \left[\sum_{ij} \left(\mathcal{A}^{a,SM}\right)^* \mathcal{A}^{a,ALP}_{ij} c_i c_j \right] + \text{Re} \left[\sum_{ijkl} \left(\mathcal{A}^{a,ALP}_{ij}\right)^* \mathcal{A}^{a,ALP}_{kl} c_i c_j c_k c_l \right]  + \ldots
\label{eq:observable_expansion}
\end{equation}
In this article we focus on the quadratic new physics contribution and do not consider any interference terms with the SM, a choice which we comment on in App.\ \ref{sec:interference}. 
Under this assumption, the quartic term in $c$, i.e.\ the last term in Eq.\ (\ref{eq:observable_expansion}), provides the dominant ALP contribution to the observables. In particular, the linear term in $c_i$ and the Fisher matrix, which is commonly used to quantify the constraining power of different observables on each Wilson coefficient, are identical zero. Physically speaking, the minimum of the $\chi^2$ distribution is found for all $c_i = 0$, but the slope and curvature around the minimum are zero, and the deviation from the minimum is a quartic polynomial in $c_i$. This also implies that the fit can not conclusively fix the sign of any Wilson coefficient, a fact that is reflected by the exclusion plots being symmetric under $c_i \rightarrow -c_i$ in quadratic terms or $c_i \rightarrow -c_i, c_j \rightarrow -c_j$ in mixed terms.\\
With 
\begin{equation}
    S_{ijkl}^a \equiv \text{Re} \left[\left(\mathcal{A}^{a,ALP}_{ij}\right)^* \mathcal{A}^{a,ALP}_{kl}\right]
\end{equation}
we define for each observable
\begin{equation}
    I^a_{ijkl} = \sum_b S_{ijkl}^a \Sigma_{ab}^{-1} S_{ijkl}^b
\end{equation}
with the experimental covariance matrix $\Sigma_{ab}$ between different observables, which is relevant e.g.\ for a distribution in various bins when the correlation between the bins is reported.
For uncorrelated observables, this reduces to
\begin{equation}
        I^a_{ijkl} = \left(\frac{S_{ijkl}^a}{\sigma_{a}}\right)^2
\end{equation}
for the experimental uncertainty $\sigma_a$ for observable $\mathcal{O}_a$. All observables considered in this article belong to one of the  sectors $\gamma \gamma$, $ZZ$, $WW$, dijet and VBF. For a given sector $A$, we define
\begin{equation}
I^A_{ijkl} = \sum_{a \in A} I^a_{ijkl} \,,
\end{equation}
which holds for uncorrelated observables in each sector. 
This allows us to calculate the \textit{relative constraining power of a dataset A on a set of Wilson coefficients} $\{c_i,c_j,c_k, c_l\}$ as
\begin{equation}
    f_{ijkl} = \frac{I_{ijkl}^A}{\sum_B I_{ijkl}^B} \,.
\label{eq:rel_constraining_power}
\end{equation}

\begin{table}[h!]
    \centering
    \begin{tabular}{|c||c|c|c|c|c|}
    \hline
    $\left(c_{\Gt}, c_{\Bt}, c_{\Wt}\right)$ & $\gamma \gamma$ & $ZZ$ & $WW$ & dijet & VBF \\ 
    \hline
    (1,0,0) & 0 & 0 & 0 & 1 & 0 \\
    (0,1,0) & 0 & 0 & 0 & 0 & 1 \\
    (0,0,1) & 0 & 0 & 0 & 0 & 1 \\
    \hline
    (1,1,0) & 0.303 & 0.103 & 0 & 0.594 & $8.83\cdot 10^{-6}$ \\
    (1,0,1) & $9.02\cdot 10^{-6}$ & 0.0662 & 0.931 & 0.003 & $1.25 \cdot 10^{-6}$ \\
    (0,1,1) & 0 & 0 & 0 & 0 & 1 \\
    \hline
    (1,1,1) & 0.003 & 0.162 & 0.832 & 0.00233 & $1.09 \cdot 10^{-6}$ \\
    \hline
    \end{tabular}
    \caption{The relative constraining power of the five datasets on combinations of Wilson coefficients according to Eq.
     (\ref{eq:rel_constraining_power}). The first three rows show scenarios with individual Wilson coefficients, the next three rows the constraining power on pairs of Wilson coefficients and the last row all three Wilson coefficients at a time. 
}
\label{tab:rel_constraining_power}
\end{table}

These relative constraining powers for various combinations of Wilson coefficients are shown in Tab.\ \ref{tab:rel_constraining_power}. Since the new physics enters via quartic terms in the Wilson coefficients, the one-coefficient-at-a-time assumption does not reveal much new information: Diboson production processes, i.e.\ $\gamma \gamma$, $ZZ$ and $WW$, depend on both $c_{\Gt}$ and a combination of $c_{\Bt}$ and $c_{\Wt}$. Hence, they can not constrain any Wilson coefficient on its own and the individual constraints on $c_{\Gt}$ exclusively come from dijet measurements and the individual constraints on $c_{\Bt}$ and $c_{\Wt}$ from VBF. The next three rows constrain pairs of Wilson coefficients. $c_{\Gt} = c_{\Bt} =1$ shows an interplay between $\gamma \gamma$ and dijet while $c_{\Gt} = c_{\Wt} =1$ is dominated by WW measurements. Only for $c_{\Bt} = c_{\Wt} =1$ VBF becomes crucial to break the degeneracy in $c_{\Gt}$. The last row shows the importance of each dataset for $c_{\Gt} = c_{\Bt} = c_{\Wt} =1$, with the largest contribution coming from WW. This analysis puts the relative constraints into context and also highlights the interplay of the different channels to break degenerate directions.

\subsection{EFT validity}
Our analysis relies on an ALP EFT with dimension–five operators \( (4/f_a)\,c_i\,a\,X\tilde X \).
For off–shell exchange the relevant hard scale is \(\sqrt{\hat s}\simeq m_{VV}\) (or \(m_{jj}\) in dijets).
A conservative way to assess EFT validity is to compare \(\sqrt{\hat s}\) with the EFT cutoff estimated by Naive Dimensional Analysis (NDA),
\(\Lambda_{\rm NDA}\sim 4\pi f_a/\max|c_i|\). Requiring \(\sqrt{\hat s}_{\max} \ll \Lambda_{\rm NDA}\) yields
\[
|c_i| \;\simeq\; \frac{4\pi\,f_a}{\sqrt{\hat s}_{\max}}\,.
\]
With \(f_a=1~\text{TeV}\) and the kinematic reaches used here
(dibosons \(m_{VV}\!\lesssim\!0.5\!-\!1.5~\text{TeV}\), VBF \(m_{VV}\!\lesssim\!1~\text{TeV}\), dijets \(m_{jj}\!\lesssim\!8~\text{TeV}\)),
this translates into indicative EFT–consistency windows:
\[
\begin{aligned}
&\text{dibosons (}\sim 1.5~\text{TeV})\!: && |c_i|\;\simeq\; \frac{4\pi}{1.5}\;\approx\; 8.4,\\
&\text{VBF (}\sim 1~\text{TeV})\!:          && |c_i|\;\simeq\; 4\pi\;\approx\; 12.6,\\
&\text{dijets (}\sim 8~\text{TeV})\!:        && |c_i|\;\simeq\; \frac{4\pi}{8}\;\approx\; 1.6.
\end{aligned}
\]
Our 95\%\,CL bounds (Table~\ref{tab:limits}) lie comfortably within these NDA ranges:
for example \(c_{\Gt}\!\lesssim\!0.4\) from dijets is well below the \(\sim\!1.6\) EFT ceiling implied by \(m_{jj}\!\sim\!8~\text{TeV}\),
and the electroweak coefficients remain \(\mathcal{O}(1)\), far below their diboson/VBF ceilings.
For an ultra–conservative criterion one may take the cutoff as \(\Lambda\sim f_a/\max|c_i|\),
which would instead require \(|c_i|\lesssim f_a/\sqrt{\hat s}_{\max}\) (giving \(|c_i|\lesssim 0.67,1.0,0.125\) for the three channels above);
our limits then remain safely within the diboson/VBF ranges, while dijets become closer to the edge.
In practice one can enforce EFT robustness by truncating the fit to bins with \(\sqrt{\hat s}<\kappa f_a\) (e.g.\ \(\kappa=1\!-\!2\));
we have checked that our conclusions are stable under this replacement of the high–energy tail by an explicit truncation.

Note that, rather than imposing a hard cut on the partonic centre-of-mass energy
$\hat{s}$, we retain the full experimental information and accompany our
results with an explicit discussion of the energy scales probed by the most
sensitive bins. Introducing a sharp truncation would be highly
process-dependent and would require additional assumptions on the structure
of the ultraviolet completion. Our approach follows standard practice in
global EFT analyses, where high-energy data are included but the resulting
constraints are interpreted as bounds on effective operators, whose UV
completion must restore perturbative unitarity above the probed energies.
In the most sensitive channels, the dominant constraints arise from bins with
typical invariant masses of order a few TeV, which we indicate explicitly
in the discussion above.

In parallel to these EFT–consistency estimates based on the size of the expansion parameter, recent work \cite{Bresciani:2025ojh} has derived substantially stronger constraints by applying partial-wave unitarity and positivity bounds to ALP–gauge amplitudes. These bounds rely on different theoretical assumptions—most notably the on-shell formalism and coupled-channel unitarity—and therefore are not directly comparable to our conservative NDA criteria, but they provide a valuable and highly complementary perspective that we now incorporate and cite in the updated version.

%%%%%%%%%%%%%%%%%%%%%%%%%%%%%%%%%%%%%%%%%%%%%%%%%%%%%%%%%%%%%%%%%%%%%%%%%%%%%%%%%%%%%%%%
\section{Conclusions}
\label{sec:conclusions}

We have presented a global analysis of ALP–mediated multiboson production at the LHC in an
EFT with bosonic, dimension-five couplings. A key feature of our approach is that the ALP is
exchanged \emph{off-shell}, so sensitivity arises from distortions of high-energy tails rather than
resonant bumps. This makes the analysis largely insensitive to the ALP mass provided
$m_a \ll \sqrt{s}$, and allows us to combine information across channels with different kinematics
and backgrounds.

On the theory side, we assumed ALP couplings exclusively to the SM gauge sector and motivated this pattern from a UV perspective, then classified the irreducible ALP–multiboson vertices and mapped the setup onto the low-energy couplings $(c_{\Gt},c_{\Wt},c_{\Bt})$. On the phenomenology side,
we confronted the EFT with a broad set of LHC measurements: inclusive $\gamma\gamma$, $ZZ\to 4\ell$,
$W^+W^-\to \ell\ell\nu\nu$, double-differential dijets, and VBF topologies ($Z\gamma jj$ and $VVjj$ with
resolved/merged categories). Each channel probes a different linear combination of Wilson coefficients:
dijets fix $c_{\Gt}$ through $g_{agg}$; $WW$ and $ZZ$ in turn constrain $c_{\Wt}$ via $g_{aWW}$ and $g_{aZZ}$; and
$Z\gamma jj$/$VVjj$ access both $c_{\Wt}$ and the difference $(c_{\Wt}-c_{\Bt})$ through $aZ\gamma$ and $aZZ$
interference. Their complementarity is essential to break flat directions.

We then performed a simultaneous three-parameter fit to $(c_{\Gt},c_{\Bt},c_{\Wt})$ at $f_a=1~\text{TeV}$
(with all limits rescaling as $1/f_a$), reporting projected, profiled, and marginalised constraints.
Numerically (Table~\ref{tab:limits}), the dijet spectrum yields the strongest bound on the gluonic
coefficient, $c_{\Gt}\lesssim 0.4$ at $2\sigma$, while $WW$/$ZZ$ lead the constraints on $c_{\Wt}$ and the
VBF channels close the $(c_{\Wt},c_{\Bt})$ plane, particularly in the photophobic-like direction
$c_{\Wt}\simeq c_{\Bt}$. Profiling improves the limits by $\sim$15\% ($c_{\Gt}$), $\sim$10\% ($c_{\Bt}$),
and nearly $\sim$25\% ($c_{\Wt}$) relative to simple projection, underscoring the need to treat correlations
consistently in a multi-parameter fit. We also checked EFT robustness using the relevant hard scales
($m_{VV}$, $m_{jj}$): our bounds lie comfortably within the NDA windows for diboson and VBF, and remain
below the conservative ceiling implied by the $m_{jj}\!\sim\!8$~TeV reach of the dijet analysis.

Several improvements are straightforward. On the experimental side, fuller use of unfolded
covariance matrices (where available), inclusion of the highest-luminosity $WW/ZZ$ measurements, and
additional electroweak channels ($Z\gamma$, $W\gamma$, $\gamma\gamma{+}$jet, tribosons, and polarized VBS)
will sharpen the $(c_{\Wt},c_{\Bt})$ constraints. On the theory side, incorporating NLO QCD/EW corrections
and matched predictions for the most sensitive tails would reduce theory systematics; extending the fit to
include loop-induced fermionic effects (notably the top) will turn our discussion of
$c_t$ into a quantitative four-parameter analysis. Finally, repeating the fit with an explicit energy truncation
$\sqrt{\hat s}<\kappa f_a$ provides a conservative cross-check that our conclusions are stable against the
highest-energy bins.

In summary, off-shell multiboson production offers a coherent and competitive programme to test ALP–gauge
interactions. The present global fit delivers stringent bounds on $(c_{\Gt},c_{\Wt},c_{\Bt})$ from existing
LHC data and clarifies the unique role of VBF and dijet channels in closing the parameter space. 
With HL-LHC
statistics and improved measurements, this strategy can decisively probe $\mathcal{O}(1)$ electroweak ALP
couplings and push the sensitivity to gluonic interactions well into the EFT-safe regime. Finally, looking further ahead, channels with three gauge bosons are not yet competitive, but as triboson and aQGC measurements improve with HL-LHC  they will provide a powerful complementary handle on ALP interactions.

\section*{Acknowledgements}
%We would like to thank \MUB{Anyone to thank?} for preliminary studies and discussions. 
The work
of V.S. is supported by the Spanish grants PID2023-148162NB-C21, CNS-2022-135688,
and CEX2023-001292-S. The work of M.U. is supported by the European Research Council
under the European Union’s Horizon 2020 research and innovation Programme (grant agreement n.950246) and in part by STFC consolidated grant ST/X000664/1. 
F.E. is supported from Charles University through project PRIMUS/24/SCI/013. A.S.B. acknowledges support from grants PID2023-148162NB-C21 and PID2022-137003NB-I00 from Spanish MCIN/AEI/
10.13039/501100011033/ and EU FEDER.

\appendix

\section{On the interference with SM amplitudes}
\label{sec:interference}
As commented in Sec.\ \ref{sec:leading_contribution}, for all limits given in this article, we focus on the pure quadratic ALP contributions to the total cross section and do not take into account interference terms with the SM. 
Since we can analytically trace the scaling of the quadratic ALP contribution to the cross section in all channels, i.e.\ the last term in Eq.\ (\ref{eq:observable_expansion}), it suffices to generate signal events for one finite choice of $c_i$. Therefore, we can model the resulting $\chi^2$ distribution as an analytic function in $c_{\tilde{G}}$, $c_{\tilde{B}}$ and $c_{\tilde{W}}$ and determine precise contours. If we included interference effects, we could not determine the scaling of the total cross section with the Wilson coefficients and would instead have to sample events for many different choices and interpolate the signal on this grid simulations. 

Here, we justify the choice to focus on the quadratic ALP contribution in hindsight by estimating the importance of the interference terms for the two channels that, according to e.g.\ Table \ref{tab:rel_constraining_power}, have the largest relative constraining power, namely WW and dijet.

\begin{figure}[h!]
    \centering
    \includegraphics[width=0.8\linewidth]{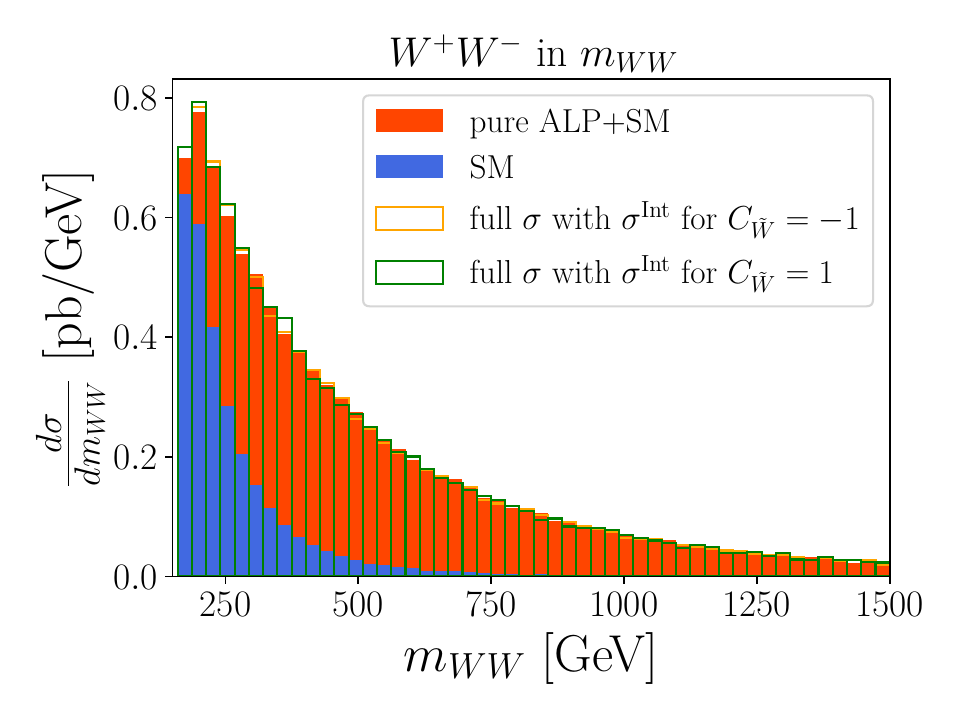}
    \caption{Differential cross section in the $WW$ final state as a function of the diboson invariant mass. The
    signal is shown for $c_{\tilde G}=1$, $c_{\tilde W}=1$ and $c_{\tilde B}=0$, compared to the ATLAS data
    and the SM background simulation in \cite{ATLAS:2019rob}. We add the contours of the full differential cross section including interference terms with the SM for $c_{\tilde W}=-1$ (yellow) and $c_{\tilde W}=1$ (green).}
    \label{fig:WW_interference}
\end{figure}

In the WW channel, and similarly in the other diboson channels $\gamma \gamma$ and ZZ, the new physics amplitudes do not interfere with the SM model at the partonic level, as there is no SM contribution to $gg \rightarrow W^+ W^-$. Hence, the interference with the SM arises only at the hadronic level after convoluting the partonic amplitudes with the parton distribution functions in the proton, and are accordingly small. 
The new physics amplitude, which is independent of $c_{\tilde{B}}$, scales with $c_{\tilde{G}} c_{\tilde{W}}$ and the interference term depends on the sign of $c_{\tilde{G}}$ and $c_{\tilde{W}}=1$. In Fig.\ \ref{fig:WW_interference}, we show the differential cross section for the pure quadratic ALP signal (red) and the SM background (blue) in the diboson invariant mass, similar to Fig.\ \ref{fig:WW}, and add contours for the full cross section including the interference terms for $c_{\tilde{G}}=1$ and the two choices $c_{\tilde{W}}=-1$ (yellow) and $c_{\tilde{W}}=1$ (green). It is striking that the full cross section including the interference agrees well with the signal without interference. 

To quantify this statement at the level of the total cross section, for $c_{\tilde{G}}=1$ and $c_{\tilde{W}}=1$, we find
\begin{equation}
    \sigma_{WW}^{\rm SM} = \unit[76.74]{pb},\quad \sigma_{WW}^{\rm ALP} = \unit[181.13]{pb},\quad \sigma _{WW}^{\rm Int} = \unit[-0.02]{pb}\,,
\end{equation}
meaning that for values of the Wilson coefficients in the range of the limits found in this article, the interference is indeed negligible with respect to the pure quadratic ALP signal.

\begin{figure}[h!]
    \centering
    \includegraphics[width=0.48\linewidth]{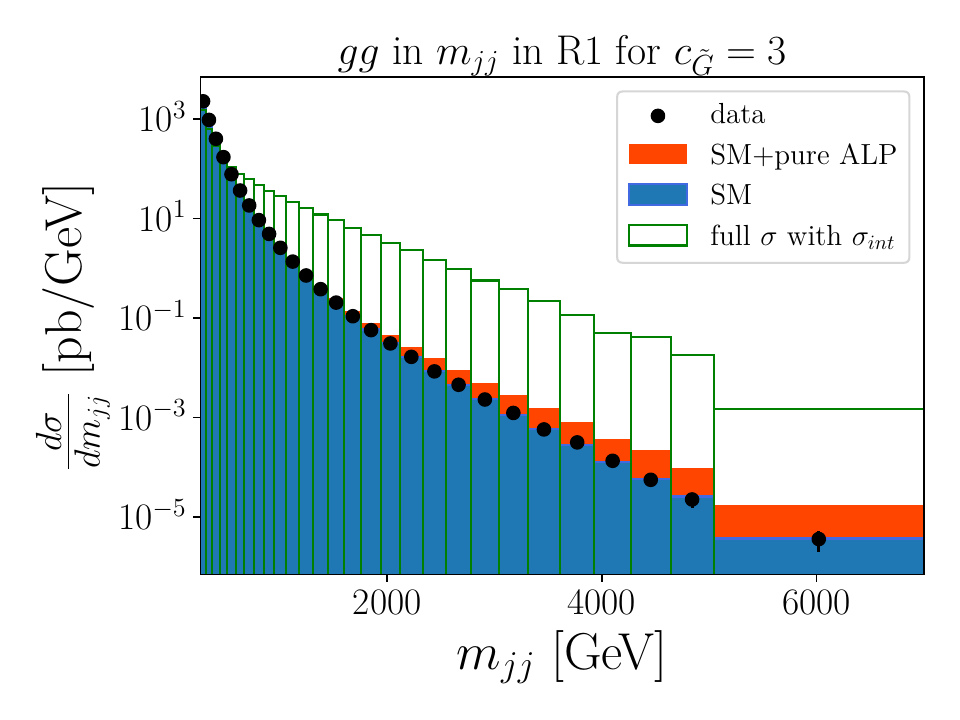}
    \includegraphics[width=0.48\linewidth]{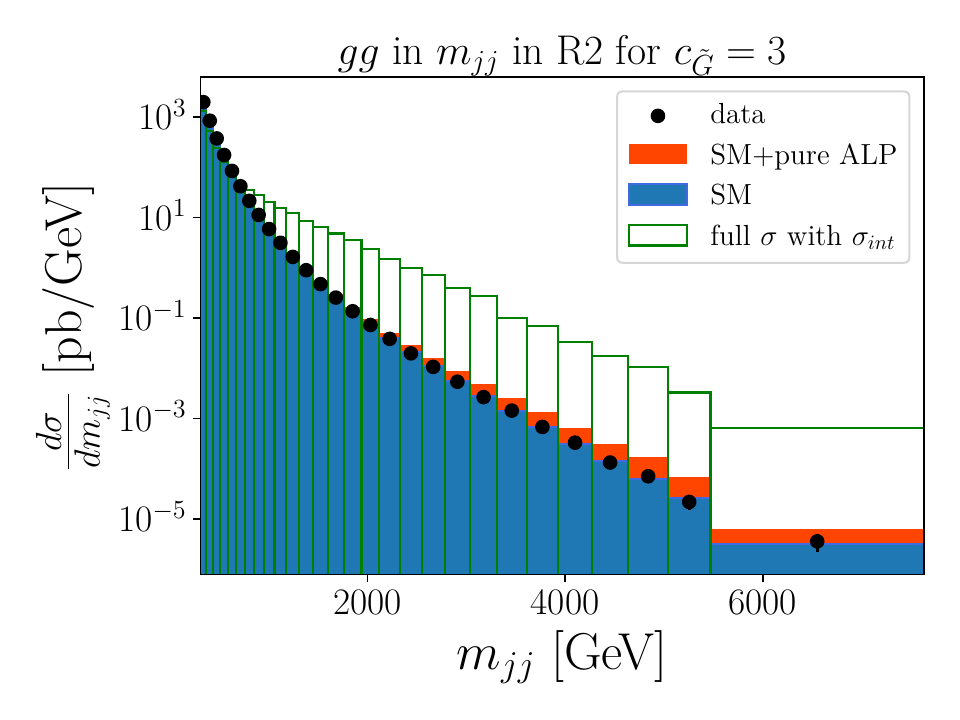}
    \vspace{10pt}
    \includegraphics[width=0.48\linewidth]{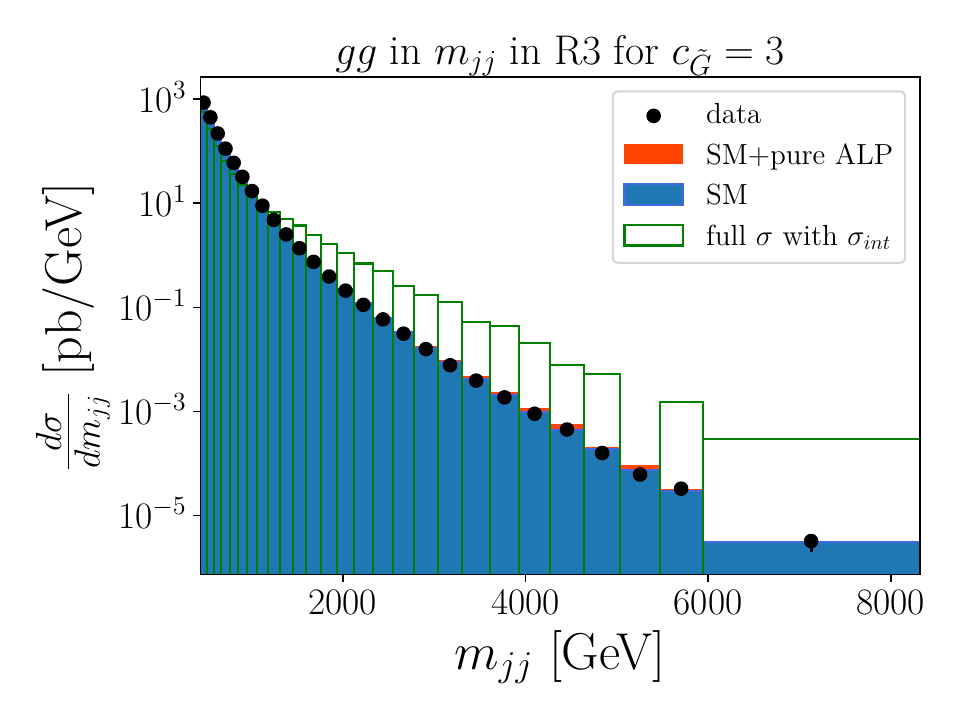}
    \includegraphics[width=0.48\linewidth]{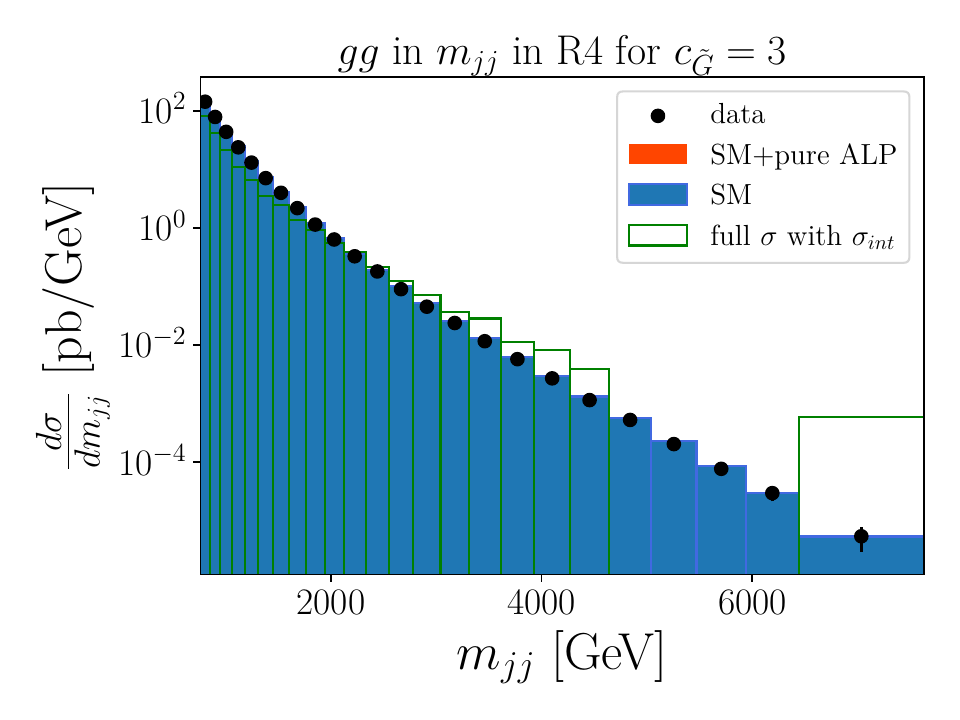}
    \vspace{10pt}
    \includegraphics[width=0.48\linewidth]{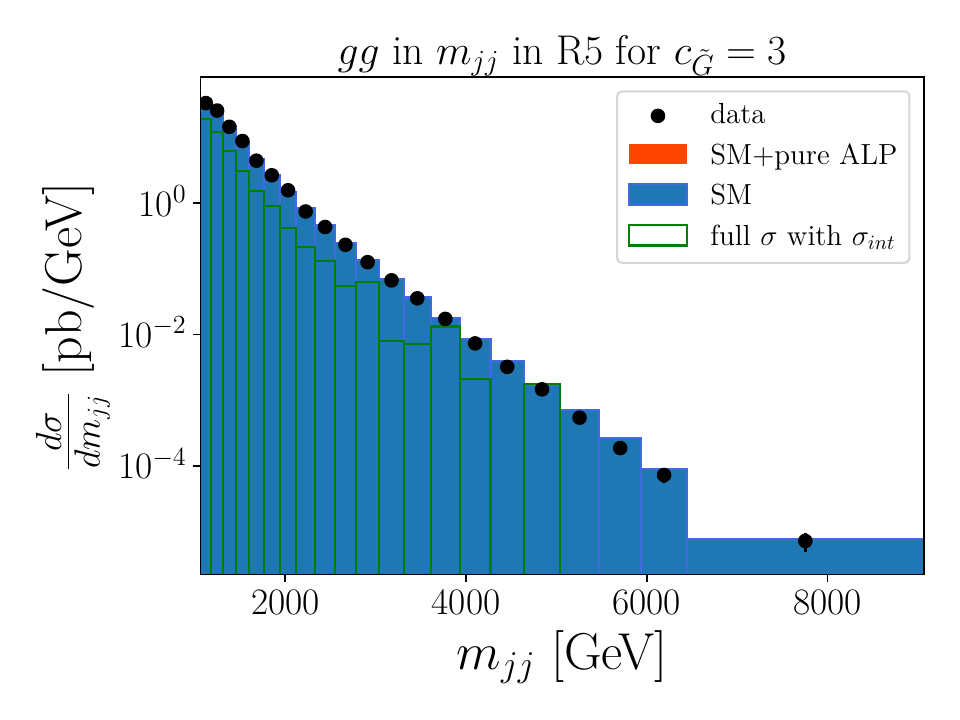}
    \includegraphics[width=0.48\linewidth]{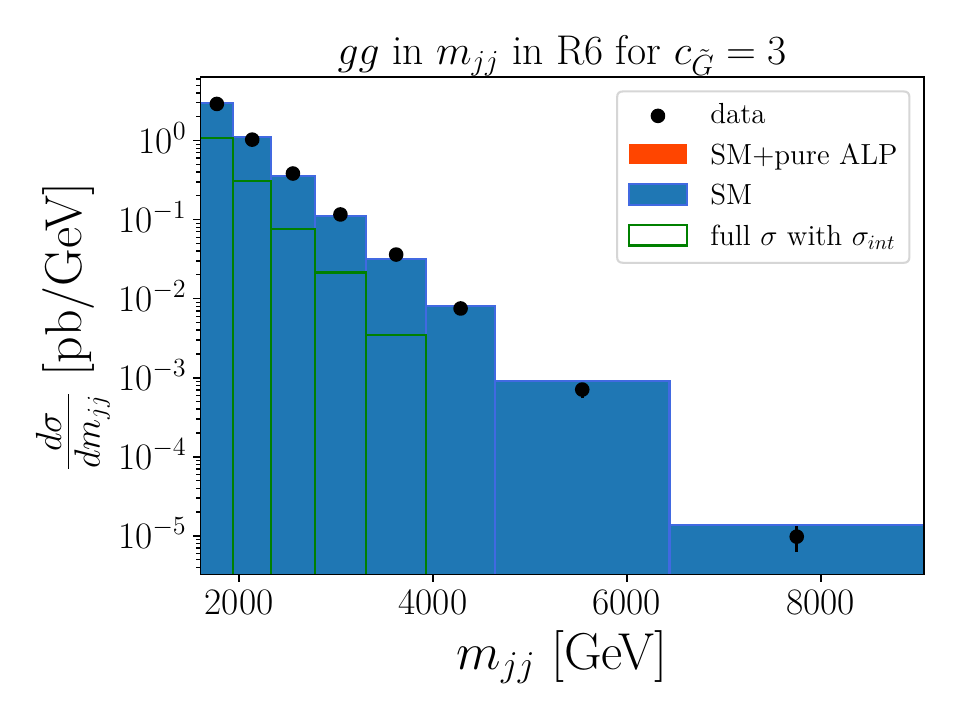}
    \vspace{10pt}
    \caption{Double-differential distribution in $m_{jj}$ and $y^*$ for $pp\to a^\ast\to gg$ for pure ALP signal (red) and including interference terms (green), compared to 
    ATLAS data~\cite{ATLAS:2017ble} and the SM background. For better visibility, the signal is shown for $c_{\tilde G}=10$, $c_{\tilde B}=c_{\tilde W}=0$.}
    \label{fig:gg_dijet_interference}
\end{figure}

Next, we consider the dijet final, for which the new physics amplitude depends quadratically on $c_{\tilde{G}}$, which implies that the interference does not depend on the sign of $c_{\tilde{G}}$. The process $gg \rightarrow gg$ can be mediated in the SM via s-, t- and u-channel exchange of a gluon. We find that the SM cross section is one order of magnitude larger than the pure ALP cross section and the interference effects are of the same order as the pure ALP cross section. 
In Fig.\ \ref{fig:gg_dijet_interference} we show the double differential cross section in the invariant dijet mass and $y*$, as in Fig.\ \ref{fig:gg_dijet}, for the background distribution and the signal, assuming $c_{\tilde{G}}=3$ for better visibility. In green we add the contour for the full cross section with the interference term for $c_{\tilde{G}}=3$.

For most bins, in particular in the signal regions with small $y*$, in which many events are observed, the new physics amplitudes interfere positively with the SM cross section and including the interference terms predicts a higher signal. In principle, one would expect that including the interference effects we could get better limits on $c_{\tilde{G}}$. However, apart from the observation that we can not analytically trace the scaling of the interference term with $c_{\tilde{G}}$, there is another caveat: In Sec.\ \ref{sec:dijet} we use the SM background estimation provided by ATLAS, which includes higher order calculation including QCD effects to an accuracy that we can not match with our MG5 simulation. As long as we consider only the pure ALP signal, we can estimate the total number of events by adding our signal estimation to the provided background. Once we include interference effects, the MG5 simulation will inevitably also contain the SM background estimation that can no longer be disentangled, and we can not use the ATLAS SM background. Therefore, the calculated $\chi^2$ difference will be huge due to the worse background estimation which overshadows the actual contribution of the interference term, and all limits derived from this distribution are worse than the limits given for the pure ALP signal in this article.

We conclude that in the dijet channel, interference effects can be important and could in principle be used to derive stronger limits on $c_{\tilde{G}}$, but to match the sensitivity of our analysis, an improved background estimation in MG5 would be necessary. The limits assuming a pure quadratic ALP signal that are derived in this article should therefore be seen as conservative bounds on the couplings.

\section{Fiducial Cuts}\label{app:cuts}
\subsection{$\gamma\gamma$ final state}
The fiducial cuts used for our signal estimation are taken from \cite{ATLAS:2021mbt}, which require the presence of at least two clusters of energy depositions
with $|\eta|< 2.47$, with transverse momentum  $p_T>30$ GeV and requiring that at least one photon has $p_T>40$ GeV. The angular separation of the photons is required to be $\Delta R_{\gamma\gamma} > 0.4$ and we require that  invariant mass of the photon to be $m_{\gamma\gamma}>200$ GeV. The integrated luminosity is 139 fb$^{-1}$.

The binning used for the differential cross section with respect to the invariant mass of the photon pair is shown only up to 500 GeV with bin limits at (13.5, 20, 22.6, 25, 27, 29.1, 31, 33.2, 35.4, 37.8, 40.4, 43, 46, 49, 52.4, 55.8, 59.6, 63.4, 67.6, 72.2, 77, 82, 87.5, 93, 99, 105, 112, 118.5, 126, 133.5, 141, 149.5, 158, 166.5, 175.5, 185, 194.5, 205, 215, 226, 238, 24, 262, 274, 287, 301, 315, 330, 345, 361, 377, 394, 411, 429, 448, 467, 487, 508) GeV.

\subsection{$ZZ$ final state}
The fiducial cuts based on the ATLAS measurement of $pp\to ZZ\to 4\ell$ at 
$\sqrt{s}=13$~TeV with 29~fb$^{-1}$ integrated luminosity~\cite{ATLAS:2023dew} are:
the cuts for leptons in the final state are $p_T>5$ GeV and $|\eta|<2.47$ with a minimal angular separation of $\Delta R>0.05$. In addition, the two same flavour opposite charge
lepton pairs invariant masses are required to satisfy $m_{ll} \in (66,116)$ GeV. Finally, the invariant mass of the four leptons in
the two pairs is required to be $m_{4l} > 180$ GeV.

The binning used for the differential cross section with respect to the invariant mass of the $Z$-boson pair is  defined by the bin limits: (180,  200,  220,  240,  265,  290,  320,  350,  390,440,  520,  620, 1200) in GeV.

\subsection{$W^+W^-$ final state}

Fiducial cuts for the ALP mediated  $pp\to W^+W^-\to \ell^+\nu \ell^-\bar\nu$ are based on the ATLAS measurement at 
$\sqrt{s}=13$~TeV with an integrated luminosity of 36.1~fb$^{-1}$~\cite{ATLAS:2019rob}. These are:
Leptons are required to
have $p_T > 27$ GeV and $|\eta| < 2.5$. The minimum missing transverse energy corresponding to the neutrinos is $E_T^{\text{miss}}>20$ GeV. The minimum angular separation for leptons is $\Delta R > 0.4$. The minimum invariant mass and transverse momentum for all leptons and neutrinos is $m_{T,4l}>55$ GeV and $p_{T, 4l}> 30$ GeV.

The binning used for the differential cross section with respect to the invariant mass of the $W$-boson pair is defined by the bin limits: $(55,75,85,95,110,125,140,160,185,220$  \\$160,185,220,280,380,600,1500)$ in GeV and all signal events fall in the binning region.

\subsection{Dijet final state}
The fiducial cuts for simulating our ALP mediated $pp\to jj$ signal are based on the ATLAS measurement \cite{ATLAS:2017ble} with integrated luminosity of $3.2$ fb$^{-1}$. These are:  The minimum transverse momentum of individual jets is $p_T> 75$ GeV with angular separation $\Delta R > 0.4$ and maximum rapidity of $|\eta|<3$. The total invariant mass for the two jets satisfies $m_{T, 2j}\in (260,7000)$ GeV. Finally, the sum of magnitudes of transverse momenta of the two jets fulfils $H_{T,2j}>200$ GeV.

The binning used for the double-differential cross section with respect to the invariant mass of the jet pair and the half-rapidity separation, $\frac{d^2 \sigma}{d m_{jj} d y*}$, is
\begin{itemize}
    \item For the half-rapidity region R1, $0 < y* < 0.5$, there are 29 bins from 260 GeV to 7000 GeV.
    \item For the region R2, $0.5 < y* < 1$, there are 29 bins from 310 GeV to 7630 GeV.
    \item For the region R3, $1 < y* < 1.5$, there are 28 bins from 440 GeV to 8317 GeV.
    \item For the region R4 $1.5 < y* < 2$, there are 25 bins from 760 GeV to 7630 GeV.
    \item For the region R5, $2 < y* < 2.5$, there are 22 bins from 1060 GeV to 9066 GeV.
    \item For the region R6, $2.5 < y* < 3$, there are 9 bins from 1600 GeV to 9066 GeV.
\end{itemize}
For the different bin widths we refer to the ATLAS data repository of \cite{ATLAS:2017ble}. 
Of our generated signal events, 93.0781\% fall into the given ATLAS bins.

\subsection{Vector Boson Fusion events}
The fiducial cuts for our VBF ALP-mediated signal are:
\paragraph{$Z\gamma$ final state:}

For this final state, the selection cuts for our signal estimation are the same as in the CMS measurement of $pp\to Z\gamma jj$ at $\sqrt{s}=13$~TeV with 
137~fb$^{-1}$~\cite{CMS:2021gme}, where the $Z$ boson is reconstructed in the 
dilepton channel. 
The cuts in the transverse momentum are $p_T>30$ GeV for the jets and  $p_T>20$ GeV for the photon; in the pseudorapidity $|\eta|<4.7$ for the jets and $|\eta|<2.5$ for the photons. The minimum angular distance is $\Delta R>0.5$ and imposed both between two jets and between the photon and a jet. The minimum invariant mass for the jet pair is  $m_{jj}>500$ GeV. 

The binning  has bin limits at (20,  80, 120, 200, 400) GeV for $p_T^{\gamma}$ and (30, 150, 250, 350, 800) GeV for $p_T^{j,\text{lead}}$.

\paragraph{$ZZ$ and $W^+W^-$ final states:} In accordance with
\cite{ATLAS:2025omi}, we simulate our events for ALP-mediated VBF di-Weak-boson production at $\sqrt{s}=13$ TeV with the following cuts:
\begin{itemize}
    \item {\textbf{0-lepton channel}} The cut in transverse momentum is $p_T>30$ GeV for the jets and $E^{\text{miss}}_T>200$ GeV for the missing energy coming from neutrinos. The rapidity for jets is $|\eta|<5$ and the angular distance among jets fulfils $\Delta R >0.4$ with a jet pair having invariant mass $m_{jj}>400$ GeV.

    The binning has bin limits at (0, 200, 400, 600, 800, 1000, 1200, 1400, 1600) GeV for the invariant mass of the $ZZ$ system in the merged and resolved regions.
    \item \textbf{1-lepton channel} The cut in transverse momentum is $p_T>30$ GeV for the jets and $p_T>20$ GeV for the lepton ($p_T>27$ GeV for the leading lepton). $E^{\text{miss}}_T>80$ GeV for the missing energy coming from the neutrino. The rapidity for jets is $|\eta|<5$ and the angular distance among jets fulfils $\Delta R_{jj} >0.4$ with a jet pair having invariant mass $m_{jj}>400$ GeV. Finally, the angular distance among jets and the leptons is cut to be $\Delta R_{\ell j} >0.4$ and the invariant mass for all leptons fulfils $m_{\ell \nu}>64$ GeV.

    The binning has bin limits at (0,     200, 400, 600, 800,  1000, 1200, 1400, 1600, 1800, 2000) GeV for the invariant mass of the $W^{+}W^{-}$ system in the merged and resolved regions.
    \item {\textbf{2-lepton channel}}   The cut in transverse momentum is $p_T>30$ GeV for the jets and $p_T>27$ GeV for the leptons. The rapidity for jets is $|\eta|<5$ and $|\eta|<4$ for leptons. The angular distance among all particles fulfils $\Delta R >0.4$ with a jet pair having invariant mass $m_{jj}>400$ GeV.

    The binning has bin limits at (0, 200, 400, 600, 800, 1000, 1200, 1400, 1600, 1800, 2000) GeV for the invariant mass of the $ZZ$ system in the merged and resolved regions.

\end{itemize}

\bibliographystyle{JHEP}
\bibliography{references.bib}

@article{AlonsoAlvarez:2019cgw,
  author       = {G. Alonso‐Álvarez and M. B. Gavela and P. Quílez},
  title        = {Axion couplings to electroweak gauge bosons},
  journal      = {Eur. Phys. J. C},
  volume       = {79},
  number       = {3},
  pages        = {223},
  year         = {2019},
  doi          = {10.1140/epjc/s10052-019-6732-5},
  eprint       = {1811.05466},
  archivePrefix= {arXiv},
  primaryClass = {hep-ph}
}

@article{Weinberg:1977ma,
    author = "Weinberg, Steven",
    title = "{A New Light Boson?}",
    reportNumber = "HUTP-77/A074",
    doi = "10.1103/PhysRevLett.40.223",
    journal = "Phys. Rev. Lett.",
    volume = "40",
    pages = "223--226",
    year = "1978"
}

@article{Choi:2020rgn,
    author = "Choi, Kiwoon and Im, Sang Hui and Sub Shin, Chang",
    title = "{Recent Progress in the Physics of Axions and Axion-Like Particles}",
    eprint = "2012.05029",
    archivePrefix = "arXiv",
    primaryClass = "hep-ph",
    reportNumber = "CTPU-PTC-20-28",
    doi = "10.1146/annurev-nucl-120720-031147",
    journal = "Ann. Rev. Nucl. Part. Sci.",
    volume = "71",
    pages = "225--252",
    year = "2021"
}

@article{Contino:2011np,
    author = "Contino, Roberto and Marzocca, David and Pappadopulo, Duccio and Rattazzi, Riccardo",
    title = "{On the effect of resonances in composite Higgs phenomenology}",
    eprint = "1109.1570",
    archivePrefix = "arXiv",
    primaryClass = "hep-ph",
    doi = "10.1007/JHEP10(2011)081",
    journal = "JHEP",
    volume = "10",
    pages = "081",
    year = "2011"
}

@article{Ferretti:2013kya,
    author = "Ferretti, Gabriele and Karateev, Denis",
    title = "{Fermionic UV completions of Composite Higgs models}",
    eprint = "1312.5330",
    archivePrefix = "arXiv",
    primaryClass = "hep-ph",
    doi = "10.1007/JHEP03(2014)077",
    journal = "JHEP",
    volume = "03",
    pages = "077",
    year = "2014"
}

@article{Sanz:2015sua,
    author = "Sanz, Veronica and Setford, Jack",
    title = "{Composite Higgses with seesaw EWSB}",
    eprint = "1508.06133",
    archivePrefix = "arXiv",
    primaryClass = "hep-ph",
    doi = "10.1007/JHEP12(2015)154",
    journal = "JHEP",
    volume = "12",
    pages = "154",
    year = "2015"
}

@inbook{Biekotter:2025fll,
    author = {Biek{\"o}tter, Anke and Mimasu, Ken},
    title = "{Axions and Axion-like particles: collider searches}",
    eprint = "2508.19358",
    archivePrefix = "arXiv",
    primaryClass = "hep-ph",
    month = "8",
    year = "2025"
}

@article{Alonso-Alvarez:2018irt,
    author = "Alonso-{\'A}lvarez, G. and Gavela, M. B. and Quilez, P.",
    title = "{Axion couplings to electroweak gauge bosons}",
    eprint = "1811.05466",
    archivePrefix = "arXiv",
    primaryClass = "hep-ph",
    reportNumber = "IFT-UAM/CSIC-18-110, FTUAM-18-25",
    doi = "10.1140/epjc/s10052-019-6732-5",
    journal = "Eur. Phys. J. C",
    volume = "79",
    number = "3",
    pages = "223",
    year = "2019"
}

@article{Craig:2018kne,
    author = "Craig, Nathaniel and Hook, Anson and Kasko, Skyler",
    title = "{The Photophobic ALP}",
    eprint = "1805.06538",
    archivePrefix = "arXiv",
    primaryClass = "hep-ph",
    doi = "10.1007/JHEP09(2018)028",
    journal = "JHEP",
    volume = "09",
    pages = "028",
    year = "2018"
}

@article{Agrawal:2024ejr,
    author = "Agrawal, Prateek and Nee, Michael and Reig, Mario",
    title = "{Axion couplings in heterotic string theory}",
    eprint = "2410.03820",
    archivePrefix = "arXiv",
    primaryClass = "hep-ph",
    doi = "10.1007/JHEP02(2025)188",
    journal = "JHEP",
    volume = "02",
    pages = "188",
    year = "2025"
}

@article{Agrawal:2022lsp,
    author = "Agrawal, Prateek and Nee, Michael and Reig, Mario",
    title = "{Axion couplings in grand unified theories}",
    eprint = "2206.07053",
    archivePrefix = "arXiv",
    primaryClass = "hep-ph",
    doi = "10.1007/JHEP10(2022)141",
    journal = "JHEP",
    volume = "10",
    pages = "141",
    year = "2022"
}

@article{Esser:2024pnc,
    author = "Esser, Fabian and Madigan, Maeve and Salas-Bernardez, Alexandre and Sanz, Veronica and Ubiali, Maria",
    title = "{Di-Higgs production via axion-like particles}",
    eprint = "2404.08062",
    archivePrefix = "arXiv",
    primaryClass = "hep-ph",
    doi = "10.1007/JHEP10(2024)164",
    journal = "JHEP",
    volume = "10",
    pages = "164",
    year = "2024"
}

@article{Hosseini:2024kuh,
    author = "Hosseini, Yasaman and Mohammadi Najafabadi, Mojtaba",
    title = "{Exploring axionlike particle couplings through single top tW-channel and top pair production at the LHC}",
    eprint = "2408.11588",
    archivePrefix = "arXiv",
    primaryClass = "hep-ph",
    doi = "10.1103/PhysRevD.110.055026",
    journal = "Phys. Rev. D",
    volume = "110",
    number = "5",
    pages = "055026",
    year = "2024"
}

@article{Esser:2023fdo,
    author = "Esser, Fabian and Madigan, Maeve and Sanz, Veronica and Ubiali, Maria",
    title = "{On the coupling of axion-like particles to the top quark}",
    eprint = "2303.17634",
    archivePrefix = "arXiv",
    primaryClass = "hep-ph",
    doi = "10.1007/JHEP09(2023)063",
    journal = "JHEP",
    volume = "09",
    pages = "063",
    year = "2023"
}

@article{Butterworth:2025szb,
    author = "Butterworth, Jon and Cullingworth, Matthew and Egan, Joseph and Esser, Fabian and Sanz, Veronica and Ubiali, Maria",
    title = "{Probing the coupling of axions to tops and gluons with LHC measurements}",
    eprint = "2508.21660",
    archivePrefix = "arXiv",
    primaryClass = "hep-ph",
    month = "8",
    year = "2025"
}

@article{Gavela:2019cmq,
    author = "Gavela, M. B. and No, J. M. and Sanz, V. and de Troc{\'o}niz, J. F.",
    title = "{Nonresonant Searches for Axionlike Particles at the LHC}",
    eprint = "1905.12953",
    archivePrefix = "arXiv",
    primaryClass = "hep-ph",
    doi = "10.1103/PhysRevLett.124.051802",
    journal = "Phys. Rev. Lett.",
    volume = "124",
    number = "5",
    pages = "051802",
    year = "2020"
}

@article{Bauer:2017ris,
    author = "Bauer, Martin and Neubert, Matthias and Thamm, Andrea",
    title = "{Collider Probes of Axion-Like Particles}",
    eprint = "1708.00443",
    archivePrefix = "arXiv",
    primaryClass = "hep-ph",
    reportNumber = "MITP-17-047",
    doi = "10.1007/JHEP12(2017)044",
    journal = "JHEP",
    volume = "12",
    pages = "044",
    year = "2017"
}

@misc{AxionLimits,
 author   = {Ciaran O’Hare},
 title    = {cajohare/AxionLimits: AxionLimits},
 month    = jul,
 year    = 2020,
 publisher  = {Zenodo},
 version   = {v1.0},
 doi     = {10.5281/zenodo.3932430},
 howpublished = {\url{https://cajohare.github.io/AxionLimits/}}
}

@article{Bruggisser:2023npd,
    author = "Bruggisser, Sebastian and Grabitz, Lara and Westhoff, Susanne",
    title = "{Global analysis of the ALP effective theory}",
    eprint = "2308.11703",
    archivePrefix = "arXiv",
    primaryClass = "hep-ph",
    reportNumber = "NIKHEF-2022-007",
    doi = "10.1007/JHEP01(2024)092",
    journal = "JHEP",
    volume = "01",
    pages = "092",
    year = "2024"
}

@article{Arvanitaki:2009fg,
    author = "Arvanitaki, Asimina and Dimopoulos, Savas and Dubovsky, Sergei and Kaloper, Nemanja and March-Russell, John",
    title = "{String Axiverse}",
    eprint = "0905.4720",
    archivePrefix = "arXiv",
    primaryClass = "hep-th",
    doi = "10.1103/PhysRevD.81.123530",
    journal = "Phys. Rev. D",
    volume = "81",
    pages = "123530",
    year = "2010"
}

@article{Svrcek:2006yi,
    author = "Svrcek, Peter and Witten, Edward",
    title = "{Axions In String Theory}",
    eprint = "hep-th/0605206",
    archivePrefix = "arXiv",
    reportNumber = "SLAC-PUB-11894",
    doi = "10.1088/1126-6708/2006/06/051",
    journal = "JHEP",
    volume = "06",
    pages = "051",
    year = "2006"
}

@article{Wilczek:1977pj,
    author = "Wilczek, Frank",
    title = "{Problem of Strong  $P$  and  $T$  Invariance in the Presence of Instantons}",
    reportNumber = "Print-77-0939 (COLUMBIA)",
    doi = "10.1103/PhysRevLett.40.279",
    journal = "Phys. Rev. Lett.",
    volume = "40",
    pages = "279--282",
    year = "1978"
}

@article{CMS:2017caz,
    author = "Sirunyan, Albert M and others",
    collaboration = "CMS",
    title = "{Search for new physics with dijet angular distributions in proton-proton collisions at $ \sqrt{s}=13 $ TeV}",
    eprint = "1703.09986",
    archivePrefix = "arXiv",
    primaryClass = "hep-ex",
    reportNumber = "CMS-EXO-15-009, CERN-EP-2017-047",
    doi = "10.1007/JHEP07(2017)013",
    journal = "JHEP",
    volume = "07",
    pages = "013",
    year = "2017"
}

@article{Peccei:1977hh,
    author = "Peccei, R. D. and Quinn, Helen R.",
    title = "{CP Conservation in the Presence of Instantons}",
    reportNumber = "ITP-568-STANFORD",
    doi = "10.1103/PhysRevLett.38.1440",
    journal = "Phys. Rev. Lett.",
    volume = "38",
    pages = "1440--1443",
    year = "1977"
}

@article{Feng:2025kof,
    author = "Feng, Jiaojiao and Mao, Ying-nan and Wang, Kechen",
    title = "{Discovery prospects for photophobic axion-like particles in the $WWjj$ final state at the High-Luminosity LHC}",
    eprint = "2511.21003",
    archivePrefix = "arXiv",
    primaryClass = "hep-ph",
    month = "11",
    year = "2025"
}

@article{Bresciani:2025ojh,
    author = "Bresciani, Luigi C. and Levati, Gabriele and Paradisi, Paride",
    title = "{Positivity and partial wave unitarity bounds on ALP theories via amplitude methods}",
    eprint = "2510.13953",
    archivePrefix = "arXiv",
    primaryClass = "hep-ph",
    month = "10",
    year = "2025"
}

@article{Bonilla:2022pxu,
    author = "Bonilla, J. and Brivio, I. and Machado-Rodr{\'\i}guez, J. and de Troc{\'o}niz, J. F.",
    title = "{Nonresonant searches for axion-like particles in vector boson scattering processes at the LHC}",
    eprint = "2202.03450",
    archivePrefix = "arXiv",
    primaryClass = "hep-ph",
    reportNumber = "IFT-UAM/CSIC-22-7, VBSCAN-PUB-01-22",
    doi = "10.1007/JHEP06(2022)113",
    journal = "JHEP",
    volume = "06",
    pages = "113",
    year = "2022"
}

@article{CMS:2014mvm,
    author = "Chatrchyan, Serguei and others",
    collaboration = "CMS",
    title = "{Measurement of differential cross sections for the production of a pair of isolated photons in pp collisions at $\sqrt{s}=7\,\text {TeV} $}",
    eprint = "1405.7225",
    archivePrefix = "arXiv",
    primaryClass = "hep-ex",
    reportNumber = "CMS-SMP-13-001, CERN-PH-EP-2014-067",
    doi = "10.1140/epjc/s10052-014-3129-3",
    journal = "Eur. Phys. J. C",
    volume = "74",
    number = "11",
    pages = "3129",
    year = "2014"
}

@article{CMS:2021gme,
    author = "Tumasyan, Armen and others",
    collaboration = "CMS",
    title = "{Measurement of the electroweak production of Z$\gamma$ and two jets in proton-proton collisions at $\sqrt{s} =$ 13 TeV and constraints on anomalous quartic gauge couplings}",
    eprint = "2106.11082",
    archivePrefix = "arXiv",
    primaryClass = "hep-ex",
    reportNumber = "CMS-SMP-20-016, CERN-EP-2021-095",
    doi = "10.1103/PhysRevD.104.072001",
    journal = "Phys. Rev. D",
    volume = "104",
    pages = "072001",
    year = "2021"
}

@article{ATLAS:2023dew,
    author = "Aad, Georges and others",
    collaboration = "ATLAS",
    title = "{Measurement of ZZ production cross-sections in the four-lepton final state in pp collisions at s=13.6TeV with the ATLAS experiment}",
    eprint = "2311.09715",
    archivePrefix = "arXiv",
    primaryClass = "hep-ex",
    reportNumber = "CERN-EP-2023-235",
    doi = "10.1016/j.physletb.2024.138764",
    journal = "Phys. Lett. B",
    volume = "855",
    pages = "138764",
    year = "2024"
}

@article{CMS:2020gtj,
    author = "Sirunyan, Albert M and others",
    collaboration = "CMS",
    title = "{Measurements of ${\mathrm{p}} {\mathrm{p}} \rightarrow {\mathrm{Z}} {\mathrm{Z}} $ production cross sections and constraints on anomalous triple gauge couplings at $\sqrt{s} = 13\,\text {TeV} $}",
    eprint = "2009.01186",
    archivePrefix = "arXiv",
    primaryClass = "hep-ex",
    reportNumber = "CMS-SMP-19-001, CERN-EP-2020-145",
    doi = "10.1140/epjc/s10052-020-08817-8",
    journal = "Eur. Phys. J. C",
    volume = "81",
    number = "3",
    pages = "200",
    year = "2021"
}

@article{ATLAS:2021mbt,
    author = "Aad, Georges and others",
    collaboration = "ATLAS",
    title = "{Measurement of the production cross section of pairs of isolated photons in $pp$ collisions at 13 TeV with the ATLAS detector}",
    eprint = "2107.09330",
    archivePrefix = "arXiv",
    primaryClass = "hep-ex",
    reportNumber = "CERN-EP-2021-105",
    doi = "10.1007/JHEP11(2021)169",
    journal = "JHEP",
    volume = "11",
    pages = "169",
    year = "2021"
}

@article{ATLAS:2019rob,
    author = "Aaboud, Morad and others",
    collaboration = "ATLAS",
    title = "{Measurement of fiducial and differential $W^+W^-$ production cross-sections at $\sqrt{s}=13$  TeV with the ATLAS detector}",
    eprint = "1905.04242",
    archivePrefix = "arXiv",
    primaryClass = "hep-ex",
    reportNumber = "CERN-EP-2019-055",
    doi = "10.1140/epjc/s10052-019-7371-6",
    journal = "Eur. Phys. J. C",
    volume = "79",
    number = "10",
    pages = "884",
    year = "2019"
}

@article{CMS:2019jcb,
    author = "Sirunyan, Albert M and others",
    collaboration = "CMS",
    title = "{Evidence for $\text {W}\text {W}$ production from double-parton interactions in proton\textendash{}proton collisions at $\sqrt{s} = 13 \,\text {TeV} $}",
    eprint = "1909.06265",
    archivePrefix = "arXiv",
    primaryClass = "hep-ex",
    reportNumber = "CMS-SMP-18-015, CERN-EP-2019-167",
    doi = "10.1140/epjc/s10052-019-7541-6",
    journal = "Eur. Phys. J. C",
    volume = "80",
    number = "1",
    pages = "41",
    year = "2020"
}

@article{ATLAS:2017ble,
    author = "Aaboud, M. and others",
    collaboration = "ATLAS",
    title = "{Measurement of inclusive jet and dijet cross-sections in proton-proton collisions at $\sqrt{s}=13$ TeV with the ATLAS detector}",
    eprint = "1711.02692",
    archivePrefix = "arXiv",
    primaryClass = "hep-ex",
    reportNumber = "CERN-EP-2017-157",
    doi = "10.1007/JHEP05(2018)195",
    journal = "JHEP",
    volume = "05",
    pages = "195",
    year = "2018"
}

@article{CMS:2018vzn,
    author = "Sirunyan, Albert M. and others",
    collaboration = "CMS",
    title = "{Measurements of the differential jet cross section as a function of the jet mass in dijet events from proton-proton collisions at $ \sqrt{s}=13 $ TeV}",
    eprint = "1807.05974",
    archivePrefix = "arXiv",
    primaryClass = "hep-ex",
    reportNumber = "CMS-SMP-16-010, CERN-EP-2018-180",
    doi = "10.1007/JHEP11(2018)113",
    journal = "JHEP",
    volume = "11",
    pages = "113",
    year = "2018"
}

@article{CMS:2024hey,
    author = "Hayrapetyan, Aram and others",
    collaboration = "CMS",
    title = "{Measurement of inclusive and differential cross sections for W$^+$W$^-$ production in proton-proton collisions at $\sqrt{s}$ = 13.6 TeV}",
    eprint = "2406.05101",
    archivePrefix = "arXiv",
    primaryClass = "hep-ex",
    reportNumber = "CMS-SMP-24-001, CERN-EP-2024-129",
    month = "6",
    year = "2024"
}

@article{Brivio:2017ije,
    author = "Brivio, I. and Gavela, M. B. and Merlo, L. and Mimasu, K. and No, J. M. and del Rey, R. and Sanz, V.",
    title = "{ALPs Effective Field Theory and Collider Signatures}",
    eprint = "1701.05379",
    archivePrefix = "arXiv",
    primaryClass = "hep-ph",
    reportNumber = "IFT-UAM-CSIC-16-141, KCL-PH-TH-2016-72, FTUAM-16-49, CP3-17-04",
    doi = "10.1140/epjc/s10052-017-5111-3",
    journal = "Eur. Phys. J. C",
    volume = "77",
    number = "8",
    pages = "572",
    year = "2017"
}

@article{CMS:2018ucw,
    author = "Sirunyan, Albert M and others",
    collaboration = "CMS",
    title = "{Search for new physics in dijet angular distributions using proton\textendash{}proton collisions at $\sqrt{s}=$ 13 TeV and constraints on dark matter and other models}",
    eprint = "1803.08030",
    archivePrefix = "arXiv",
    primaryClass = "hep-ex",
    reportNumber = "CMS-EXO-16-046, CERN-EP-2018-036",
    doi = "10.1140/epjc/s10052-018-6242-x",
    journal = "Eur. Phys. J. C",
    volume = "78",
    number = "9",
    pages = "789",
    year = "2018",
    note = "[Erratum: Eur.Phys.J.C 82, 379 (2022)]"
}

@article{ATLAS:2025omi,
    author = "Aad, Georges and others",
    collaboration = "ATLAS",
    title = "{Electroweak diboson production in association with a high-mass dijet system in semileptonic final states from $pp$ collisions at $\sqrt{s} = 13$ TeV with the ATLAS detector}",
    eprint = "2503.17461",
    archivePrefix = "arXiv",
    primaryClass = "hep-ex",
    reportNumber = "CERN-EP-2025-050",
    month = "3",
    year = "2025"
}

@article{NNPDF:2021uiq,
    author = "Ball, Richard D. and others",
    collaboration = "NNPDF",
    title = "{An open-source machine learning framework for global analyses of parton distributions}",
    eprint = "2109.02671",
    archivePrefix = "arXiv",
    primaryClass = "hep-ph",
    reportNumber = "Edinburgh 2021/13, Nikhef-2021-020, TIF-UNIMI-2021-12",
    doi = "10.1140/epjc/s10052-021-09747-9",
    journal = "Eur. Phys. J. C",
    volume = "81",
    number = "10",
    pages = "958",
    year = "2021"
}

@article{NNPDF:2021njg,
    author = "Ball, Richard D. and others",
    collaboration = "NNPDF",
    title = "{The path to proton structure at 1{\%} accuracy}",
    eprint = "2109.02653",
    archivePrefix = "arXiv",
    primaryClass = "hep-ph",
    reportNumber = "Edinburgh 2021/12, Nikhef-2021-013, TIF-UNIMI-2021-11",
    doi = "10.1140/epjc/s10052-022-10328-7",
    journal = "Eur. Phys. J. C",
    volume = "82",
    number = "5",
    pages = "428",
    year = "2022"
}

@article{Britzger:2022lbf,
    author = "Britzger, D. and others",
    title = "{NNLO interpolation grids for jet production at the LHC}",
    eprint = "2207.13735",
    archivePrefix = "arXiv",
    primaryClass = "hep-ph",
    reportNumber = "CERN-TH-2022-125, IPPP/22/53, MPP-2022-80, ZU-TH 34/22",
    doi = "10.1140/epjc/s10052-022-10880-2",
    journal = "Eur. Phys. J. C",
    volume = "82",
    number = "10",
    pages = "930",
    year = "2022"
}

@article{Carrazza:2020gss,
    author = "Carrazza, S. and Nocera, E. R. and Schwan, C. and Zaro, M.",
    title = "{PineAPPL: combining EW and QCD corrections for fast evaluation of LHC processes}",
    eprint = "2008.12789",
    archivePrefix = "arXiv",
    primaryClass = "hep-ph",
    doi = "10.1007/JHEP12(2020)108",
    journal = "JHEP",
    volume = "12",
    pages = "108",
    year = "2020"
}

@article{Schwan:2021txc,
    author = "Schwan, Christopher",
    title = "{PineAPPL: NLO EW corrections for PDF processes}",
    eprint = "2108.05816",
    archivePrefix = "arXiv",
    primaryClass = "hep-ph",
    doi = "10.21468/SciPostPhysProc.8.079",
    journal = "SciPost Phys. Proc.",
    volume = "8",
    pages = "079",
    year = "2022"
}

@article{NNPDF:2019ubu,
    author = "Abdul Khalek, Rabah and others",
    collaboration = "NNPDF",
    title = "{Parton Distributions with Theory Uncertainties: General Formalism and First Phenomenological Studies}",
    eprint = "1906.10698",
    archivePrefix = "arXiv",
    primaryClass = "hep-ph",
    reportNumber = "Edinburgh 2019/9, Nikhef/2019-014, TIF-UNIMI-2019-9 DAMTP-2019-24,
  CAVENDISH-HEP-19-11",
    doi = "10.1140/epjc/s10052-019-7401-4",
    journal = "Eur. Phys. J. C",
    volume = "79",
    number = "11",
    pages = "931",
    year = "2019"
}

@article{Ball:2021icz,
    author = "Ball, Richard D. and Pearson, Rosalyn L.",
    title = "{Correlation of theoretical uncertainties in PDF fits and theoretical uncertainties in predictions}",
    eprint = "2105.05114",
    archivePrefix = "arXiv",
    primaryClass = "hep-ph",
    reportNumber = "Edinburgh 2019/17",
    doi = "10.1140/epjc/s10052-021-09602-x",
    journal = "Eur. Phys. J. C",
    volume = "81",
    number = "9",
    pages = "830",
    year = "2021"
}

@misc{Ploughshareurl,
     doi = {\href{https://ploughshare.web.cern.ch/ploughshare/}{https://ploughshare.web.cern.ch/ploughshare/}},
}

@article{Chiefa:2025loi,
    author = "Chiefa, Amedeo and Costantini, Mark N. and Cruz-Martinez, Juan and Nocera, Emanuele R. and Rabemananjara, Tanjona R. and Rojo, Juan and Sharma, Tanishq and Stegeman, Roy and Ubiali, Maria",
    title = "{Parton distributions confront LHC Run II data: a quantitative appraisal}",
    eprint = "2501.10359",
    archivePrefix = "arXiv",
    primaryClass = "hep-ph",
    reportNumber = "Nikhef 2025-002, Edinburgh 2024/10, CERN-TH-2024-16, CERN-TH-2024-169, CERN-TH-2024-169",
    doi = "10.1007/JHEP07(2025)067",
    journal = "JHEP",
    volume = "07",
    pages = "067",
    year = "2025"
}

@article{Chen:2022tpk,
    author = "Chen, X. and Gehrmann, T. and Glover, E. W. N. and Huss, A. and Mo, J.",
    title = "{NNLO QCD corrections in full colour for jet production observables at the LHC}",
    eprint = "2204.10173",
    archivePrefix = "arXiv",
    primaryClass = "hep-ph",
    reportNumber = "ZU-TH 11/22, KA-TP-07-2022, IPPP/22/20, P3H-22-037, CERN-TH-2022-067",
    doi = "10.1007/JHEP09(2022)025",
    journal = "JHEP",
    volume = "09",
    pages = "025",
    year = "2022"
}

@article{Gehrmann-DeRidder:2019ibf,
    author = "Gehrmann-De Ridder, A. and Gehrmann, T. and Glover, E. W. N. and Huss, A. and Pires, J.",
    title = "{Triple Differential Dijet Cross Section at the LHC}",
    eprint = "1905.09047",
    archivePrefix = "arXiv",
    primaryClass = "hep-ph",
    reportNumber = "CERN-TH-2019-054, IPPP/19/31, ZU-TH 19/19, CFTP/19-016",
    doi = "10.1103/PhysRevLett.123.102001",
    journal = "Phys. Rev. Lett.",
    volume = "123",
    number = "10",
    pages = "102001",
    year = "2019"
}

@article{NNLOJET:2025rno,
    author = "Huss, A. and others",
    collaboration = "NNLOJET",
    title = "{NNLOJET: a parton-level event generator for jet cross sections at NNLO QCD accuracy}",
    eprint = "2503.22804",
    archivePrefix = "arXiv",
    primaryClass = "hep-ph",
    reportNumber = "CERN-TH-2025-012, IPPP/25/09, ZU-TH 11/25",
    month = "3",
    year = "2025"
}

@article{Alwall:2014hca,
    author = "Alwall, J. and Frederix, R. and Frixione, S. and Hirschi, V. and Maltoni, F. and Mattelaer, O. and Shao, H. -S. and Stelzer, T. and Torrielli, P. and Zaro, M.",
    title = "{The automated computation of tree-level and next-to-leading order differential cross sections, and their matching to parton shower simulations}",
    eprint = "1405.0301",
    archivePrefix = "arXiv",
    primaryClass = "hep-ph",
    reportNumber = "CERN-PH-TH-2014-064, CP3-14-18, LPN14-066, MCNET-14-09, ZU-TH-14-14",
    doi = "10.1007/JHEP07(2014)079",
    journal = "JHEP",
    volume = "07",
    pages = "079",
    year = "2014"
}

@misc{ALPUFO,
  title = {\texttt{ALP\_linear\_UFO}},
  author = "I. Brivio",
  howpublished = {\url{http://feynrules.irmp.ucl.ac.be/attachment/wiki/ALPsEFT/ALP_linear_UFO.tar.gz}},
  note = {Accessed: 2022-11-30}
}

@article{Bisal:2025jwv,
    author = "Bisal, Subhadip",
    title = "{Constraining ALP-Top Interaction from the Chromoelectric Dipole Moment of the Top Quark}",
    eprint = "2507.12570",
    archivePrefix = "arXiv",
    primaryClass = "hep-ph",
    month = "7",
    year = "2025"
}

@article{Barbosa:2025zyn,
    author = "Barbosa, Sergio and Coelho, Matheus and Fichet, Sylvain and da Silveira, Gustavo Gil and Machado, Magno",
    title = "{LHC as an Axion-Photon Collider}",
    eprint = "2506.10066",
    archivePrefix = "arXiv",
    primaryClass = "hep-ph",
    doi = "10.1103/t1rl-4926",
    journal = "Phys. Rev. Lett.",
    volume = "135",
    number = "18",
    pages = "181801",
    year = "2025"
}
\end{document}